\documentclass{jfm}
\usepackage{graphicx}
\usepackage{epstopdf, epsfig}
\usepackage{amssymb}
\usepackage{amsmath,bm}

\usepackage{color, soul} 
\graphicspath{/figures/} 
\DeclareGraphicsExtensions{.eps,.pdf,.png,.jpg,.jpeg,.xcf}
\usepackage[justification=centering]{caption}
\usepackage{subcaption}
\usepackage{placeins}

\usepackage{adjustbox}
\usepackage{multirow}
\usepackage{algpseudocode,algorithm2e,algorithmicx}
\usepackage{array}
\newcolumntype{P}[1]{>{\centering\arraybackslash}p{#1}}
\usepackage{cleveref}

\shorttitle{Stability of a hyperbolic critical point and three dimensionality}
\shortauthor{B. Liu and A. R. Magee}
\title{Numerical stability and three dimensionality of a streamline hyperbolic critical point in wake at low Reynolds number}
\author{B. Liu\aff{1}
	\corresp{\email{a0098961@u.nus.edu}}
	and
	A. R. Magee\aff{2}
}
\affiliation{\aff{1}Department of Mechanical Engineering, National University of Singapore, Singapore 119077, Singapore
	\aff{2}Department of Civil and Environmental Engineering, National University of Singapore, Singapore 117576, Singapore
}

\begin{document}

\maketitle

\begin{abstract}
In this article the numerical stability of a streamline singular hyperbolic/saddle critical point (HSP) and its relationship with the divergence of pressure force/fluid flux are numerically investigated at low Reynolds numbers. Three canonical configurations at different Reynolds numbers are considered: (a) an isolated cylinder; (b) a pair of side-by-side cylinders and (c) a near-wall cylinder. In Fourier stability analysis, it is proven that a HSP is intrinsically meta-stable for a balanced shear layer interaction and numerically stable for a range of spatial and temporal discretization. It is also found that a HSP forms along the (vorticity-free) shear-layer interfaces and imposes adverse pressure gradients in wake, which potentially deteriorates the near-wake stability. Furthermore, a HSP is found intrinsically associated with a net positive value of Poisson Pressure Equation (PPE), the fluid-flux divergence, and the fluid three dimensionality. A vorticity-free stagnant zone is formed around a HSP, which cuts the kinetic energy supply of shear layers in wake, projects third-dimensional fluid fluxes and develops three-dimensional streamwise braids. These findings are confirmed and explained via the quantification of the fluid-flux divergence, the hydrodynamic responses of cylinder(s) and the secondary enstrophy. The primary focus in this article is to analytically establish the subtle relationship between HSP, imbalanced shear-layer interaction, Reynolds number, fluid-flux divergence and fluid three dimensionality. To the knowledge of authors, so far these analytical relationships have not been reported in literature a prior.
\end{abstract}

\begin{keywords}
Instability, Vortex Flows, Boundary Layers, Turbulent Flows
\end{keywords}

\section{Introduction} \label{sec:intro}
It was in 1912 that~\cite{von1912uber} analyzed the stability of vortex configurations and established a theoretical link between the vortex street structure and the drag observed by the bluff body. By investigating the linear stability of pint vortex configuration, he concluded that two rows of oppositely-signed vortices were unstable in both symmetric and anti-symmetric configurations except for one specific anti-symmetric geometry which exhibited neutral stability. The shear-layer interaction is important in the analysis of hydrodynamic instabilities and other aspects of fluid mechanics. Due to its presence, the fluid dynamics becomes splendid and unpredictable. The shear layers are sensitive to non-conformity in flow field, extend these influences farther away and exhibit complex mutual interference.

In classical temporal hydrodynamic stability theorems, e.g., Kelvin-Helmholtz instability, Rayleigh's inflection point theorem and Couette-Taylor centrifugal instability, the roles of shear-layer interaction, inflectional velocity profile and the associated adverse pressure gradients are explicitly emphasized. The essence is rooted in the fluid incapability of statically sustaining shearing and curling in fluid flow, while subjecting to a non-uniform velocity field, e.g., boundary layer or shear-layer interaction. Particularly, Kelvin Helmholtz instability states that the disturbances accumulates around the inflection point of velocity profiles along the shear-layer interface, which is critical to hydrodynamic stability. Recently,~\cite{liu2018dynamics} reported that the inflectional velocity profiles, $\partial v /\partial x$ and $\partial u /\partial y$, across a HSP simultaneously. It implies the significance of a HSP in the analysis of near-wake stability behind a bluff body.~\cite{liu2018dynamics} further noticed that a HSP forms along shear-layer interface and generate third-dimensional flux via observations of velocity profiles and vorticity contours. In this article, the focus is to analytically prove the relationship between a HSP, the shear-layer interaction and third-dimensional fluid flux.   

Solutions around a singularity are normally in-determinant and irreversible. It is also noticed the formation of singularities often arise when unexpected interactions occur between systems. It is widely believed the configuration, which most likely leads to a singularity, consists of two interacting non-parallel vortex tubes, as documented by~\cite{beale1984remarks}.~\cite{liu2016interaction} reported that a HSP appears in the gap-flow middle path between side-by-side cylinders right at the moment of the gap flow switches its sides, the gap-flow flip flop (the pitchfork bifurcation of a symmetric system). Recently,~\cite{liu2018dynamics} also observed that asymmetric shear-layer interaction is detrimental to the stability of a HSP and induces third-dimensional flow structure earlier in the gap flow. The relationship between fluid stability and a (singular) hyperbolic/saddle critical point had been discussed as early as the topological investigation of turbulent flow pattern by~\cite{perry1987description}. They documented that the streamwise vortex rollers are formed long the interfaces of the counter-signed interacting shear-layers and across the saddle point regions. A high streamwise vorticity concentration of the same sign appears on both sides of the saddle point along the $z$-axis, which is consistent with the observation of \cite{zhou1994critical,liu2018dynamics}, whereby the streamwise vortical structures are inclined and crossed approximately at the saddle point region in $x$-$y$ plane. 

The aforementioned findings suggest that the presence of a HSP is intrinsically linked with the progress of flow transition and fluid three dimensionality. However, the laminar-to-turbulent transition is a hysteresis and discontinuous process. It can occur through several mechanisms and stages, e.g., natural transition, bypass transition,  G\"ortler instability and cross-flow instability.~\cite{williamson1996three} emphasized the presence of the singular HSP in wake and explained its key role in flow transition and elliptical instability in terms of three dimensionality of flow. In the context of three dimensionality associated with the elliptical instability, the hyperbolic critical points were observed around the interacting vortices~\cite{kerswell2002elliptical,le2002theoretical,meunier2005physics}. Furthermore, the hyperbolic critical points in the fluid domain had been previously reported as an unstable factor by~\cite{lifschitz1991local,leblanc1997stability,ryan2005three}, where the maximum perturbation growth was found precisely around these critical points in the vortex wake. More recently, ~\cite{liu2018dynamics} reported that the inflectional velocity profiles ($\partial v /\partial x$ and $\partial u/\partial y$) across a HSP simultaneously. In accordance to Kelvin Helmholtz instability, the disturbances will accumulate around the inflectional points of velocity profiles. Hence this finding agrees well with the aforementioned observations of a HSP, in which the disturbances are accumulating around inflectional points. In this article, we would like to analytically explain the mechanisms of a HSP and its relationship with fluid instabilities, e.g, vortex shedding and fluid three dimensionality. 

The subjects of flow transition and three dimensionality are broad. In this article, we would like to merely focus on the characteristics of a singular hyperbolic critical point in wake. We would like to address its analytical relationship with the asymmetric shear-layer interaction, fluid-flux divergence and three dimensionality. Recently,~\cite{huang2010oblique,liu2018dynamics} reported that a planar shear flow could enhance the fluid three dimensionality in wake behind a cylinder, where asymmetric shear-layer interaction is naturally induced in wake. Meanwhile~\cite{zhao2014three,liu2018dynamics} also observed a remarkable suppression of three-dimensional flow structure during VIV lock-in of an elastically mounted rigid cylinder at moderate Reynolds numbers $Re \in [300, 500]$, in which the flow feature is complete two-dimensional and no singular HSP is observed in wake. Their findings inspired the research in this article to analytically explain their mechanisms. To this end, we would like to address the following questions:\\
\begin{itemize}
	\item How does an asymmetric shear-layer interaction influence the stability of a HSP?
	\item How does Reynolds number effect influence the stability of a HSP?
	\item What is the relationship between a HSP and the net positive fluid-flux divergence?
	\item How does a HSP induce adverse pressure gradients and influence near-wake stability?
	\item How does a HSP induce the third-dimensional fluid flux in flow transition?
\end{itemize}
\hfill

The structure of this article is organized as follow. The governing equation and numerical formulations are presented first in Section~\ref{sec:goven}. Subsequently, the problem setup, key parameters and validation of numerical formulations are presented in Section~\ref{sec:problemValidation}. Following that, the numerical stability of a HSP and its relationship with Poisson Pressure Equation (PPE) are analytically proven in Section~\ref{sec:prove}. The numerical results are discussed in Section~\ref{sec:results} to support the conclusions drawn in previous sections. Finally, a concluding remark is presented in Section~\ref{sec:con} .

\section{Governing equation and numerical formulations} \label{sec:goven}

\subsection{Governing equation and its variational form}
In this article, the primary focus is to investigate the stability of a HSP in the incompressible, isothermal and Newtonian fluid flow, which is governed by the unsteady incompressible Navier-Stokes equation in Equation~\eqref{eq:ns}. 
\begin{subequations} \label{eq:ns}
	\begin{align}
	\rho \big(\partial_t(\bm{u}) + \bm{u} \cdot \nabla \bm{u} \big) - \nabla \cdot \bm{\sigma}\{\bm{u},p\} &= \rho \bm{g} \qquad \forall \bm{x} \in \Omega(t) \\
	\nabla \cdot \bm{u} &= 0 \quad \quad \;\;\forall \bm{x} \in \Omega(t) \\
	\bm{u} &= \bm{\tilde{u}}  \quad \;\;\;\;\; \forall \bm{x} \in \Gamma_D (t) \\
	\bm{\sigma}\{\bm{u},p\}\cdot\bm{n} &= \bm{\tilde{\tau}} \;\;\;\;\;\quad \forall \bm{x} \in \Gamma_N (t) \\
	\bm{u} &= \bm{u}_0 \quad \;\;\; \forall \bm{x} \in \Omega(0)
	\end{align}
\end{subequations}
where $\rho$, $\bm{u}$, $\bm{u}_0$ and $\bm{g}$ are respectively the fluid density, the fluid velocity vector, the initial fluid velocity vector and the unit body force vector of fluid. $\bm{\tilde{u}}$, $\bm{\tilde{\tau}}$ and $\bm{n}$ are respectively the prescribed fluid velocity, the prescribed fluid traction force and the outward normal vector of fluid domain. Based on the kinematics of continuum mechanics, the material time derivative $D_t (\rho \bm{u})$, which represents the kinematics of a fluid particle in Lagrangian description, in Navier Stokes equation can be formulated in terms of the spatial time derivative $\partial_t (\rho \bm{u})$ and the fluid flux term $\rho \bm{u}\cdot \nabla \bm{u}$ in Eulerian description. In Section~\ref{sec:ppe}, we will subsequently prove that a HSP is associated with the net positive divergence of pressure force/fluid flux, $ \nabla\cdot (\rho \bm{u}\cdot \nabla \bm{u})$. $\Omega$ and $\Gamma$ refer to the computational domain and computational boundary respectively, e.g., the Dirichlet boundary condition in Eq.~\eqref{eq:bc_dir} and Neumann boundary condition in Eq.~\eqref{eq:bc_neu} are applied along $\Gamma_D$ and $\Gamma_N$ respectively. $\Gamma_D$ and $\Gamma_N$ are complementary subsets of $\Gamma$, $\Gamma = \Gamma_D \cup \Gamma_N$ and $\Gamma_D \cap \Gamma_N = \emptyset$. 

\begin{subequations}
	\begin{align}
	\bm{u} &= \bm{\tilde{u}} \quad \forall \bm{x} \in \Gamma_D (t) \label{eq:bc_dir} \\
	\bm{\tau} &= \bm{\tilde{\tau}} \quad \forall \bm{x} \in \Gamma_N (t) \label{eq:bc_neu}
	\end{align}
\end{subequations}
where $\bm{\tau} = \bm{\sigma}\{\bm{u},p\} \cdot \bm{n} = [\tau_x, \tau_y, \tau_z]^T$ is the surface stress. $\bm{\sigma}\{\bm{u},p\}$ is the Cauchy stress tensor and defined as 

\begin{eqnarray}
\bm{\sigma} \{\bm{u},p\} &=& -p \bm{I} + 2 \mu D(\bm{u}) \\
D(\bm{u}) &=& \frac{1}{2} \big[\nabla \bm{u} + (\nabla \bm{u})^T\big] \nonumber
\end{eqnarray}
where p, $\mu$ and $\bm{I}$ are the fluid pressure, the dynamic viscosity of fluid and an identity matrix respectively. The superscript $T$ is a transpose operator. The stress tensor is expressed as a combination of its isotropic and deviatoric tensor ($D(\bm{u})$) parts.

The variational form of the stabilized finite element formulation of the Navier-Stokes equation in Equation~\eqref{eq:ns} is derived in Equation~\eqref{eq:weak}. 
\begin{eqnarray}
& &\underbrace{(\bm{\psi}_u,\rho\partial_t (\bm{u}))_{\Omega} + (\bm{\psi}_u,\rho\bm{u}\cdot \nabla \bm{u})_{\Omega} + (D(\bm{\psi}_u), \bm{\sigma})_{\Omega}- \langle \bm{\psi}_u, \bm{\sigma} \cdot \bm{n} \rangle_{\Gamma_N}}_{\mathcal{B}_{NS}(\{[\bm{\psi}_u, \psi_p],[\bm{u},p]\})} \nonumber\\
& &\underbrace{- (\bm{\psi}_u,\rho \bm{g})_{\Omega} + (\psi_p,\nabla \cdot \bm{u})_{\Omega}}_{\mathcal{B}_{NS}(\{[\bm{\psi}_u, \psi_p],[\bm{u},p]\})} \underbrace{-\langle \bm{\psi}_u, -\mu (\nabla \bm{u})^T \cdot \bm{n} \rangle_{\Gamma_{out}}}_{\mathcal{B}_{corr}(\{[\bm{\psi}_u, \psi_p],[\bm{u},p]\})} \nonumber\\ 
& &\underbrace{+\sum \limits^{n_{el}}_{e=1}\tau_m(\rho \bm{u}\cdot \nabla \bm{\psi}_u - \mu \nabla^2 \bm{\psi}_u + \nabla \psi_p, \hat{\partial}_t (\rho\bm{u}) + \rho\bm{u}\cdot \nabla \bm{u}-\rho\bm{g} }_{\mathcal{B}_{stab}(\{[\bm{\psi}_u, \psi_p],[\bm{u},p]\})} \nonumber\\
& &\underbrace{-\mu \nabla^2 \bm{u} + \nabla p)_{\Omega} + \sum \limits^{n_{el}}_{e=1} \tau_c( \rho \nabla \cdot \bm{\psi}_u, \nabla \cdot \bm{u})_{\Omega}}_{\mathcal{B}_{stab}(\{[\bm{\psi}_u, \psi_p],[\bm{u},p]\})} \nonumber\\ 
& &= 0 \quad \forall [\bm{\psi}_u,\psi_p] \in \hat{\bm{\mathcal{V}}}_h \times \hat{\mathcal{Q}}_h \subset \hat{\bm{\mathcal{V}}} \times \hat{\mathcal{Q}}  \label{eq:weak}
\end{eqnarray}
where $[\bm{\psi}_u, \psi_p]^T$ is the vector of test functions for the velocity and the pressure of fluid. The operators $(\cdot)_{\Omega}$ and $\langle \cdot \rangle_{\Gamma}$ respectively refer to the domain integral and the boundary integral. The correction term $\mathcal{B}_{corr}(\{[\bm{\psi}_u, \psi_p],[\bm{u},p]\})$ proposed by~\cite{Heywood1996IJfnmif} is applied along the true "do-nothing" outflow boundary to eliminate the numerical flux, while the symmetric stress tensor $\bm{\sigma} = -p \bm{I} + 2\mu D(\bm{u})$ is used. 
A residual-based stabilization technique, Petrov-Galerkin method~\citep{Brooks1982Cmiamae,Shakib1991CMiAMaE,Tezduyar1992CMiAMaEa,Franca1992CMiAMaE}, the term $\mathcal{B}_{stab}(\{[\bm{\psi}_u, \psi_p],[\bm{u},p]\})$ in Eq.~\eqref{eq:weak}, is implemented to minimize the residual of the equation system in a weak/integral sense and ensure the equal approximation function space for the velocity and pressure.
The stabilization parameter, $\tau_m$ and $\tau_c$ are defined as 
\begin{subequations}
	\begin{align}
	\tau_m &= \big[\Big(\frac{2 \rho}{\Delta t}\Big)^2 + (\rho)^2 \bm{u} \cdot \bm{G} \bm{u} + C_I (\mu)^2 \bm{G}:\bm{G}\big]^{-0.5} \\
	\tau_c &= \big(tr(\bm{G})\tau_m\big)^{-1}; \quad \bm{G} = \Big(\frac{\partial \bm{\zeta}}{\partial \bm{x}}\Big)^T\frac{\partial \bm{\zeta}}{\partial \bm{x}}
	\end{align}
\end{subequations}
where $\bm{G}$ and $C_I$ are respectively element cotravariant metric tensor and a positive constant independent upon mesh size~\cite{harari1992c}. $\bm{\zeta}$ is the natural coordinates. The vector-valued trial and test function spaces $\boldsymbol{\mathcal{V}}$ and $\hat{\boldsymbol{\mathcal{V}}}$ of velocity are defined as 
\begin{eqnarray}
\boldsymbol{\mathcal{V}} &=& \{\bm{\psi}_u \in \bm{\mathcal{H}}^1 (\Omega(t)): \bm{\psi}_u=\tilde{\bm{\psi}}_u \quad \forall \bm{x} \in \Gamma_D(t)\} \nonumber\\
\hat{\boldsymbol{\mathcal{V}}} &=& \{\bm{\psi}_u  \in \bm{\mathcal{H}}^1 (\Omega(t)): \bm{\psi}_u =\mathbf{0} \quad\;\; \forall \bm{x} \in \Gamma_D(t) \}
\end{eqnarray} 
On the other hand, the scalar-valued trial and test function spaces $\mathcal{Q}$ and $\hat{\mathcal{Q}}$ of pressure are defined as 
\begin{eqnarray}
\mathcal{Q} &=& \{\psi_p \in \mathcal{H}^1 (\Omega(t)): \psi_p=\tilde{\psi}_p \quad \forall \bm{x} \in \Gamma_D(t)\} \nonumber\\
\hat{\mathcal{Q}} &=& \{\psi_p \in \mathcal{H}^1 (\Omega(t)): \psi_p=0 \quad \forall \bm{x} \in \Gamma_D(t) \}
\end{eqnarray} 
where $\boldsymbol{\mathcal{H}}^1$ is the Sobolev space, in which $[(\bm{\psi}_u)^2,\psi_p^2]$ and  $[|\nabla\bm{\psi}_u|^2,|\nabla \psi_p|^2]$ have finite integrals within $\Omega(t)$ and allows discontinuous derivatives. Their corresponding discrete function spaces are denoted with subscript ($h$), e.g., $\hat{\mathcal{Q}}_h$. 

In this formulation, a body-fitted fitted mesh is employed to precisely align the interface between the fluid and structural domains, $\Gamma^{fsi}$. The objective to ensure accurate approximation of boundary layer dynamics and facilitate the investigations of hydrodynamic stability and flow transition around a HSP in wake. The basic first-order Euler temporal integration is implemented, which is consistent with the numerical formulation in the Fourier stability analysis in Section~\ref{sec:vonNeumman}. The detailed setups of the computational domain, the mesh and the key parameters are presented in Section~\ref{sec:problemValidation}.

\section{Problem setup and validation} \label{sec:problemValidation}
The stability of a HSP in wake and its relationship with flow transition are investigated in three canonical configurations: (a) an isolated cylinder, (b) a pair of side-by-side cylinders and (c) a near-wall cylinder. In the wake behind a cylinder, the shear-layer interaction is prominent and intensified. Most of the time, the shear layers are asymmetric in the wake and interact in complex patterns, e.g., after the onset of vortex shedding, flow transition and gap-flow instabilities. In the cases of an isolated cylinder, we are particularly interested in analysis of Reynolds number effect and flow transition. Hence, a three-dimensional computational domain is employed in the case of an isolated cylinder to investigate the three dimensionality of a HSP. It will be shown later on in Section~\ref{sec:vonNeumman} that the asymmetric shear-layer interaction, which is prominent in the side-by-side and near-wall arrangements, deteriorates the stability of a HSP in wake.  
\subsection{Problem setup and key parameters} \label{sec:problem}
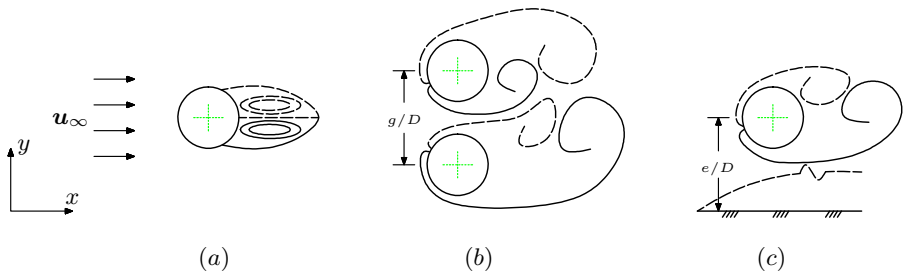
\begin{figure}
	\centering
	\includegraphics[trim=0.1cm 10cm 1cm 13cm,scale=0.6,clip]{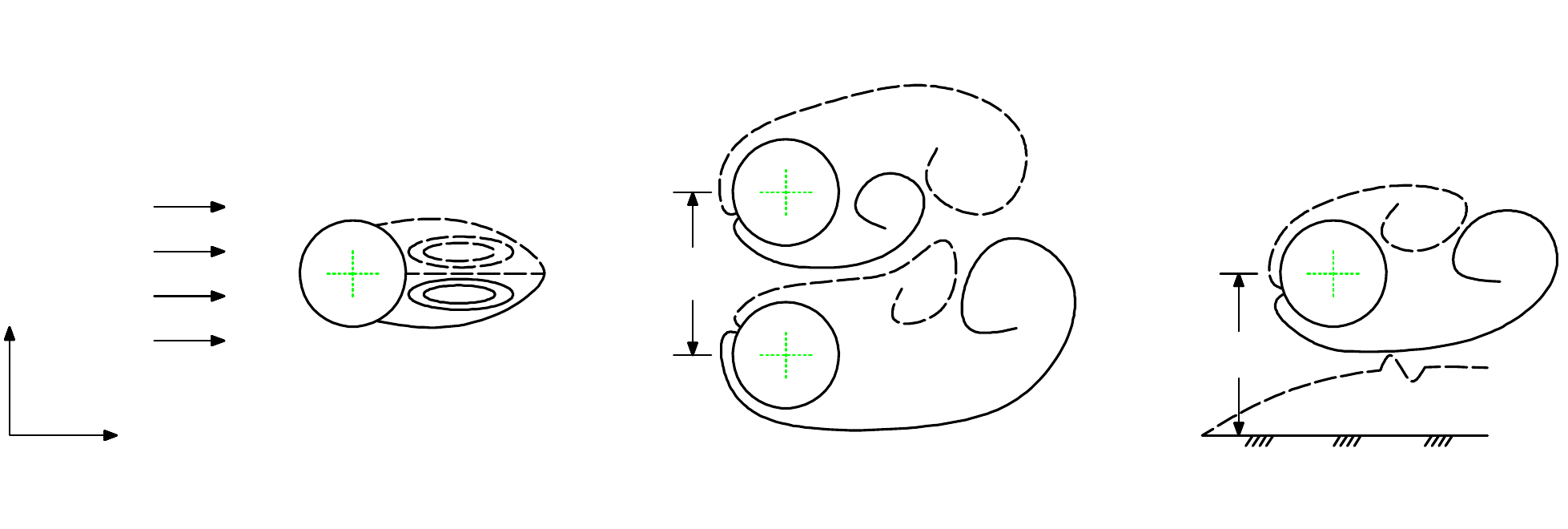}
	\begin{picture}(0,0)
	\put(-270,0){$(a)$}	
	\put(-170,0){$(b)$}
	\put(-60,0){$(c)$}
	\put(-320,23){$x$}
	\put(-338,43){$y$}
	\put(-325,53){$\bm{u}_{\infty}$}
	\put(-199,53){\tiny$g/D$}
	\put(-81,35){\tiny$e/D$}
	\end{picture}
	\caption{Schematic diagrams of the canonical arrangements of circular cylinder(s): (a) an isolated cylinder; (b) a pair of side-by-side cylinders; (c) a near-wall cylinder}
	\label{fig:schem}
\end{figure}
\begin{table*}
	\setlength{\tabcolsep}{12pt}
	\renewcommand{\arraystretch}{2}
	\begin{center}
		\begin{tabular}{l l l}
			\hline
			Parameter & Value & Description\\
			\hline
			$D$ & 1.0 & Diameter of cylinder\\
			$L_{u}/D$ & 10.0-15.0 & Upstream distance\\
			$L_{d}/D$ & 40.0-50.0 & Downstream distance\\
			$H/D$ & 50.0 & Height of domain\\
			$L/D$ & 10.0 & Spanwise length of cylinder\\
			$g/D$ & 0.2-5.0 & Center-to-center gap distance\\
			$e/D$ & 0.2-6.0 & Center-to-wall gap distance\\ 
			$Re = \frac{\rho U D}{\mu}$ & 20.0-500.0 & Reynolds number\\
			$C_d = \frac{\int \tau_x d\Gamma}{0.5 \rho U^2 D L}$ & -- & Drag coefficient\\
			$C_l = \frac{\int \tau_y d\Gamma}{0.5 \rho U^2 D L}$ & -- & Lift coefficient\\
			$\xi_i (t) = \frac{D^2}{U^2V}\int \omega^2_i (\boldsymbol{x},t) dV$ & -- & Enstrophy\\
			$-\nabla^2 p = \rho \nabla \cdot (\bm{u} \cdot \nabla \bm{u})$ & -- & Divergence of pressure force/fluid flux\\
			\hline
		\end{tabular}
		\caption{The key dimensionless parameters in the problem setup and simulations, where $\omega_i$ and $V$ are the i-component vorticity and integration volume respectively.}
		\label{tab:para}
	\end{center}
\end{table*}
\begin{table*}
	\setlength{\tabcolsep}{12pt}
	\renewcommand{\arraystretch}{2}
	\begin{center}
		\def\arraystretch{1.0}
		\begin{tabular}{l p{1.5cm} p{1.5cm} p{1.5cm}}
			\hline
			Number of grids & $C^{m}_{d}$ &  $C^{rms}_{l}$ & $St$ \\
			\hline
			$1.4$ $\times$ $10^5$ ($\Delta z = 0.40$) & 1.391 (10.9\%) &  0.961 (101.5\%) & 0.219 (6.8\%)\\
			$5.5$ $\times$ $10^5$ ($\Delta z = 0.15$) &  1.261 (0.6\%)&  0.484 (1.5\%)& 0.205 (0.0\%)\\
			$1.3$ $\times$ $10^6$ ($\Delta z = 0.10$) & 1.254 & 0.477 &  0.205\\
			\hline
		\end{tabular}
	\end{center}
	\caption{Hydrodynamic responses of a circular circular cylinder at $Re=300$, $\Delta t = 0.05$ and different mesh resolutions, where $C^m_d$, $C^{rms}_l$ and $St$ respectively are the mean drag coefficient, the root-mean-squared lift coefficient and the Strouhal number.}
	\label{tab:mesh_con}
\end{table*}
The investigation is first carried out for the cases of flow over an isolated circular cylinder, as shown in Figure~\ref{fig:schem}(a). The free-stream flows from the left of the computational domain to the right at a constant velocity, $\bm{u}_{\infty} = [U, 0.0, 0.0]^T$, where $U = 1.0$. The diameter of cylinder is taken as the characteristic length $D=1.0$. The upstream ($L_{u}$), downstream distance ($L_d$) and the height of the computational domain ($H$) respectively are $10D$, $40D$ and $50D$ to minimize the influence of artificially-imposed boundary conditions.  The cylinder is situated at the location $(0D, 0D)$, where is $10D$ downstream the inlet. The traction-free boundary condition is imposed along the computational boundaries, except the inlet. A reference pressure value $p=0$ is pinned at the outlet. 
Resemble to the configuration of an isolated circular cylinder in Figure~\ref{fig:schem}(a), two identical cylinders are respectively placed side-by-side at locations $[0.0,0.5g/D]$ and $[0.0,-0.5g/D]$ in Figure~\ref{fig:schem}(b), where are $10D$ downstream the inlet. The value of $g/D$ is the dimensionless center-to-center distance between cylinders. In Figure~\ref{fig:schem}(b), the values of the upstream, the downstream and the height of domain are identical to those of an isolated cylinder in Figure~\ref{fig:schem}(a). All boundaries of the computational domain are imposed with the traction-free boundary condition, except the inlet. A reference pressure $p=0.0$ is also pinned at the outlet.

In the configuration of a near-wall cylinder, the flat boundary beneath the cylinder is imposed with "no-slip" boundary condition, where the wall boundary layer develops from the upstream to the downstream, as shown in Figure~\ref{fig:schem}(c). The cylinder is placed at the location $(0.0, e/D)$ and $15D$ downstream the inlet. The upstream, the downstream and the height of the computational domain are $15D$, $50D$ and $50D$ respectively in a near-wall configuration. The value of $e/D$ is the dimensionless transverse distance between the center of cylinder to the wall. Similar to the configurations of an isolated cylinder and side-by-side cylinders, the inlet is imposed with the freestream velocity $\bm{u}_{\infty} = [U, 0.0, 0.0]^T$ and the outlet is pinned with a reference pressure. The rest of the boundaries are imposed with the traction-free boundary condition too. 

\begin{table*}
	\setlength{\tabcolsep}{12pt}
	\renewcommand{\arraystretch}{2}
	\begin{center}
		\def\arraystretch{1.0}
		\begin{tabular}{l p{1.5cm} p{1.5cm} p{1.5cm}}
			\hline
			Time step & $C^{m}_{d}$ &  $C^{rms}_{l}$ & $St$\\
			\hline
			$\Delta t = 0.100$ & 1.374 (10.3\%)& 0.853  (79.6\%)& 0.209 (2.1\%)\\
			$\Delta t = 0.050$ & 1.261 (1.2\%)& 0.484 (1.9\%)& 0.205 (0.9\%)\\
			$\Delta t = 0.025$ & 1.251 (0.48\%)& 0.480 (1.1\%)& 0.203 (0.0\%)\\
			$\Delta t = 0.010$ & 1.246& 0.475& 0.203\\
			\hline
		\end{tabular}
	\end{center}
	\caption{Hydrodynamic responses of a circular circular cylinder at $Re=300$, $\Delta z= 0.15$ and different time steps}
	\label{tab:time_con}
\end{table*}
\begin{table*}
	\setlength{\tabcolsep}{12pt}
	\renewcommand{\arraystretch}{2}
	\begin{center}
		\def\arraystretch{1.5}
		\begin{tabular}{l p{1cm} p{1cm} p{1cm} p{1cm}}
			\hline
			&  & $C^{m}_d$ &  $C^{max}_l$ & $S_t$\\
			\hline
			\multirow{4}{*}{$Re=100$} & \multicolumn{1}{l}{\cite{liu1998preconditioned}} &  \multicolumn{1}{l}{1.350} & \multicolumn{1}{l}{$\pm$0.339} & \multicolumn{1}{l}{0.164}  \\
			& \multicolumn{1}{l}{\cite{calhoun2002cartesian}} &  \multicolumn{1}{l}{1.330} & \multicolumn{1}{l}{$\pm$0.298} & \multicolumn{1}{l}{0.175}  \\
			& \multicolumn{1}{l}{\cite{russell2003cartesian}} &  \multicolumn{1}{l}{1.380} & \multicolumn{1}{l}{$\pm$0.300} & \multicolumn{1}{l}{0.169}  \\
			& \multicolumn{1}{l}{\cite{liu2020nitsche}} &  \multicolumn{1}{l}{1.365} & \multicolumn{1}{l}{$\pm$0.301} & \multicolumn{1}{l}{0.164}  \\
			& \parbox[t]{5cm}{Present} & \multicolumn{1}{l}{1.326} &\multicolumn{1}{l}{$\pm$0.311} & \multicolumn{1}{l}{0.164}  \\
			\cline{1-5}
			\multirow{4}{*}{$Re=200$} & \multicolumn{1}{l}{\cite{liu1998preconditioned}} &  \multicolumn{1}{l}{1.310} & \multicolumn{1}{l}{$\pm$0.690} & \multicolumn{1}{l}{0.192}  \\
			& \multicolumn{1}{l}{\cite{calhoun2002cartesian}} &  \multicolumn{1}{l}{1.172} & \multicolumn{1}{l}{$\pm$0.594} & \multicolumn{1}{l}{0.202}  \\
			& \multicolumn{1}{l}{\cite{russell2003cartesian}} &  \multicolumn{1}{l}{1.390} & \multicolumn{1}{l}{$\pm$0.50} & \multicolumn{1}{l}{0.195}  \\
			& \multicolumn{1}{l}{\cite{liu2020nitsche}} &  \multicolumn{1}{l}{1.372} & \multicolumn{1}{l}{$\pm$0.648} & \multicolumn{1}{l}{0.194}  \\
			& \parbox[t]{5cm}{Present} & \multicolumn{1}{l}{1.366} &\multicolumn{1}{l}{$\pm$0.654} & \multicolumn{1}{l}{0.194}  \\
			\hline
		\end{tabular}
	\end{center}
	\caption{Numerical results of hydrodynamic responses of a circular circular cylinder at $Re \in [100, 200]$, $\Delta t = 0.01$ and the total number of $x$-$y$ plane grid $4.9 \times 10^4$}
	\label{tab:2D_val}
\end{table*}
On the $x$-$y$ plane, the first layers of structured mesh around the cylinder(s) and the wall are controlled within the linear viscous sub-layer ($y^+ < 1.0$) at various Reynolds numbers to ensure an accurate approximation of boundary layer dynamics. Similar to the mesh discretization strategy in~\cite{liu2016interaction}, the mesh elements radiates away the cylinder(s) and the wall at a growth rate less than $1.05$, to avoid the issue of mesh skewness. To investigate the flow transition around a streamline hyperbolic critical point in wake, the computational domain is extended in the spanwise direction ($z$-axis) until $10D$ for the cases of an isolated cylinder. The element length along the spanwise direction is taken at $dz = 0.15D$, as recommended by~\cite{liu2018dynamics}.

The key dimensionless parameters are summarized in Table~\ref{tab:para}. The drag and lift coefficients are computed from the resultant traction forces in $x$ and $y$ axes along the cylinder(s). The value of Reynolds number ranges from $20.0$ to $500.0$, which cover the flow regimes with the onset of vortex shedding and the laminar-to-turbulent flow transition. The cylinder-cylinder ($g/D$) and cylinder-wall ($e/D$) distances are chosen in a way, such that their shear/boundary layers start interacting and influencing each other. The objective is to investigate the effect of asymmetric shear-layer interaction and Reynolds number effect on the stability of HSP in wake. In accordance to the investigation of~\cite{papaioannou2006three,liu2018dynamics}, the dimensionless enstrophy $\xi(t)$ is introduced to quantify the effect of three-dimensional flow feature in wake via the value of secondary enstrophy $\xi_{xy}(t) = \xi_x(t) + \xi_y(t)$, because the value of enstrophy is directly linked with the dissipation of fluid kinetic energy and the generation \& breakdown of coherent flow structures. 

\subsection{Convergence and validations of numerical formulation} \label{sec:validation}
\begin{table*}
	\setlength{\tabcolsep}{12pt}
	\renewcommand{\arraystretch}{2}
	\begin{center}
		\def\arraystretch{1.5}
		\begin{tabular}{l p{1cm} p{1cm} p{1cm} p{1cm}}
			\hline
			&  & $C^{m}_d$ &  $C^{rms}_l$ & $St$\\
			\hline
			\multirow{4}{*}{Simulation}  & \multicolumn{1}{l}{\cite{Zhang1995POF}} & \multicolumn{1}{l}{1.292} & \multicolumn{1}{l}{0.447} & \multicolumn{1}{l}{0.210}\\
			& \multicolumn{1}{l}{\cite{persillon1998physical}} & \multicolumn{1}{l}{1.366} & \multicolumn{1}{l}{0.477} & \multicolumn{1}{l}{0.206} \\
			& \multicolumn{1}{l}{\cite{Norberg2001JoFaSa}} & \multicolumn{1}{l}{-} & \multicolumn{1}{l}{0.44} & \multicolumn{1}{l}{0.203} \\
			& \multicolumn{1}{l}{\cite{liu2018dynamics}} & \multicolumn{1}{l}{1.26} & \multicolumn{1}{l}{0.50} & \multicolumn{1}{l}{0.205} \\	
			& \parbox[t]{5cm}{Present} & \multicolumn{1}{l}{1.261} &\multicolumn{1}{l}{0.484} & \multicolumn{1}{l}{0.205}  \\
			\cline{1-5}
			\multirow{3}{*}{Experiment} & \multicolumn{1}{l}{Wieselsberger (1921)} & \multicolumn{1}{l}{1.25} & \multicolumn{1}{l}{-} & \multicolumn{1}{l}{-} \\
			& \multicolumn{1}{l}{\cite{williamson1996three}} & \multicolumn{1}{l}{-} & \multicolumn{1}{l}{-} & \multicolumn{1}{l}{0.203} \\
			\hline
		\end{tabular}
	\end{center}
	\caption{Numerical and experimental results of hydrodynamic responses of a circular circular cylinder at $Re = 300$, $\Delta z = 0.15$ and $\Delta t = 0.05$}
	\label{tab:3D_val}
\end{table*}

The finite element mesh employed is made up of four-noded quadrilateral element in two-dimensional simulations and eight-noded hexahedral element in three-dimensional simulations. The case of an isolated circular cylinder in three-dimensional flow is selected to investigate the effect of mesh resolution and time step. Three mesh configurations are employed for the spatial convergence analysis in Table~\ref{tab:mesh_con}. The mesh resolution on the $x$-$y$ plane and the spanwise direction ($z$ axis) are systematically refined from one to another. It is noticed that the errors of hydrodynamic responses obtained with the mesh resolution of spanwise element length $\Delta z = 0.15$ is within $2 \%$, in which the value of $C_l$ has the largest error. Hence the mesh resolution of $\Delta z = 0.15$ is selected for the three-dimensional numerical simulations. 
The temporal convergence analysis in Table~\ref{tab:time_con} shows that the obtained hydrodynamic responses at time step $\Delta t = 0.05$ is within 2 \%, with respect to the reference value at $\Delta t = 0.01$. On the other hand, the error of the obtained numerical solution at $\Delta t = 0.05$ is about 2 \% compared with the reference values. Hence, the time step $\Delta t = 0.01$ is employed in two-dimensional simulations, and $\Delta t = 0.05$ is used in the three-dimensional simulations. 

The hydrodynamic responses of an isolated cylinder at $Re=100$ and $Re=200$ in two-dimensional flow are compared with the other numerical results in literature in Table~\ref{tab:2D_val}. The mesh resolution in two-dimensional simulation ($4.9 \times 10^4$) is taken from the three-dimensional mesh configuration at $\Delta z = 0.15$. It can be seen the mean drag coefficient ($C^m_d$), the maximum lift coefficient ($C^{max}_l$) and the Strouhal number ($St$) obtained from the derived numerical formulation match well with literature in two-dimensional simulations. The obtained numerical results based on the derived numerical formulation are also compared with literature. The three-dimensional simulation of an isolated circular cylinder at $Re=300$, the time step $\Delta t = 0.05$ and the mesh resolution of $\Delta z = 0.15$ is chosen to validate the implemented numerical formulation of Navier-Stokes equation with the numerical and experimental results. The obtained numerical solutions of three-dimensional simulations agree well with literature, as presented in Table~\ref{tab:3D_val}.

The above validation and convergence analyses show that the numerical results obtained by the derived numerical formulation can accurately approximate the flow field. In the next section, the numerical stability of a HSP and its relationship with three dimensionality are presented. We are not only interested in the physical insights of a HSP, but also its impact on the fluid stability analysis and the engineering design in the real-word problems.

\section{Numerical stability of a streamline hyperbolic/saddle point in wake} \label{sec:prove}
In this section, the numerical stability of a HSP is numerically investigated. We particularly aim at identifying the key parameters influencing the stability of a HSP in numerical simulations and the relationship of a HSP with the fluid-flux divergence. To this end, we employ the Fourier stability analysis to investigate the dynamics of errors around a HSP. In the Fourier stability analysis, a full spectrum of the initial disturbances is transformed via Fourier expansion and its dynamics/growth is investigated following the evolution of the discretized dynamical system.

\subsection{Fourier stability analysis} \label{sec:vonNeumman}  
In accordance to the second derivative test, a planar streamline singular hyperbolic/saddle critical point can be defined as the form in Equation~\eqref{eq:ssp} with respect to the stream function.
\begin{subequations}\label{eq:ssp}
	\begin{align}
	\frac{\partial \psi}{\partial x} &= 0 \label{eq:ssp1}\\
	\frac{\partial \psi}{\partial y} &= 0 \label{eq:ssp2}\\
	\frac{\partial^2 \psi}{\partial x^2} \frac{\partial^2 \psi}{\partial y^2} - \big(\frac{\partial^2 \psi}{\partial x \partial y}\big)^2 &< 0 \label{eq:ssp3}
	\end{align}
\end{subequations}
where $\psi$ is the stream function and defined as $u = \partial \psi / \partial y$ and $v = -\partial \psi / \partial x$. 
In the classical temporal instability theorem, e.g., Kelvin-Helmholtz instability and Rayleigh inflection point theorem~\footnote{Although classical temporal instabilities are derived in inviscid flow, their fundamental concepts of a mechanical instability are still valid in viscous flow, e.g.,~\cite{lindsay1984kelvin}. The viscous action not only dissipates energy and attenuates instabilities, but delays the phase of responses and possibly excites the initial linear transient growth of non-orthogonal modes.}, the inflectional velocity profiles, $\frac{\partial u}{\partial y}$ or $\frac{\partial v}{\partial x}$, had been emphasized and analytically proven to be detrimental to the hydrodynamic stability. In the stability analysis of a hyperbolic point in this work, we are particularly interested in the cases in which the shear stresses originating from the shear strains in two independent dimensions, $\frac{\partial u}{\partial y}$ and $\frac{\partial v}{\partial x}$, are imbalanced, because they are directly linked with the inflectional velocity profiles and hydrodynamic instability. 
In fact, their difference ($\frac{\partial v}{\partial x}-\frac{\partial u}{\partial y}$) quantifies the $x$-$y$ planar vorticity component ($\omega_z$). Its non-zero value indicates a rotational fluid flow region, where a fluid element rotates about its own axis and its angular motion is significant. On the other hand, its zero value refers to a vorticity-free zone in two-dimensional cases, where the rotation of fluid about its axis is inhibited. Hence we introduce a positive scalar (P), a scaling factor, to quantify the ratio between the planar shear stress, as shown in Equation~\eqref{eq:scale}. 
\begin{eqnarray}
\frac{\partial^2 \psi}{\partial x^2} = -P \frac{\partial^2 \psi}{\partial y^2} \implies \frac{\partial v}{\partial x} = P \frac{\partial u}{\partial y} \label{eq:scale}
\end{eqnarray}
which automatically satisfies the constraint of a hyperbolic point in Equation~\eqref{eq:ssp3}.

Because the hyperbolic point is defined with respect to the stream function, the governing equation of the two-dimensional unsteady incompressible Navier-Stokes equation can be re-casted in the stream function-vorticity form in Equation~\eqref{eq:ns_stream}.
\begin{subequations} \label{eq:ns_stream}
	\begin{align}
\frac{\partial^2 \psi}{\partial x^2} + \frac{\partial^2 \psi}{\partial y^2} &= - \omega \label{eq:AG1} \\
\frac{\partial \omega}{\partial t} + u \frac{\partial \omega}{\partial x} + v \frac{\partial \omega}{\partial y} &= \frac{1}{Re} \big(\frac{\partial^2 \omega}{\partial x^2} + \frac{\partial^2 \omega}{\partial y^2}\big)
	\end{align}
\end{subequations}
where $\omega$, $\psi$ and $Re$ are the vorticity, the stream function and the Reynolds number respectively. Subsequently, imposing the constraints in Equation~\eqref{eq:ssp1},~\eqref{eq:ssp1} and~\eqref{eq:scale} on Equation~\eqref{eq:ns_stream}, the constraint Navier-Stokes equation can be expressed as
\begin{eqnarray}
\frac{\partial \omega}{\partial t} = \frac{1}{Re} \big[ \big(\frac{1}{P} -1\big)\frac{\partial^4 \psi}{\partial x^4} + \big(P-1\big) \frac{\partial^4 \psi}{\partial y^4}\big] \label{eq:ns_stream_con}
\end{eqnarray}

The spatial domain is discretized with the uniformed structured mesh, $\Delta x = \Delta y = h$, in the second-order central accurate scheme, as shown in Equation~\eqref{eq:fd_space}. 
\begin{subequations} \label{eq:fd_space}
	\begin{align}
\frac{\partial^2 \psi}{\partial x^2} &= \frac{1}{h^2} \big(\psi_{i+1,j} - 2\psi_{i,j} + \psi_{i-1,j}\big)+\mathcal{O}(h^2) \\
\frac{\partial^2 \psi}{\partial y^2} &= \frac{1}{h^2} \big(\psi_{i,j+1} - 2\psi_{i,j} + \psi_{i,j-1}\big)+\mathcal{O}(h^2) \\
\frac{\partial^4 \psi}{\partial x^4} &= \frac{1}{h^4} \big(\psi_{i-2,j}-4\psi_{i-1,j}+6\psi_{i,j}-4\psi_{i+1,j}+\psi_{i+2,j}\big)+\mathcal{O}(h^2) \\
\frac{\partial^4 \psi}{\partial y^4} &= \frac{1}{h^4} \big(\psi_{i,j-2}-4\psi_{i,j-1}+6\psi_{i,j}-4\psi_{i,j+1}+\psi_{i,j+2}\big)+\mathcal{O}(h^2) 
	\end{align}
\end{subequations}

On the other hand, the time domain is discretized with the first-order Euler forward scheme. Hence, the constraint Navier-Stokes equation can be expressed in the form of a linear ordinary differential equation, as shown in Equation~\eqref{eq:ns_stream_con_dis}.
\begin{subequations} \label{eq:ns_stream_con_dis}
	\begin{align}
	& \frac{P-1}{\Delta t h^2} \big[(\psi^{n+1}_{i,j+1} - 2 \psi^{n+1}_{i,j} + \psi^{n+1}_{i,j-1}) - (\psi^n_{i,j+1} - 2 \psi^{n}_{i,j} + \psi^n_{i,j-1})\big]\ \nonumber\\
	&=\frac{1}{Re h^4} \big[\big(\frac{1}{P}-1\big)\big(\psi^n_{i-2,j} - 4\psi^n_{i-1,j} - 4 \psi^n_{i+1,j} + \psi^n_{i+2,j} + 6 \psi^n_{i,j}\big) \nonumber \\
	&+\big(P-1\big)\big(\psi^n_{i,j-2} - 4\psi^n_{i,j-1}-4\psi^n_{i,j+1}+\psi^n_{i,j+2}+6\psi^n_{i,j}\big)\big] 
	\end{align}
\end{subequations}
In the Fourier stability analysis, the following parameters are defined
\begin{subequations} \label{eq:NL}
	\begin{align}
	\psi^n_{i,j} &= V^n e^{ik_x x_i} e^{ik_y y_j} \label{eq:fourier}\\
	 G &= \frac{V^{n+1}}{V^n} \\
	 \alpha &= k_x h; \quad \beta = k_y h
	\end{align}
\end{subequations}
where Equation~\eqref{eq:fourier} is the Fourier expansion of the stream function and $V^n$ is its amplitude at time step (n). G is defined as the amplification factor to represent the temporal evolution of the initial disturbance. Substituting Equation~\eqref{eq:NL} into Equation~\eqref{eq:ns_stream_con_dis}, it becomes

\begin{eqnarray}
& & \frac{P-1}{\Delta t h^2} G \big(e^{i\beta}-2+e^{-i\beta}\big) - \frac{P-1}{\Delta t h^2}\big(e^{i\beta} -2+e^{-i\beta}\big) \nonumber \\
&=& \frac{1}{Re h^4} \big[ \big(\frac{1}{P}-1\big)\big(e^{-i2\alpha}-4e^{-i\alpha}-4e^{i\alpha}+e^{i2\alpha} + 6\big) \nonumber\\
&+& \big(P-1\big)\big(e^{-i2\beta} - 4e^{-i\beta} - 4e^{i\beta} + e^{i2\beta} + 6\big)\big] \label{eq:AN2}
\end{eqnarray}

By $e^{i\beta}+e^{-i\beta} = 2 cos \beta$ and $cos 2\beta = 2cos^2 \beta -1$, Equation~\eqref{eq:AN2} can be re-casted in the form in Equation~\eqref{eq:AN6}, and $G$ should satisfy the criterion in Equation~\eqref{eq:AN3} for the stability of a HSP.

\begin{eqnarray}
G\big(cos \beta -1\big) - \big(cos \beta -1\big) &=& \frac{\Delta t}{(P-1)Re h^2} \big[\frac{1-P}{P}\big(cos 2\alpha - 4 cos \alpha + 3\big)\nonumber\\
& & + \big(P-1\big) \big(cos 2\beta - 4 cos \beta +3\big)\big] \nonumber\\
&=& \underbrace{\frac{\Delta t}{Re h^2 P}}_{\theta_1} \big(4cos \alpha - cos 2 \alpha -3\big) \nonumber\\
&+& \underbrace{\frac{\Delta t}{Re h^2}}_{\theta_2} \big(cos 2\beta - 4cos \beta +3\big) \label{eq:AN6}
\end{eqnarray}
\begin{eqnarray}
& & |G| \leqslant 1 \nonumber\\
\implies & & -2 \leqslant \frac{1}{cos \beta -1} \big[\theta_1 \big(4cos \alpha  -cos 2\alpha -3\big) \nonumber\\
& &  + \theta_2 \big(cos 2\beta - 4 cos \beta +3\big) \big] \leqslant 0 \label{eq:AN3}
\end{eqnarray}
To avoid trivial solution, we have $cos \beta \neq 1$ and $-2 \leqslant cos \beta - 1 < 0$. Since a uniform discretization is assumed, e.g, $h = \Delta x = \Delta y$, hence $\alpha = k_x h = k_y h = \beta$. Therefore, Equation~\eqref{eq:AN3} can be expressed in the form of Equation~\eqref{eq:AN4} to guarantee stability.

\begin{eqnarray}
-1 \leqslant (\theta_2 - \theta_1) (cos \beta -1) \leqslant 0 \label{eq:AN4}
\end{eqnarray}
where $\theta_1=\frac{\Delta t}{Reh^2P}$ and $\theta_2 = \frac{\Delta t}{Re h^2}$. Therefore, the stability criteria in Equation~\eqref{eq:AN4} can be re-casted in the form below.
\begin{eqnarray}
0 \leqslant (1-cos\beta)(1- \frac{1}{P}) \leqslant \frac{Re h^2}{\Delta t}  < 2.0 \label{eq:AN5}
\end{eqnarray}
It is noteworthy that the stable solution condition of Equation~\eqref{eq:AN5} is dependent upon the numerical discretizations in simulation ($h$ and $\Delta t$), the Reynolds number ($Re$) and the spectrum of errors/noises ($1-cos\beta$). Let $\Psi = Re h^2/\Delta t$. To eliminate the spatial and temporal discretization errors, the limit of $\Psi$ is taken as the values of $h$ and $\Delta t$ approach to zero along their positive axes, and its limit is shown in Equation~\eqref{eq:limit}.
\begin{eqnarray}
\lim \limits_{h, \Delta t \to +0} \Psi = 0 \label{eq:limit}
\end{eqnarray}
Consequently we can find that a HSP is intrinsically meta-stable at $P=1$ at a location ($\bm{x}_0$) in the space~\footnote{If $P\neq 1.0$, the HSP is unstable at $\bm{x}_0$, e.g., move away from its original location.}, where $\partial v / \partial x - \partial u / \partial y = 0$. However, based on Equation~\eqref{eq:AN5}, we realize a HSP might be numerically 'stable' at $P \approx 1$, due to the temporal and spatial discretizations, e.g., in some cases of $\Psi < 2.0$. It is also noticed as Reynolds number increases, it is possible $\Psi > 2.0$ and a HSP becomes unstable at $P \approx 1.0$. In particular, if one manipulates Equation~\eqref{eq:AN5} and considers Reynolds number and errors together as $Re/(1-cos\beta)$, it is noticed that the increment of Reynolds number can make a HSP more sensitive to the errors of a wide spectrum, which is detrimental to the numerical stability of a HSP too.  

To sum up, a HSP is merely meta-stable at $P=1$, a perfect balance of shear-layer interaction. In other words, it refers to the vorticity-free region where the vortcity component on this phase plane is zero, e.g., the shear-layer interface or the outer potential flow region. However, if the fluid planar velocity is zero and it is also not allowed to rotate about its axis normal to its phase plane, where does the fluid flow subsequently? The only plausible path is the third dimension normal to its phase plane. In Section~\ref{sec:ppe} and Section~\ref{sec:three}, we will prove and discuss that a HSP is intrinsically associated with fluid three dimensionality. It is known that the disturbances of a wide spectrum are inevitable, e.g., background noises, human and machinery activities in experiments or the numerical dispersion in simulations, e.g.,~\cite{Sengupta2007JoCP,sengupta2013high}. Furthermore, the high value of Reynolds number in the real-world turbulent flow is also detrimental to the stability of a HSP. Consequently, a stable HSP really becomes a special case; whereas it generally manifests itself as a potential unstable factor in the flow field. 

\subsection{Poisson pressure equation and flow transition} \label{sec:ppe}
It is well known that the pressure scalar is a secondary variable in flow system, because the only thing matters is its gradient in the Navier-Stokes equation, instead of its absolute value. The negative pressure gradient, $-\nabla p$, is one of the typical driving forces of fluid flow and prominently correlated to the hydrodynamic instabilities, e.g., the adverse pressure gradients and inflectional velocity profiles. 

In many applications, the Poisson pressure equation (PPE) is derived to reveal the subtle conversion between the velocity field and the pressure field, in the terms of the primal variables or the stream functions, as shown in Equation~\eqref{eq:ppe1} and~\eqref{eq:ppe2} respectively.
\begin{subequations} \label{eq:ppe}
	\begin{align}
	\nabla \cdot (-\nabla p) = -\nabla^2 p &= \rho \nabla \cdot (\bm{u} \cdot \nabla \bm{u}) \label{eq:ppe1}\\
			   &= -2\rho\Big(\frac{\partial^2 \psi}{\partial x^2} \frac{\partial^2 \psi}{\partial y^2} - \big[\frac{\partial^2 \psi}{\partial x \partial y}\big]^2\Big) \label{eq:ppe2}
	\end{align}
\end{subequations}   
By comparing the definition of a hyperbolic point in Equation~\eqref{eq:ssp3} and PPE in Equation~\eqref{eq:ppe2}, it shows that a HSP in wake is in fact a pressure source point, where a relative high pressure value is diverging away from its neighborhood in the low pressure wake. Observing Equation~\eqref{eq:ppe1}, it is also realized that the divergence of pressure force in fact quantifies the divergence of the fluid flux. Moreover, the comparison of Equation~\eqref{eq:ppe1} and~\eqref{eq:ppe2} reveals that a HSP in flow field, as defined in Equation~\eqref{eq:ssp3}, is intrinsically associated with a net positive divergence of fluid flux. Pressure is a scalar, so this positive pressure source at HSP potentially induces fluid flux in all directions around it, including the third dimension normal to its phase plane. In Section~\ref{sec:results}, we show that this relationship with PPE is subtly linked with the fluid three dimensionality in flow transition. In accordance to Rayleigh inflection theorem and the boundary layer dynamics, it is well known that the adverse pressure gradient is detrimental to the hydrodynamic stability in the shear layers. It possibly induces the inflectional velocity profile and flow separation. Consequently, the presence of a HSP in the wake, a typical low pressure region, is potentially critical to the fluid stability and deserves attention. The onset of flow transition is really a broad and intriguing topic. In this article, we merely focus on the intrinsic characteristics of a HSP and its relationship with fluid three-dimensionality. The quantitative and qualitative discussions about the characteristics of a HSP in wake and its relationship with the imbalanced shear layer interaction, the Reynolds number effect, the fluid-flux divergence and fluid three dimensionality will be subsequently elaborated in Section~\ref{sec:results}.

\section{Results and discussions} \label{sec:results}
The key characteristics of a HSP are discussed in this section. These discussions confirm the analytical proof presented in Section~\ref{sec:prove}: (1) the stability of a HSP with respect to the asymmetric shear-layer interaction and (2) its relationship with fluid-flux divergence and fluid three dimensionality. We primarily focus on three aspects: (a) where does a HSP form in the wake; (b) what are the influences of asymmetric shear-layer interaction and Reynolds number on the stability of a HSP and (c) what is its relationship with three dimensionality in flow transition. To this end, three canonical configurations are considered, an isolated cylinder, side-by-side cylinders and a near-wall cylinder, in which asymmetric shear-layer interaction is induced in wake.  
\subsection{Characteristics of a singular streamline hyperbolic/saddle point in wake} \label{sec:cylinder}
\begin{figure} \centering
	\begin{subfigure}[b]{0.5\textwidth}	
		\centering
		\hspace{-25pt}\includegraphics[trim=0.1cm 0.1cm 0.1cm 0.1cm,scale=0.25,clip]{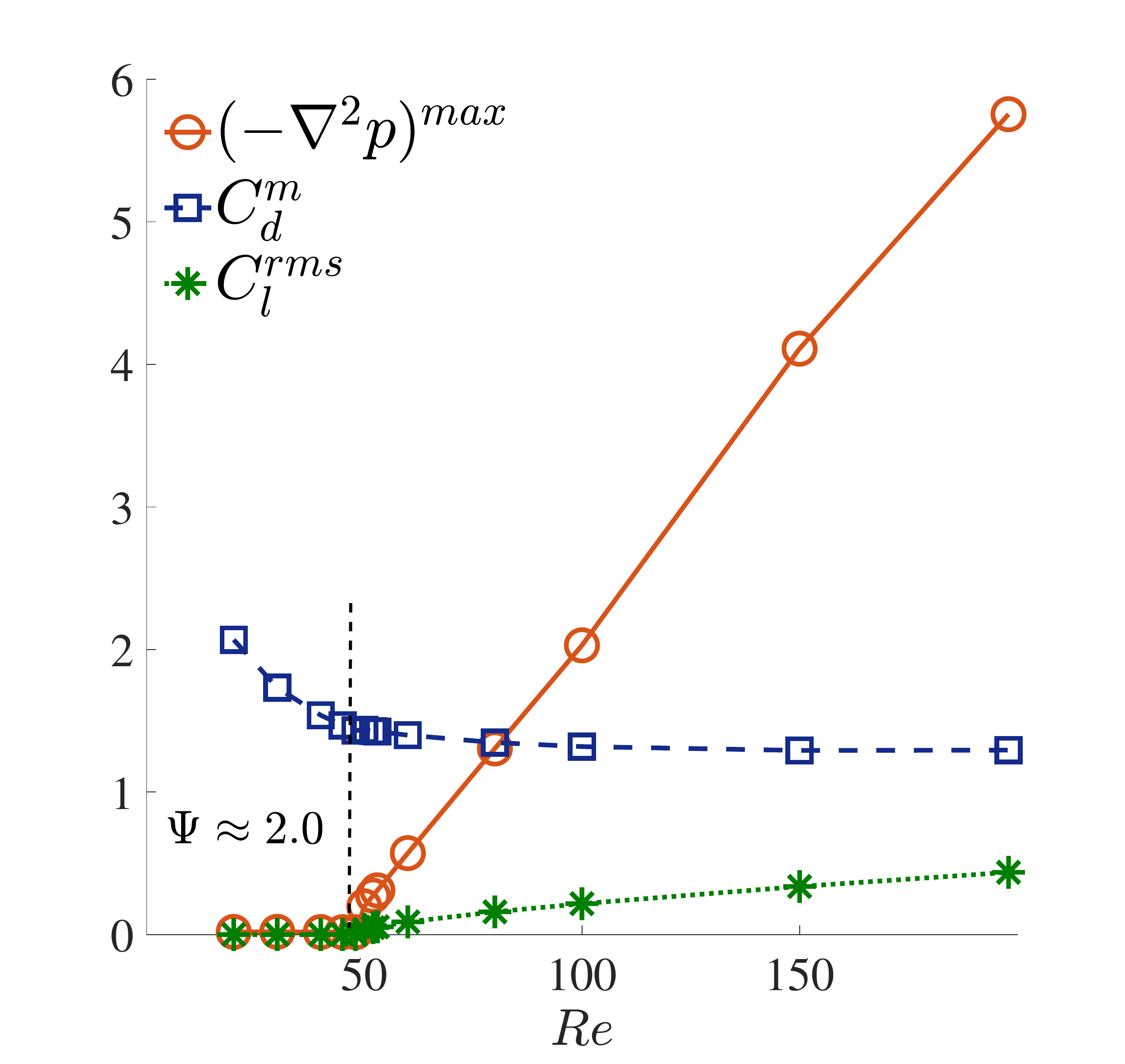}
		\begin{picture}(0,0)
		\put(-150,80){\includegraphics[trim=0.1cm 0.1cm 0.1cm 0.1cm,scale=0.2,clip]{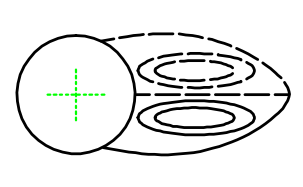}}
		\end{picture}
		\caption{}
		\label{fig:iso_flux}
	\end{subfigure}%
	\begin{subfigure}[b]{0.5\textwidth}
		\centering
		\hspace{-25pt}\includegraphics[trim=0.1cm 0.1cm 0.1cm 0.1cm,scale=0.25,clip]{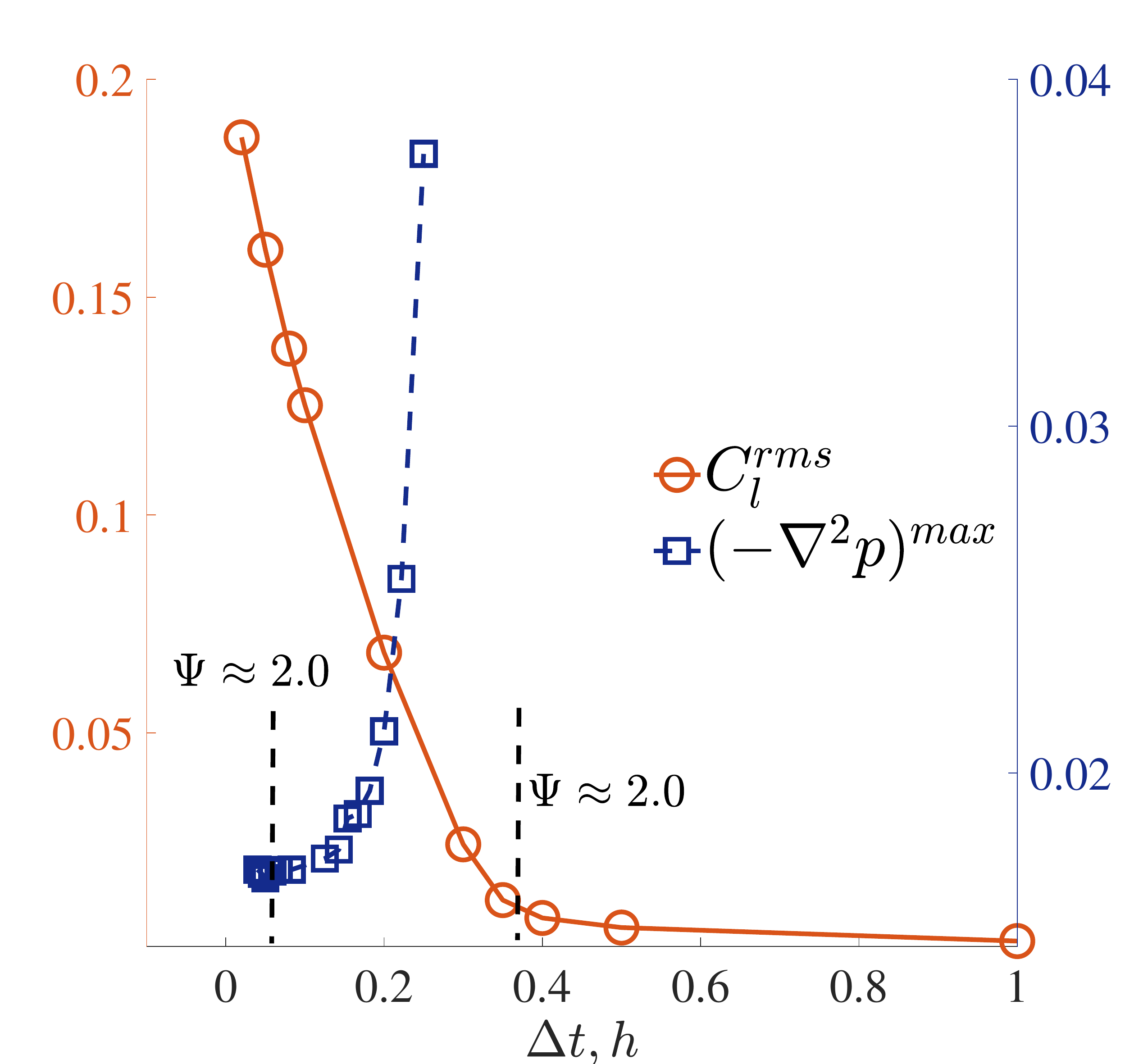}
		\caption{}
		\label{fig:time_space}
	\end{subfigure}
	\caption{Numerical instability of a HSP in the wake of a cylinder at $Re \in [20, 198]$: (a) evolution of instability with respect to Reynolds number; (b) evolution of instability with respect to temporal ($\Delta t$) and spatial discretizations ($h$). In Figure~\ref{fig:time_space}, the left and the right y axes respectively are $C^{rms}_l$ and $(-\nabla^2 p)^{max}$. }
	\label{fig:iso}
\end{figure}
\begin{figure} \centering
	\begin{subfigure}[b]{0.5\textwidth}	
		\centering
		\hspace{-25pt}\includegraphics[trim=1cm 0.1cm 0.1cm 0.1cm,scale=0.33,clip]{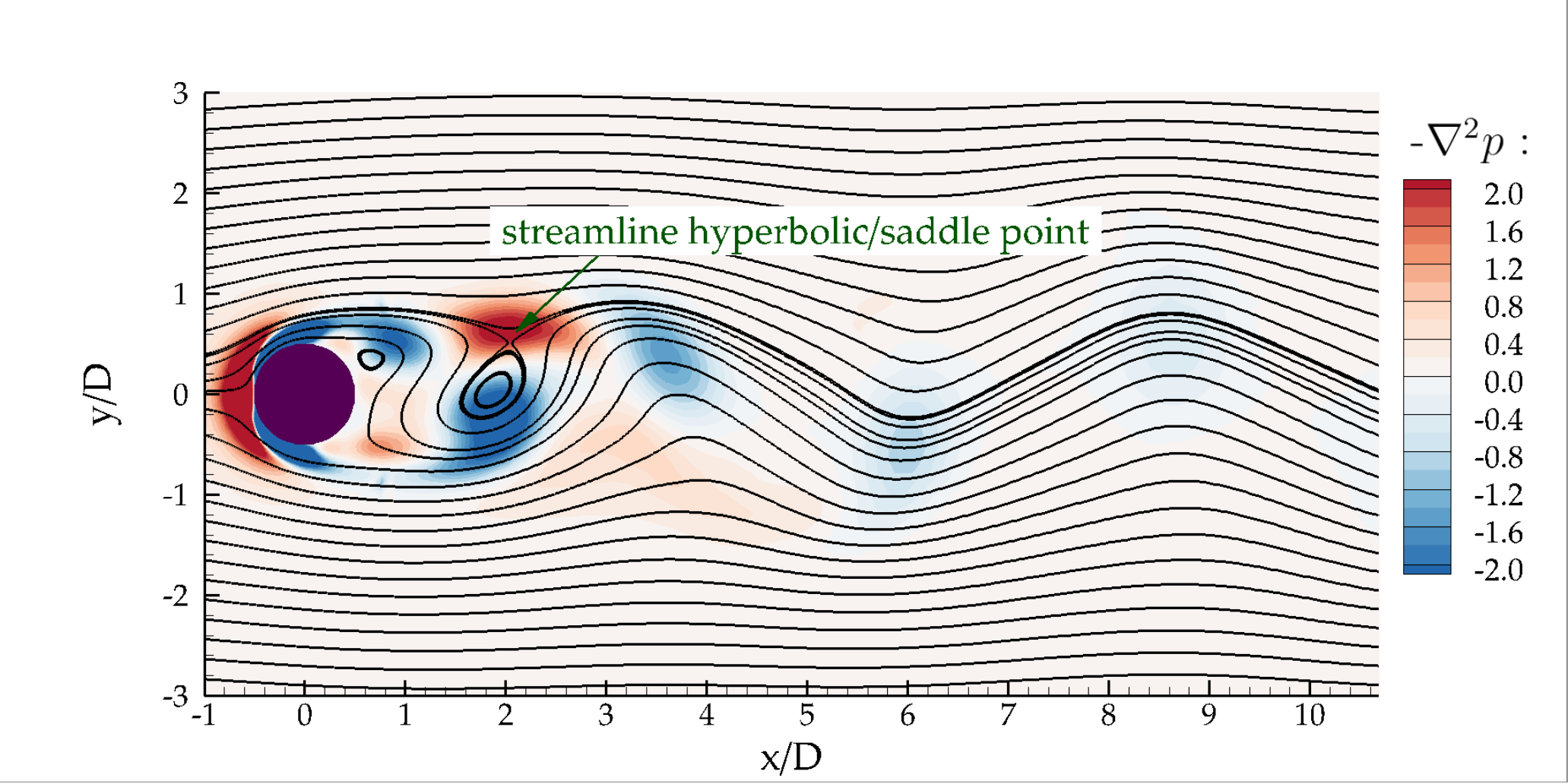}
		\caption{$\qquad\quad$}
		\label{fig:con_flux_re100}
	\end{subfigure}%
	\begin{subfigure}[b]{0.5\textwidth}
		\centering
		\hspace{-25pt}\includegraphics[trim=1cm 0.1cm 0.1cm 0.1cm,scale=0.33,clip]{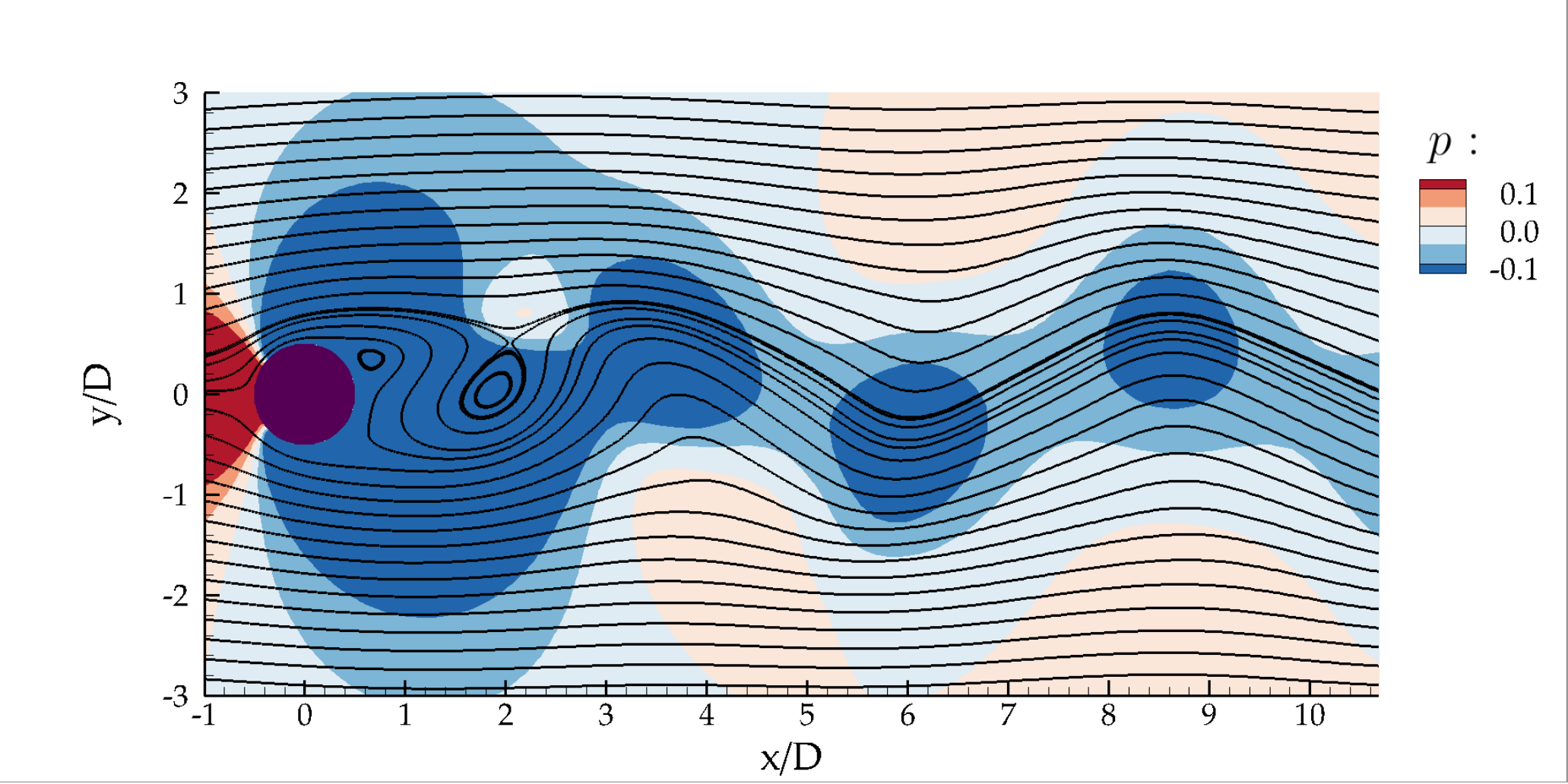}
		\caption{$\qquad\quad$}
		\label{fig:con_p_re100}
	\end{subfigure}
	\begin{subfigure}[b]{0.5\textwidth}	
		\centering
		\hspace{-25pt}\includegraphics[trim=1cm 0.1cm 0.1cm 0.1cm,scale=0.33,clip]{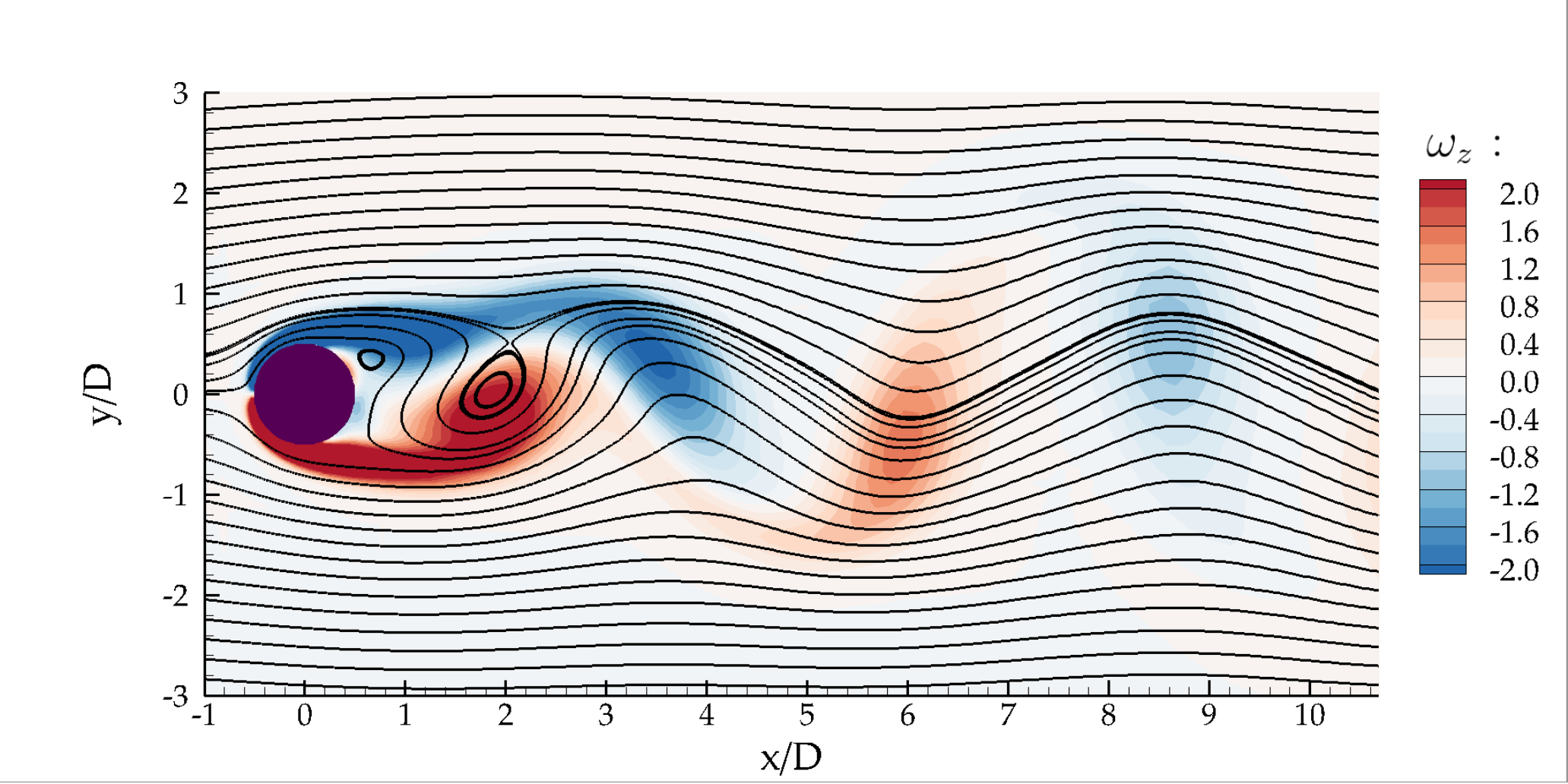}
		\caption{$\qquad\quad$}
		\label{fig:con_wz_re100}
	\end{subfigure}%
	\begin{subfigure}[b]{0.5\textwidth}
		\centering
		\hspace{-25pt}\includegraphics[trim=1cm 0.1cm 0.1cm 0.1cm,scale=0.33,clip]{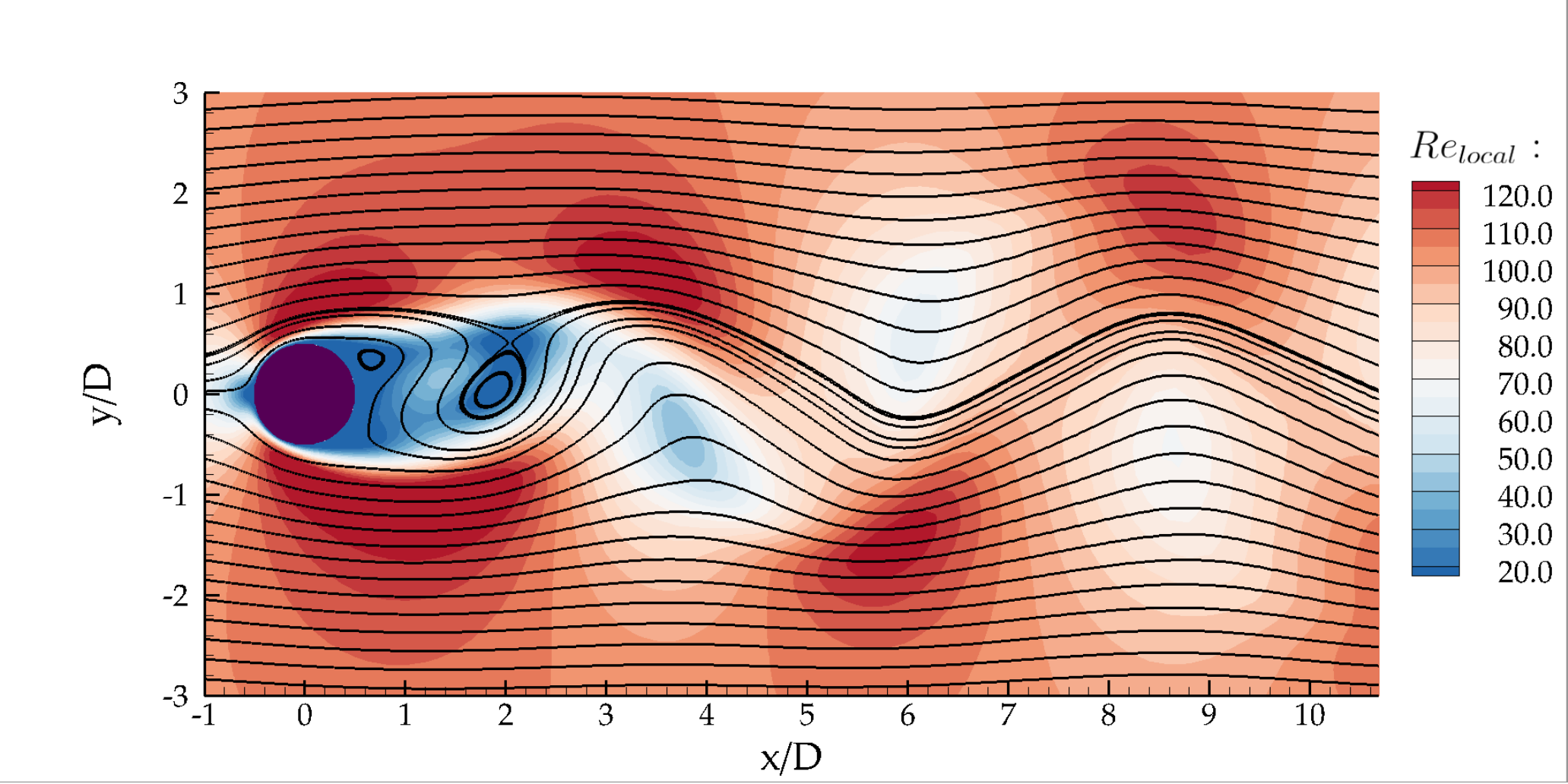}
		\caption{$\qquad\quad$}
		\label{fig:con_local_Re_re100}
	\end{subfigure}
	\caption{Wake behind a circular cylinder at $Re=100$ and $tU/D = 275$: (a) divergence of pressure force; (b) pressure field; (c) spanwise vorticity ($\omega_z$) and (d) local Reynolds number ($Re_{local}$). The value of $Re_{local}$ is defined as $Re_{local} = (\rho |\bm{u}(\bm{x})|D)/\mu$, where $|\bm{u}(\bm{x})|$ is the velocity magnitude at location $\bm{x}$.}
	\label{fig:con_re100}
\end{figure}
\begin{figure} \centering
	\begin{subfigure}[b]{0.5\textwidth}	
		\centering
		\hspace{-25pt}\includegraphics[trim=0.1cm 0.1cm 0.1cm 0.1cm,scale=0.25,clip]{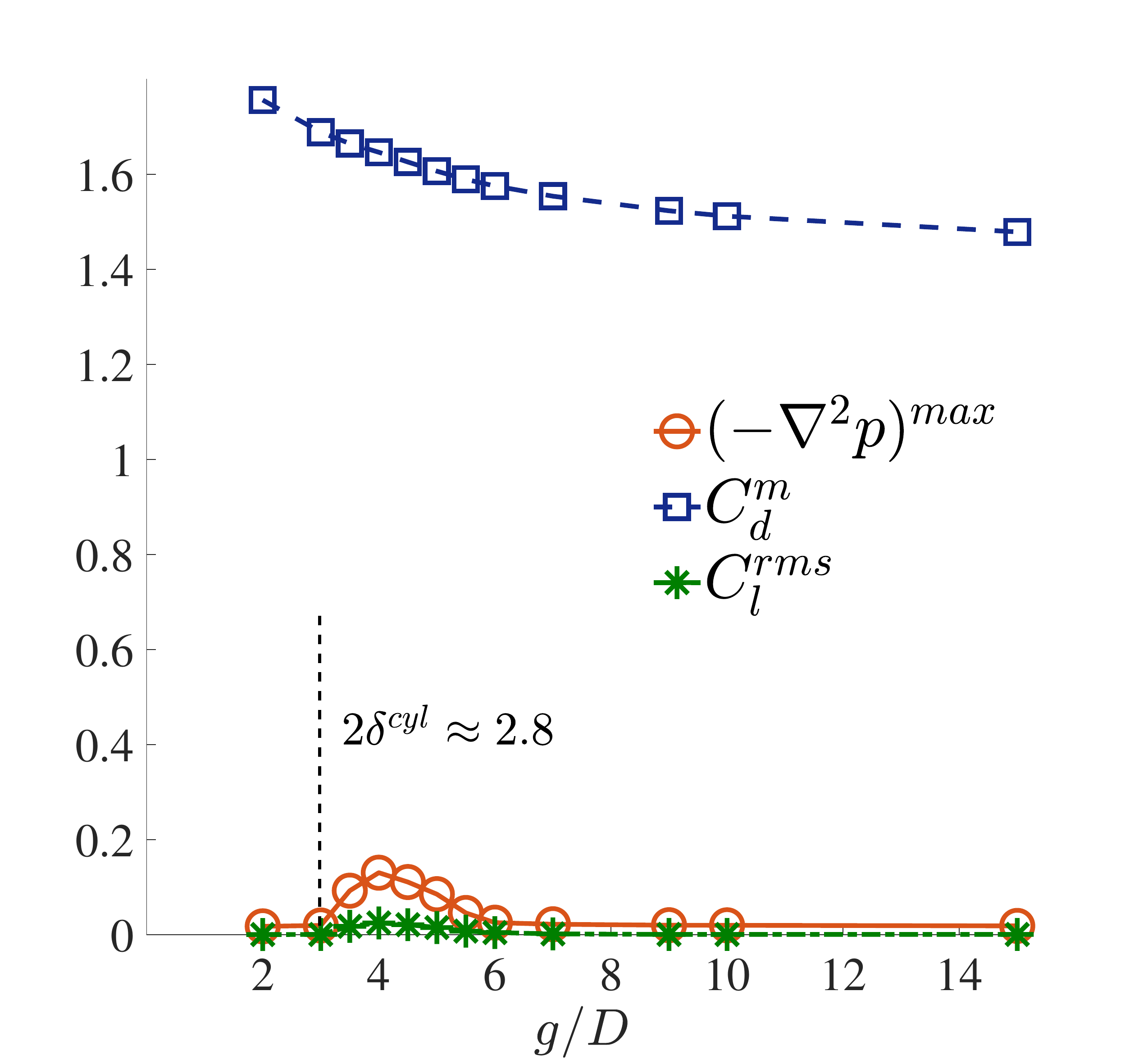}
		\begin{picture}(0,0)
		\put(-150,80){\includegraphics[trim=0.1cm 0.1cm 0.1cm 0.1cm,scale=0.12,clip]{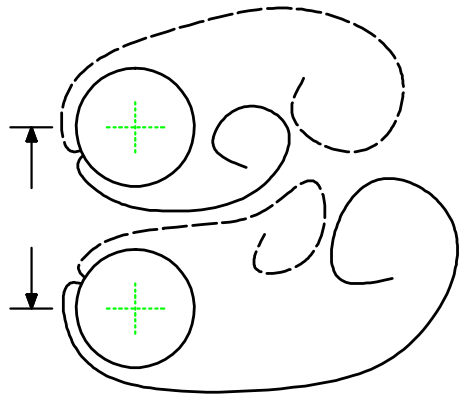}}
		\end{picture}
		\caption{}
		\label{fig:flux_sbs_re47}
	\end{subfigure}%
	\begin{subfigure}[b]{0.5\textwidth}
		\centering
		\hspace{-25pt}\includegraphics[trim=0.1cm 0.1cm 0.1cm 0.1cm,scale=0.25,clip]{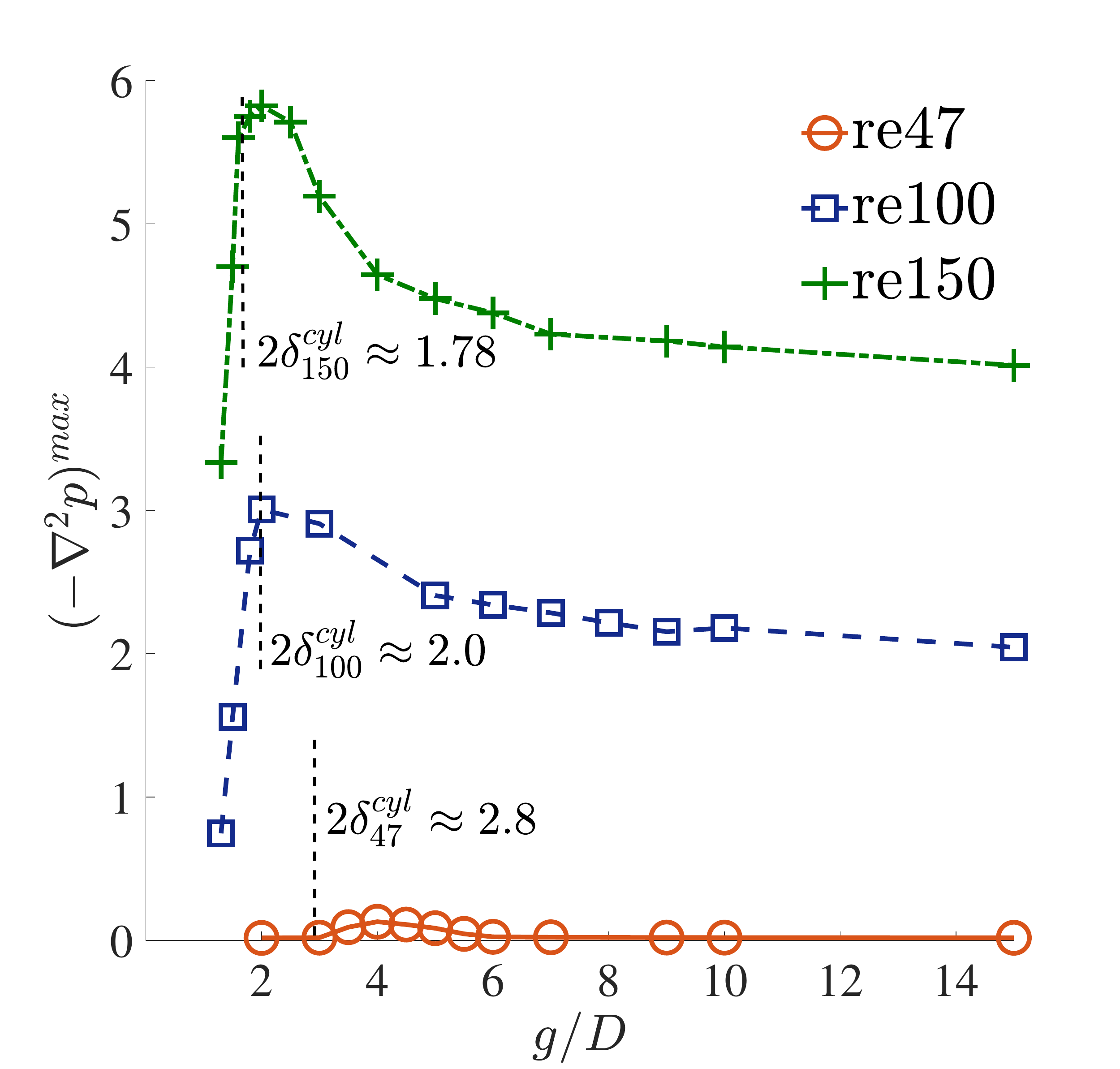}
		\caption{}
		\label{fig:flux_sbs_re47_re150}
	\end{subfigure}
	\caption{Evolution of maximum fluid-flux divergence at a HSP and hydrodynamic responses for a pair of side-by-side cylinders subjecting the gap-flow induced proximity interference: (a) side-by-side cylinders at $Re=47$ and (b) side-by-side cylinders at $Re \in [47, 150]$ and $g/D \in [1.8,15.0]$. The value of $\delta$ is the boundary layer thickness of a cylinder, and its subscript refers to the Reynolds number of this cylinder.}
	\label{fig:flux_sbs}
\end{figure}
\begin{figure} \centering
	\begin{subfigure}[b]{0.5\textwidth}	
		\centering
		\hspace{-25pt}\includegraphics[trim=1cm 0.1cm 0.05cm 0.1cm,scale=0.33,clip]{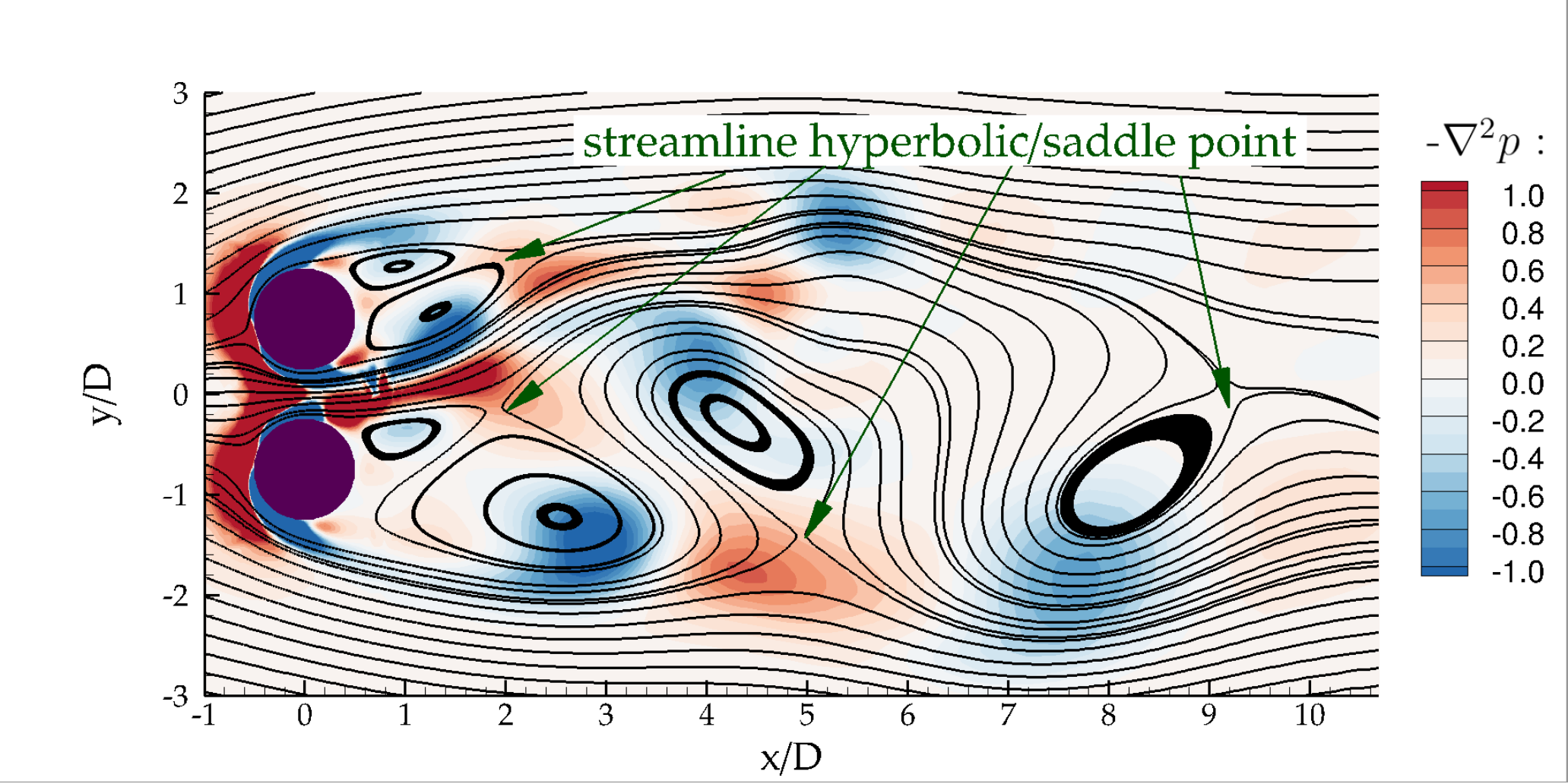}
		\caption{$\qquad\quad$}
		\label{fig:con_flux_re100td15}
	\end{subfigure}%
	\begin{subfigure}[b]{0.5\textwidth}
		\centering
		\hspace{-25pt}\includegraphics[trim=1cm 0.1cm 0.05cm 0.1cm,scale=0.33,clip]{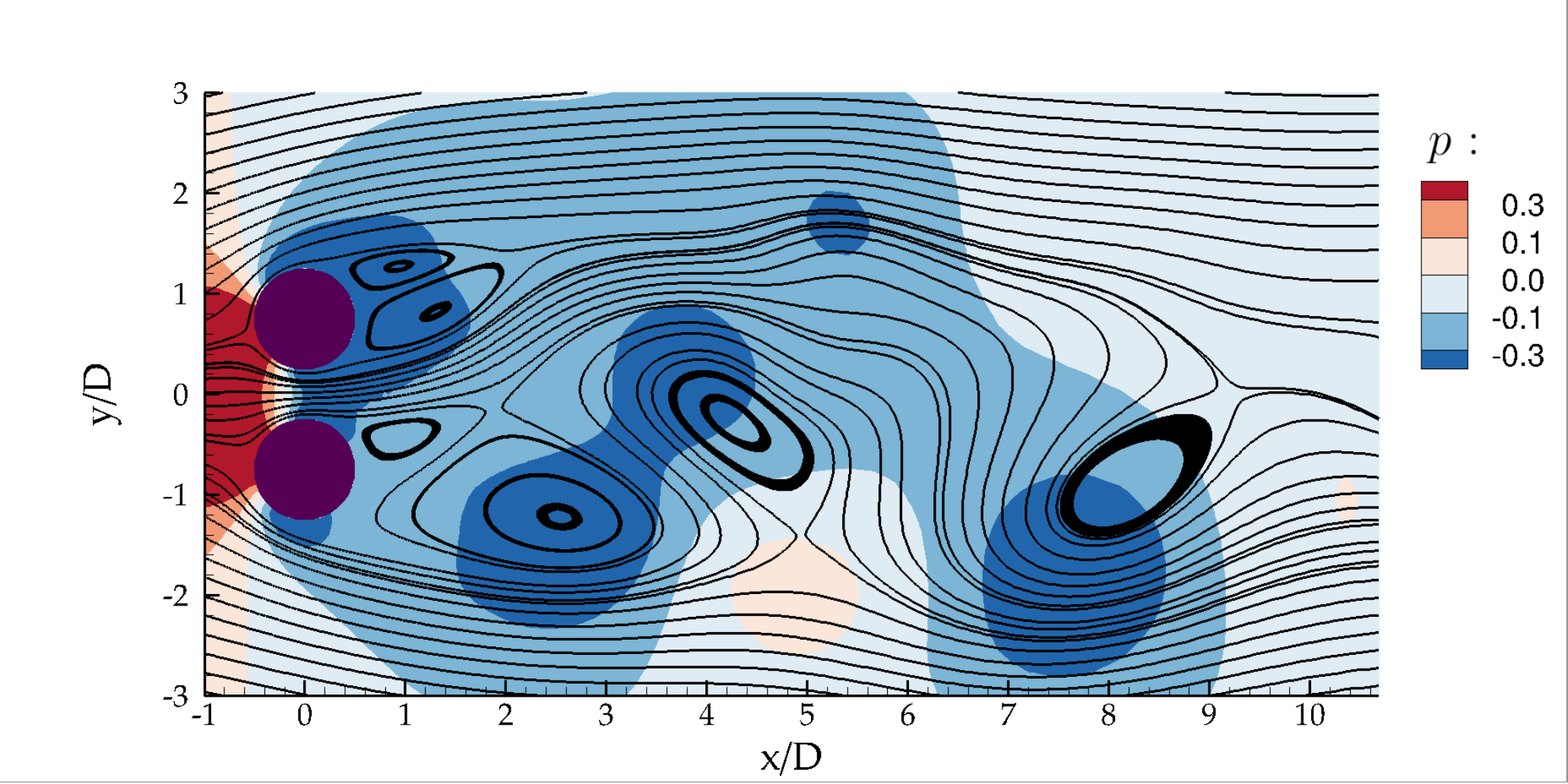}
		\caption{$\qquad\quad$}
		\label{fig:con_p_re100td15}
	\end{subfigure}
	\begin{subfigure}[b]{0.5\textwidth}	
		\centering
		\hspace{-25pt}\includegraphics[trim=1cm 0.1cm 0.05cm 0.1cm,scale=0.33,clip]{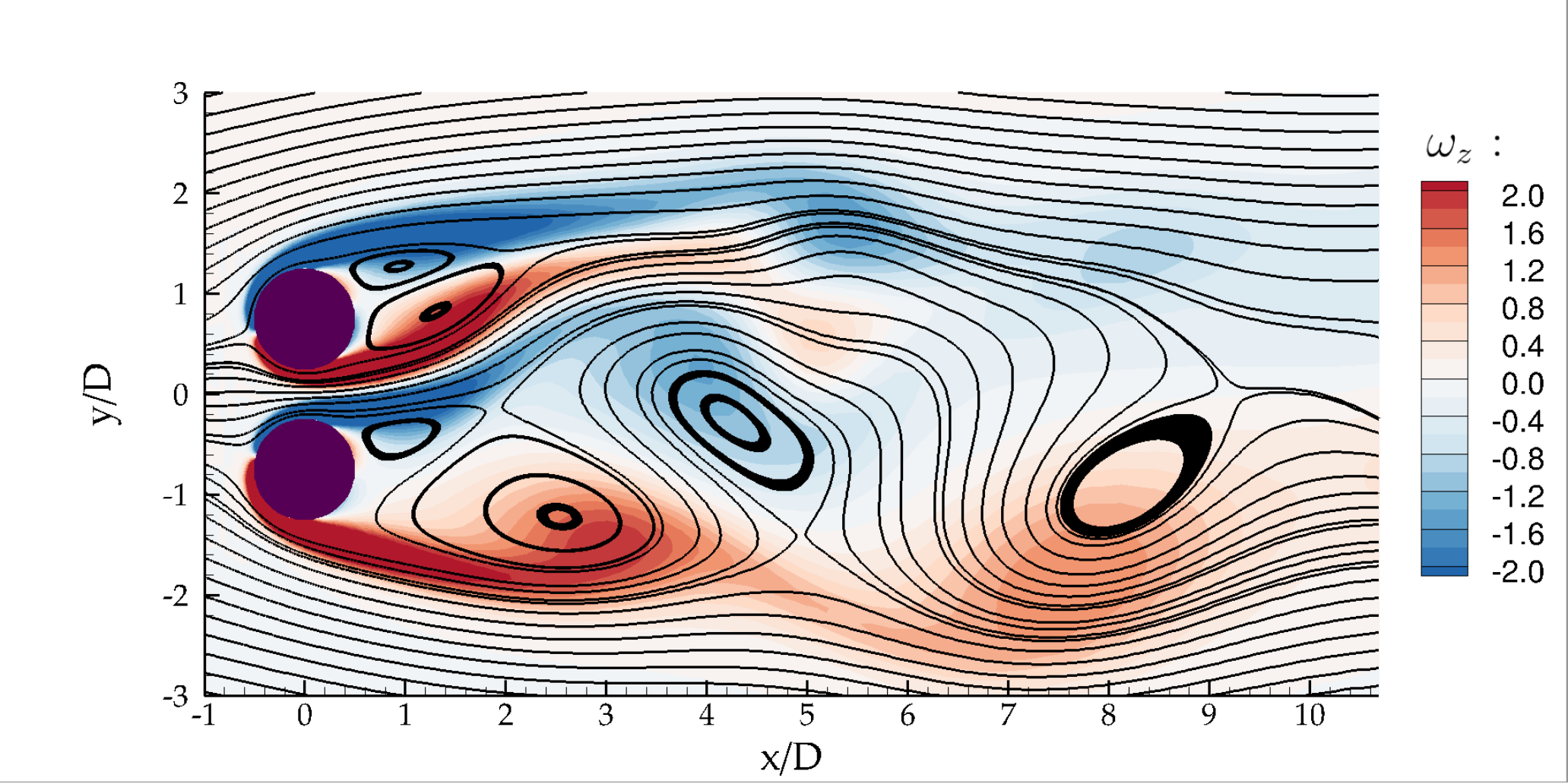}
		\caption{$\qquad\quad$}
		\label{fig:con_wz_re100td15}
	\end{subfigure}%
	\begin{subfigure}[b]{0.5\textwidth}
		\centering
		\hspace{-25pt}\includegraphics[trim=1cm 0.1cm 0.05cm 0.1cm,scale=0.33,clip]{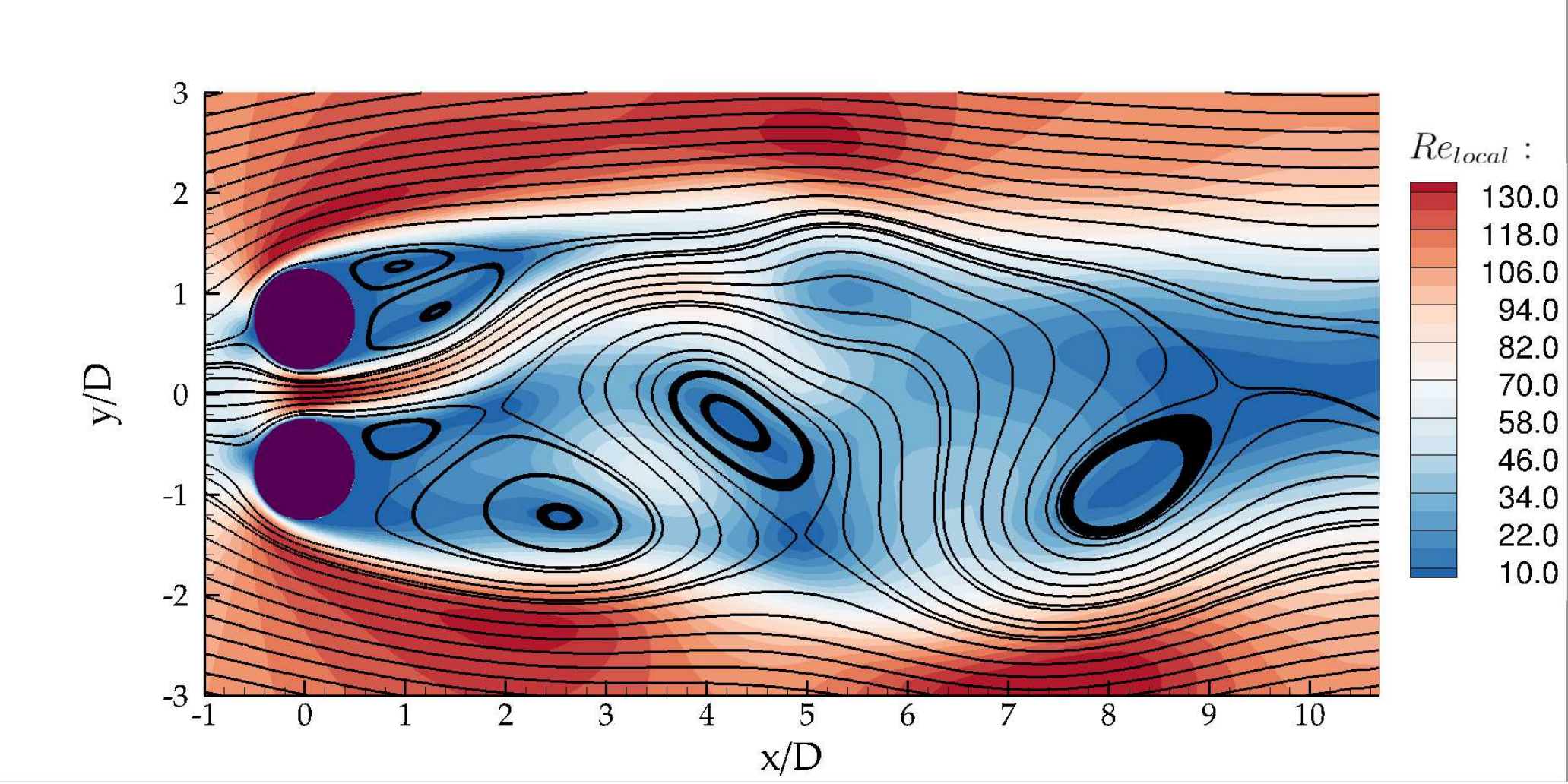}
		\caption{$\qquad\quad$}
		\label{fig:con_local_Re_re100td15}
	\end{subfigure}
	\caption{Wake behind a pair of side-by-side cylinders at $Re=100$, $g/D = 1.5$ and $tU/D = 225$: (a) divergence of pressure force; (b) pressure field; (c) spanwise vorticity ($\omega_z$) and (d) local Reynolds number ($Re_{local}$)}
	\label{fig:con_re100td15}
\end{figure}
\begin{figure} \centering
	\begin{subfigure}[b]{0.5\textwidth}	
		\centering
		\hspace{-25pt}\includegraphics[trim=0.1cm 0.1cm 0.1cm 0.1cm,scale=0.25,clip]{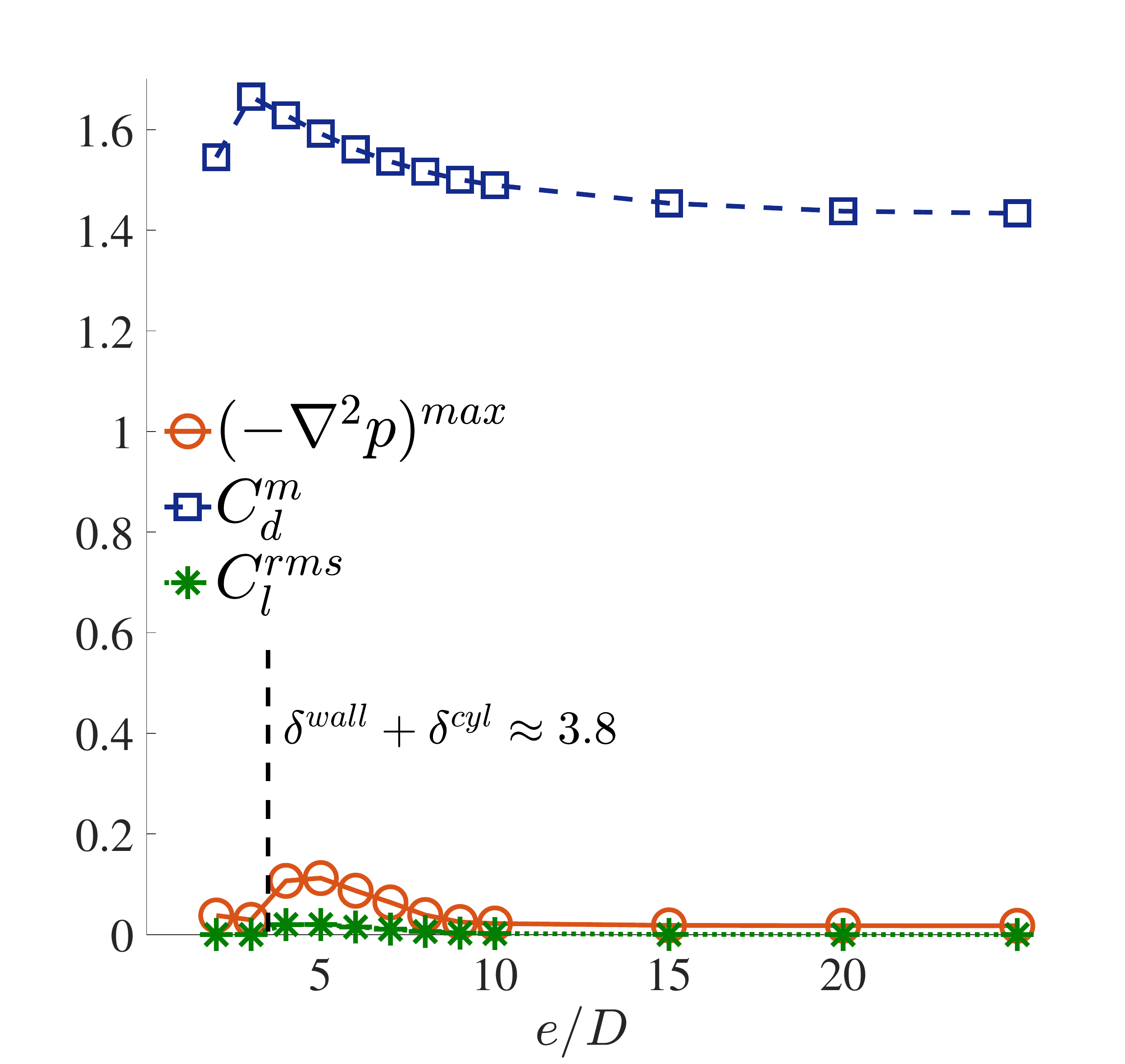}
		\begin{picture}(0,0)
		\put(-70,80){\includegraphics[trim=0.1cm 0.1cm 0.1cm 0.1cm,scale=0.17,clip]{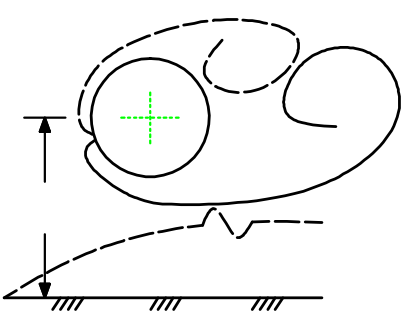}}
		\end{picture}
		\caption{}
		\label{fig:flux_nearWall_re47}
	\end{subfigure}%
	\begin{subfigure}[b]{0.5\textwidth}
		\centering
		\hspace{-25pt}\includegraphics[trim=0.1cm 0.1cm 0.1cm 0.1cm,scale=0.25,clip]{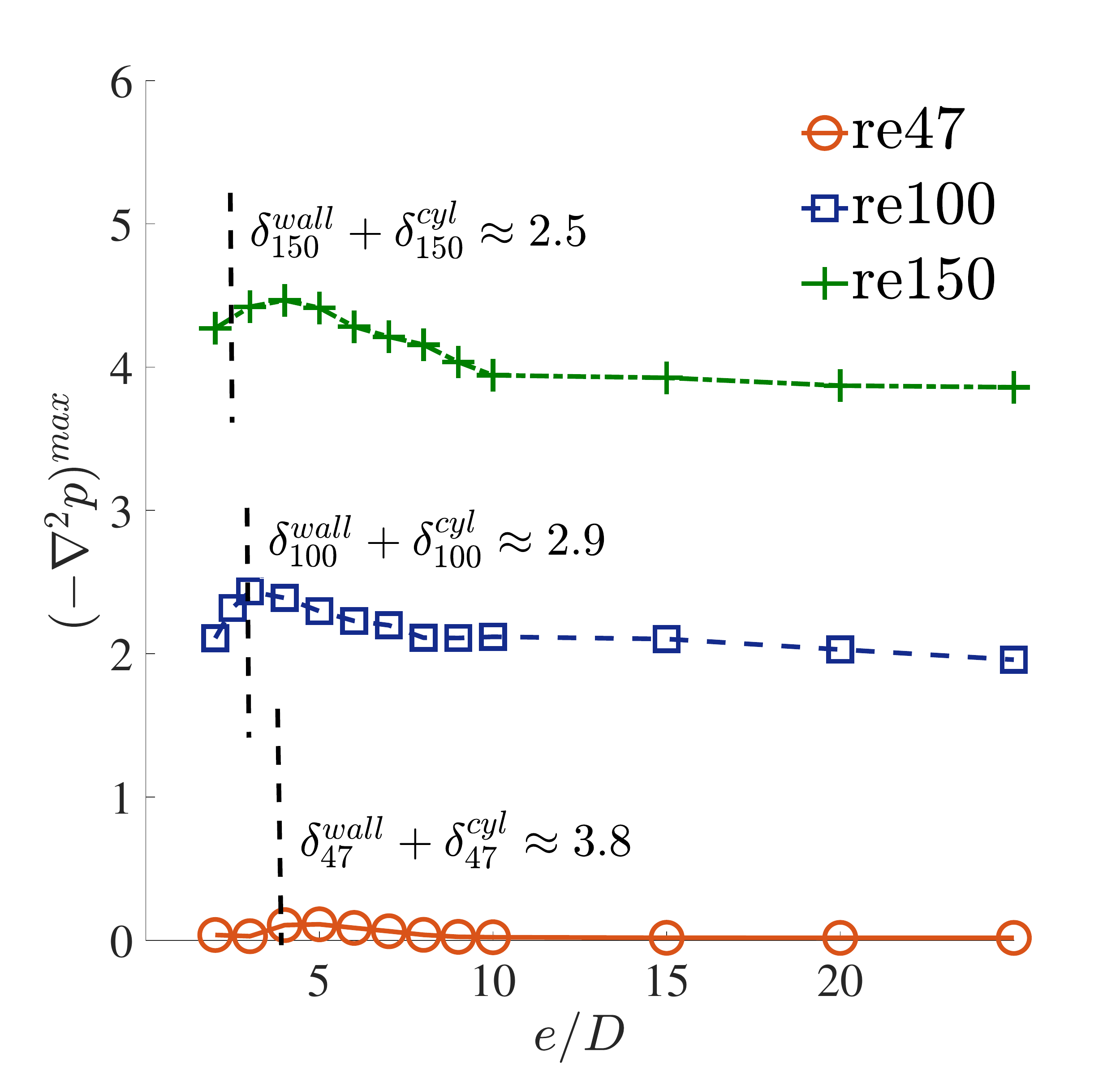}
		\caption{}
		\label{fig:flux_nearWall_re47_re150}
	\end{subfigure}
	\caption{Evolution of maximum fluid-flux divergence at a HSP and hydrodynamic responses for a near-wall cylinder subjecting to the wall-induced proximity interference: (a) a near-wall cylinder at $Re=47$ and (b) a near-wall cylinder at $Re \in [47, 150]$ and $e/D \in [2,15]$. The value of $\delta^{wall}$ and $\delta^{cyl}$ are the boundary layer thicknesses of the wall and cylinder respectively, and their subscripts refer to Reynolds number of cylinder.}
	\label{fig:flux_nearWall}
\end{figure}
\begin{figure} \centering
	\begin{subfigure}[b]{0.5\textwidth}	
		\centering
		\hspace{-25pt}\includegraphics[trim=1cm 0.1cm 0.05cm 0.1cm,scale=0.33,clip]{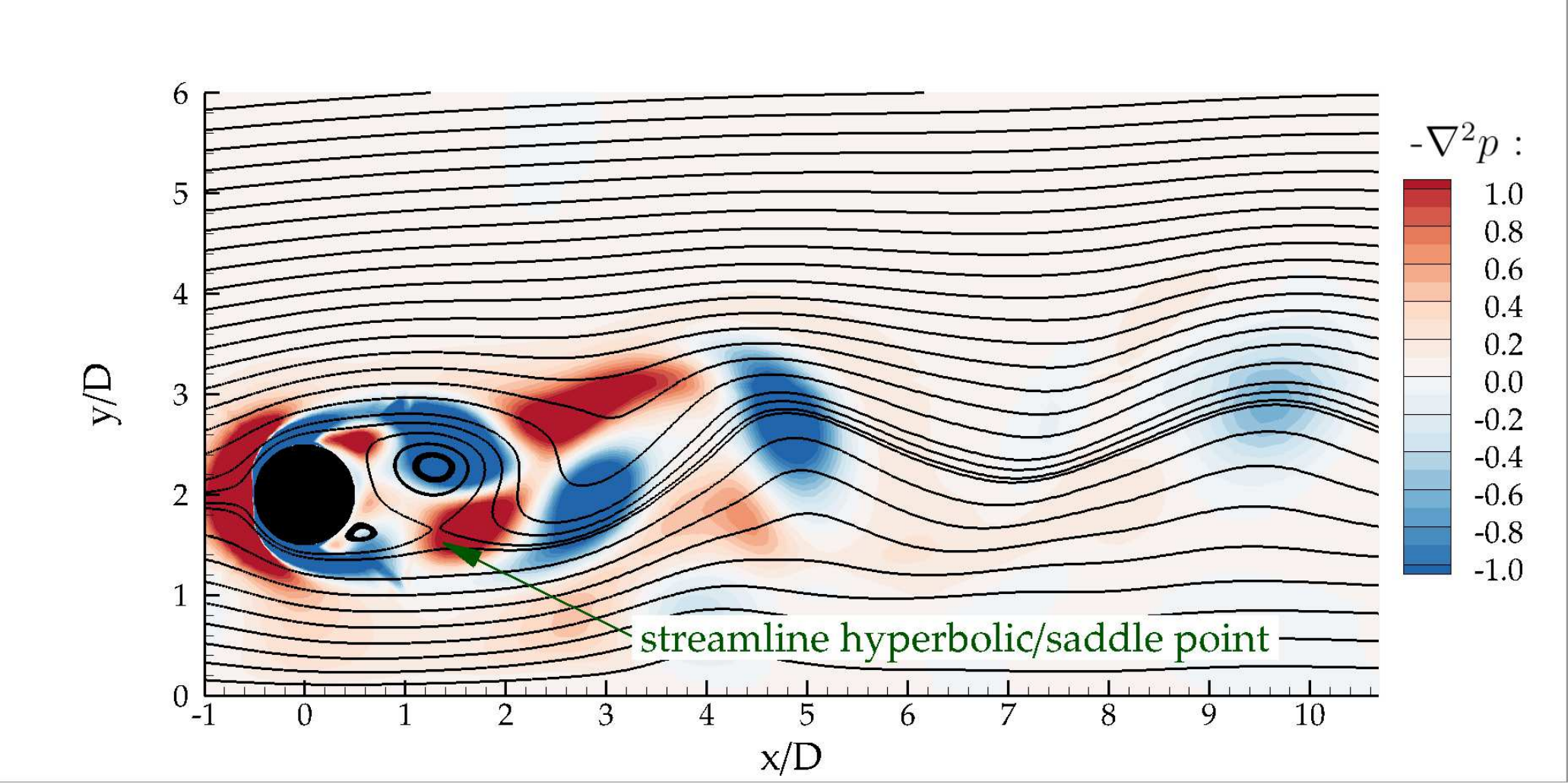}
		\caption{$\qquad\quad$}
		\label{fig:con_flux_re100e20}
	\end{subfigure}%
	\begin{subfigure}[b]{0.5\textwidth}
		\centering
		\hspace{-25pt}\includegraphics[trim=1cm 0.1cm 0.05cm 0.1cm,scale=0.33,clip]{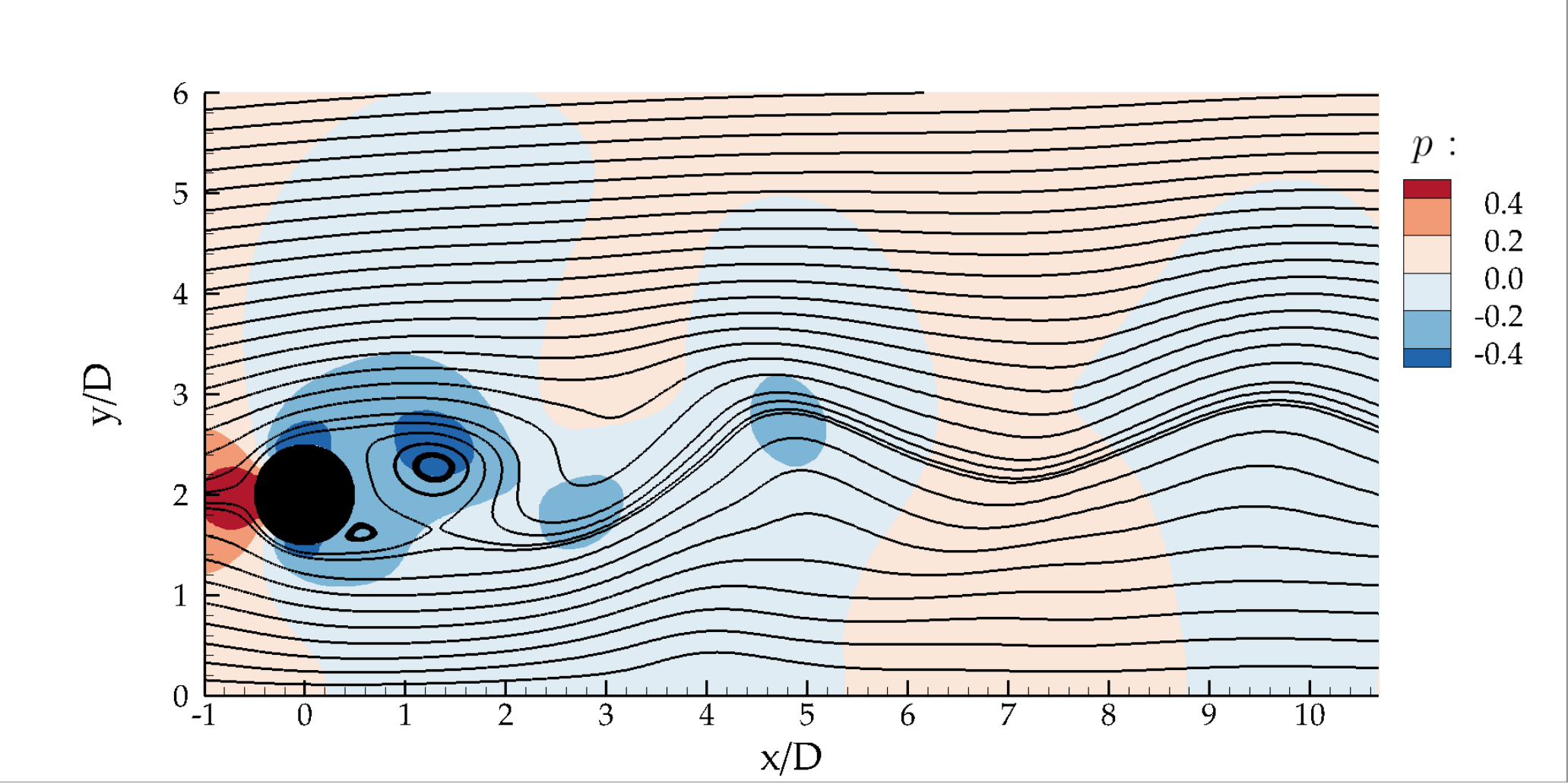}
		\caption{$\qquad\quad$}
		\label{fig:con_p_re100e20}
	\end{subfigure}
	\begin{subfigure}[b]{0.5\textwidth}	
		\centering
		\hspace{-25pt}\includegraphics[trim=1cm 0.1cm 0.05cm 0.1cm,scale=0.33,clip]{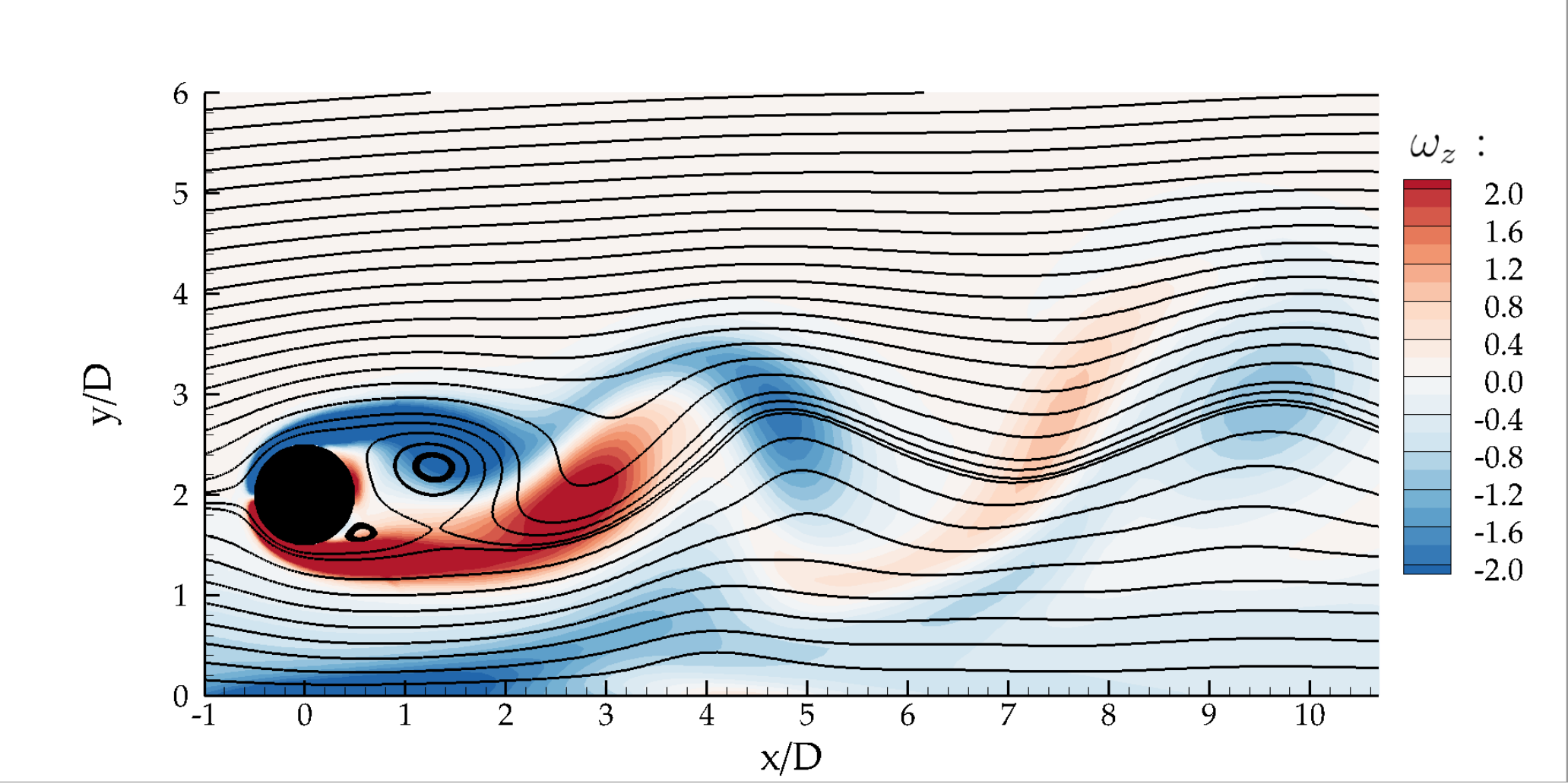}
		\caption{$\qquad\quad$}
		\label{fig:con_wz_re100e20}
	\end{subfigure}%
	\begin{subfigure}[b]{0.5\textwidth}
		\centering
		\hspace{-25pt}\includegraphics[trim=1cm 0.1cm 0.05cm 0.1cm,scale=0.33,clip]{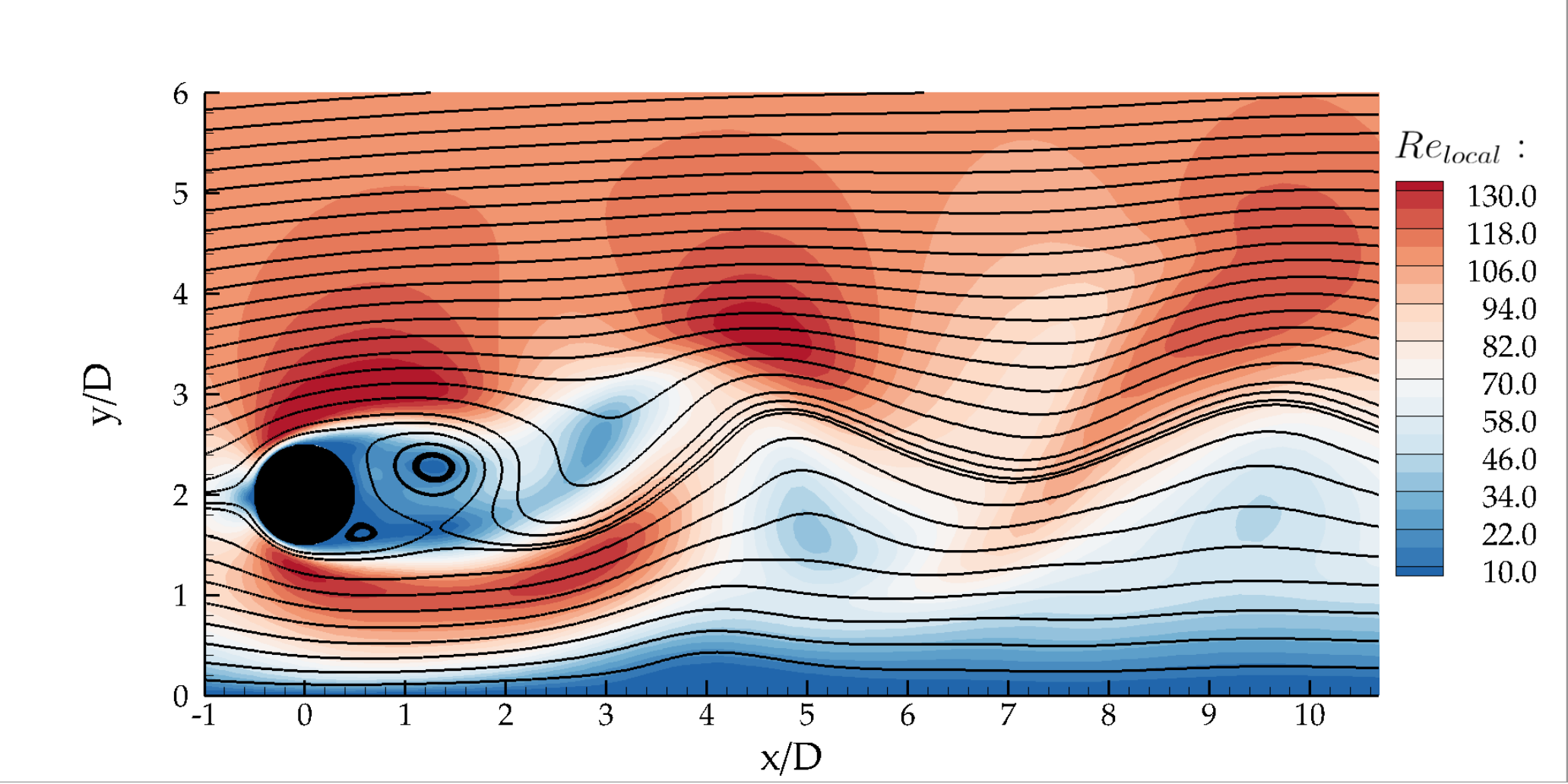}
		\caption{$\qquad\quad$}
		\label{fig:con_local_Re_re100e20}
	\end{subfigure}
	\caption{Wake behind a near-wall cylinder at $Re=100$, $e/D = 2.0$ and $tU/D = 275$: (a) divergence of pressure force; (b) pressure field; (c) spanwise vorticity ($\omega_z$) and (d) local Reynolds number ($Re_{local}$)}
	\label{fig:con_re100e20}
\end{figure}
\begin{figure} \centering
	\begin{subfigure}[b]{0.5\textwidth}	
		\centering
		\hspace{-25pt}\includegraphics[trim=0.1cm 0.1cm 0.1cm 0.1cm,scale=0.3,clip]{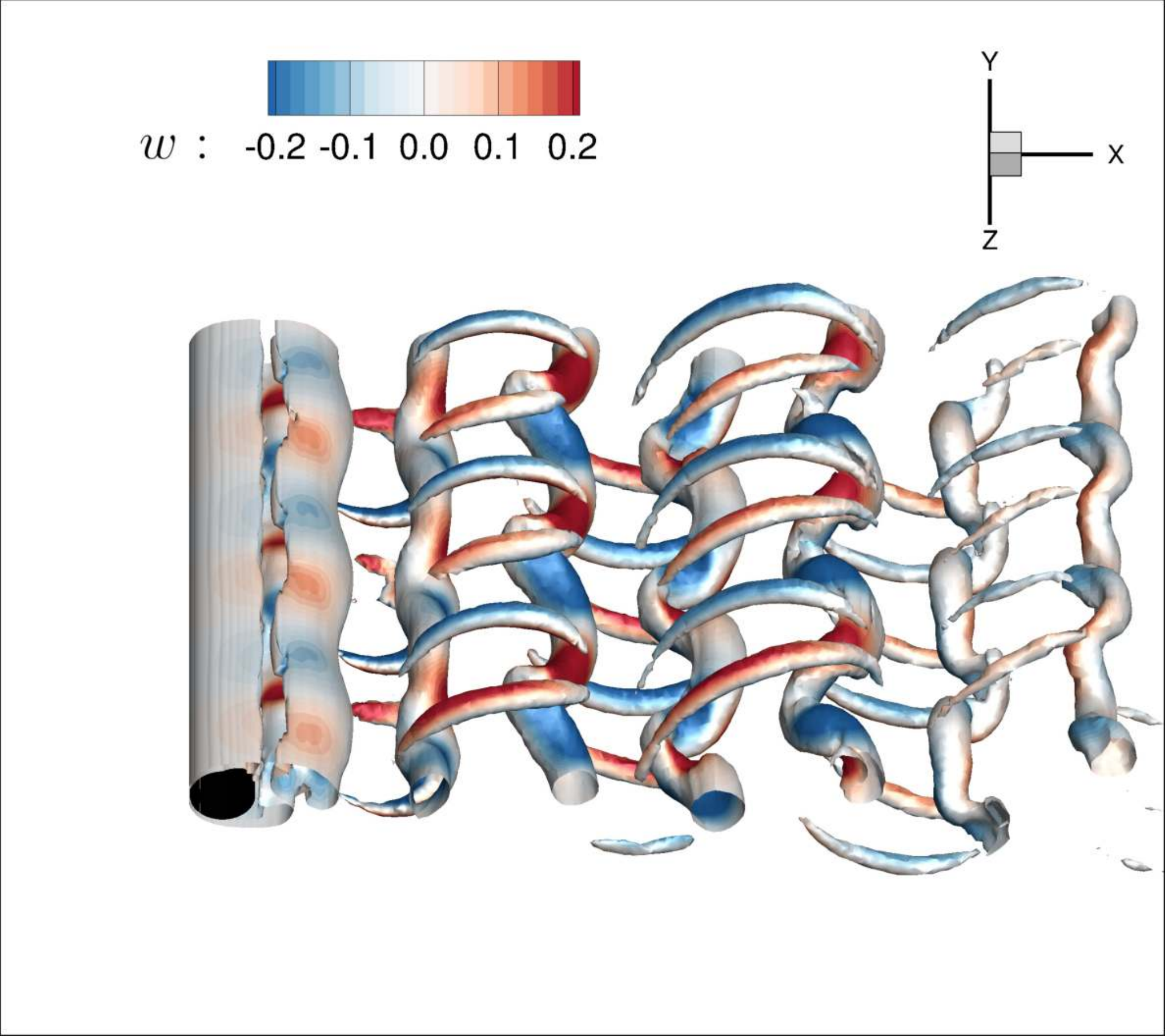}
		\caption{}
		\label{fig:Q_re500_275}
	\end{subfigure}%
	\begin{subfigure}[b]{0.5\textwidth}
		\centering
		\hspace{-25pt}\includegraphics[trim=0.1cm 0.1cm 0.1cm 0.1cm,scale=0.3,clip]{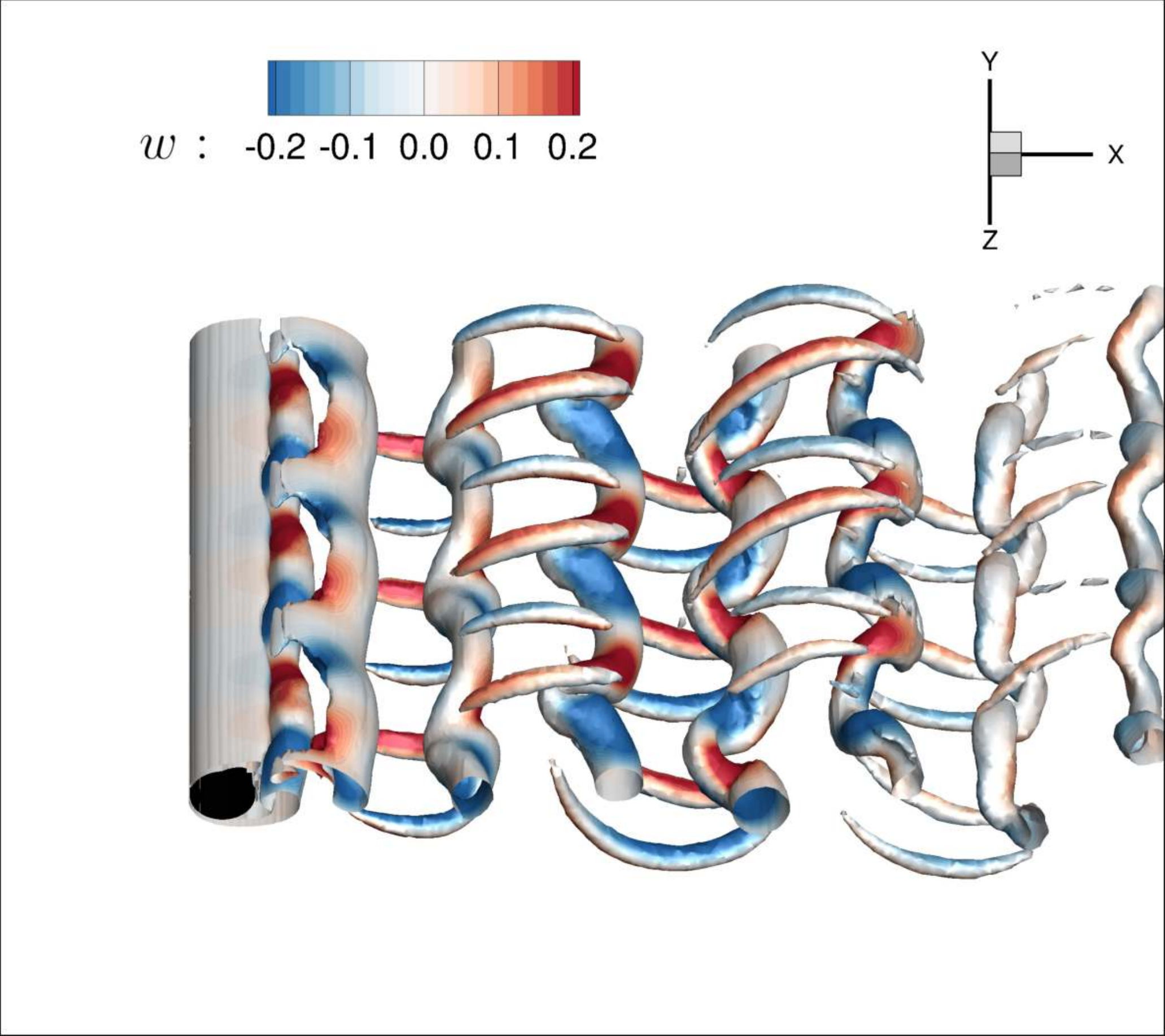}
		\caption{}
		\label{fig:Q_re500_300}
	\end{subfigure}
	\begin{subfigure}[b]{0.5\textwidth}	
		\centering
		\hspace{-25pt}\includegraphics[trim=0.1cm 0.1cm 0.1cm 0.1cm,scale=0.3,clip]{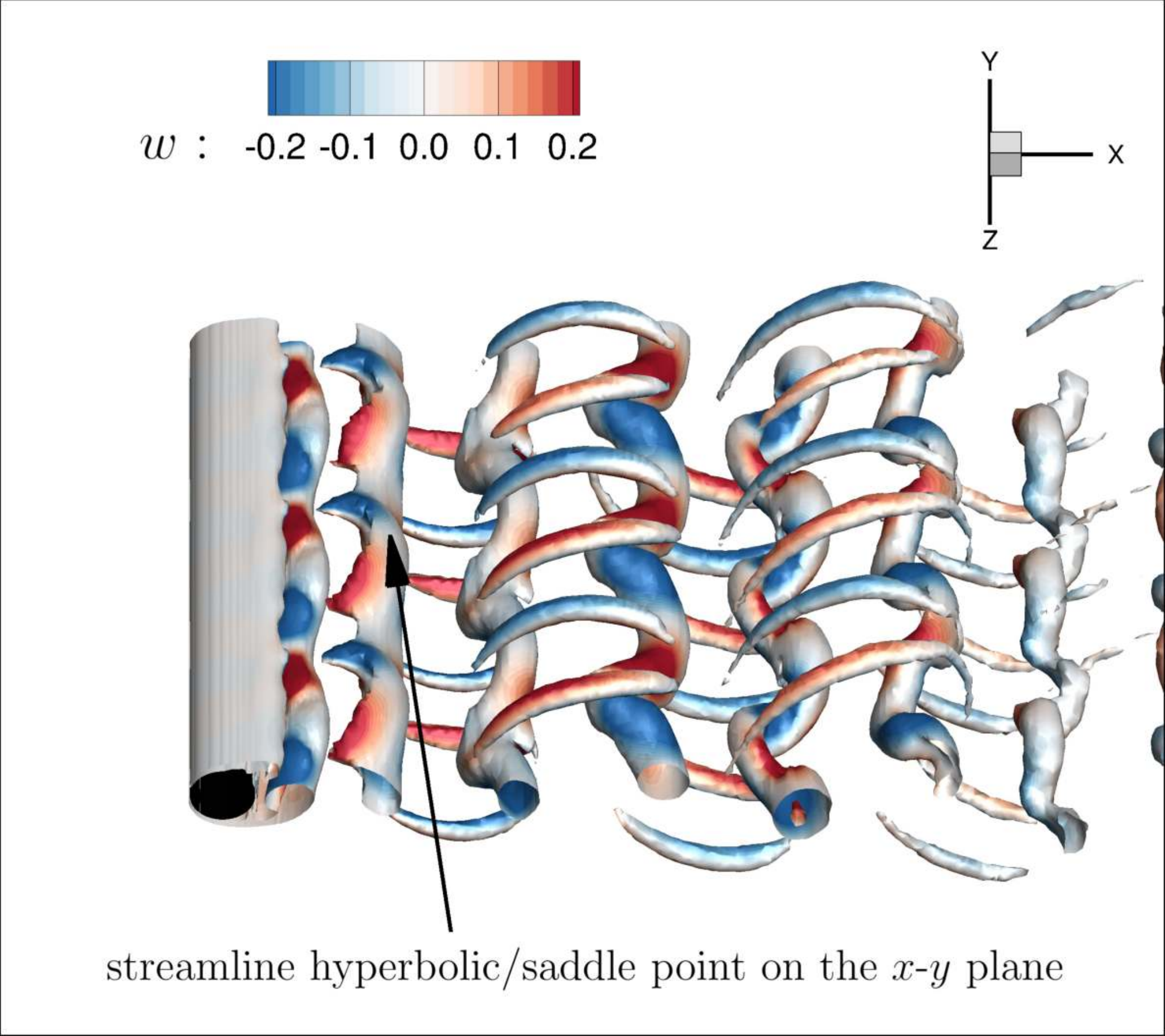}
		\caption{}
		\label{fig:Q_re500_325}
	\end{subfigure}%
	\begin{subfigure}[b]{0.5\textwidth}
		\centering
		\hspace{-25pt}\includegraphics[trim=0.1cm 0.1cm 0.1cm 0.1cm,scale=0.3,clip]{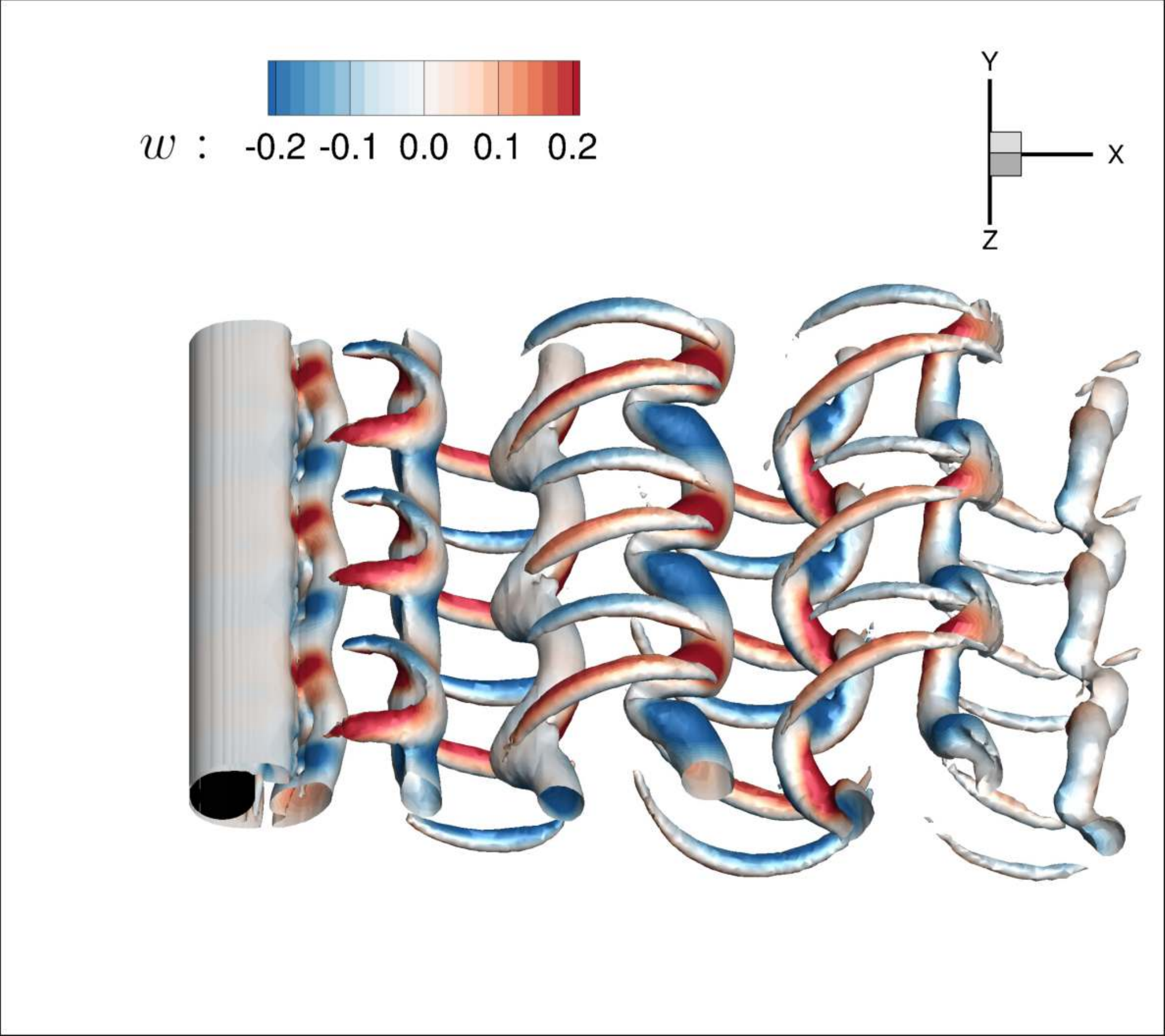}
		\caption{}
		\label{fig:Q_re500_350}
	\end{subfigure}
	\caption{Wake behind a circular cylinder at $Re=500$, $L/D = 10$, $\Delta t=0.05$ visualized using Q criterion ($Q=0.5$) and colored with $z$-component velocity ($w$): (a) $tU/D = 1/4 T$; (b) $tU/D = 2/4 T$; (c) $ tU/D =3/4 T$ and (d) $tU/D = 4/4 T$. The value of $T$ refers to the time interval of one shedding period of the primary vortex shedding roll. A $x$-$y$ planar HSP appears at $tU/D = 3/4 T$ and $z/D=1.5, 5$ and $8.5$ in Figure~\ref{fig:Q_re500_325}}
	\label{fig:Q_re500}
\end{figure}
In the classical benchmark case of a uniform flow over a cylinder, a HSP forms stably at the far-end tip of the re-circulating bubble at $Re \lessapprox 48$. If the Reynolds number exceeds this critical value, as shown by the value of $C^{rms}_l$ in Figure~\ref{fig:iso_flux}, the wake loses its stability and starts developing the periodic vortices downstream, the well-known von K\'arm\'an vortex street. In Figure~\ref{fig:iso_flux}, it can seen that the onset of instability at a HSP appears for $\Psi > 2.0$ approximately. The maximum fluid-flux divergence $(-\nabla^2 p)^{max}$ and the rooted-mean-squared lift coefficient $C^{rms}_l$ grow linearly with respect to Reynolds number. This observation agrees well with the correlation $\Psi = Re h^2 /\Delta t$ in Section~\ref{sec:vonNeumman}. Whereas the mean drag force is reduced as Reynolds number increases until $Re \approx 48$. Unlike the values of $-\nabla^2 p$ and $C^{rms}_l$, after the onset of instability at HSP, the values of $C^m_d$ do not alter significantly, instead. On the other hand, it is also found in Figure~\ref{fig:time_space} that the instability of a HSP shows a quadratic growth (the dotted line with square marks) and a linear growth (the solid line with circle marks) with respect to $h$ and $\Delta t$ respectively, based on the implemented numerical formulation in this article. For instance, if the time step ($\Delta t$) is altered, the Reynolds number ($Re=100$) and the mesh size ($h=0.08834$) are held constant. On the other hand, if the mesh size ($h$) is changed, the Reynolds number ($Re=45$) and the time step ($\Delta t = 0.036$) are keep constant. The results in Figure~\ref{fig:time_space} also confirm that the instability of a HSP becomes excited at $\Psi > 2.0$ approximately. The increasing positive magnitude of  $(-\nabla^2 p)^{max}$ in Figure~\ref{fig:iso} indicates a relatively strong net positive fluid flux and pressure value around a HSP in wake. This relative high pressure value within the low pressure wake of cylinder induces an adverse pressure gradient, as shown in Figure~\ref{fig:con_p_re100}, deteriorating the hydrodynamic stability of shear layers. Furthermore, the strong positive fluid flux could induces local third-dimensional fluid flux as well. If it becomes strong enough at high Reynolds number, it is possible to exceed the local critical Reynolds number and cause the instabilities, e.g., the onset of vortex shedding or three-dimensional braids/ribs. The detailed discussion of fluid three dimensionality will be presented in Section~\ref{sec:three}.

As proven in Section~\ref{sec:ppe}, Figure~\ref{fig:con_flux_re100} clearly shows that a HSP is indeed associated with a strong fluid-flux divergence, $\nabla \cdot (\rho \bm{u}\cdot \nabla \bm{u}) > 0.0$. The relatively high pressure zone is formed around a HSP, as shown in Figure~\ref{fig:con_p_re100}. It induces adverse pressure gradients to the shedding shear layers and detrimental to the near-wake stability. As analytically proven in Section~\ref{sec:vonNeumman}, a HSP is meta-stable with respect to balanced shear strains $\partial v/\partial x = \partial u/\partial y$, which refers to a region of zero spanwise vorticity $\omega_z = 0$ in this case, e.g., the shear-layer interfaces. As result, it is found that a HSP always forms along the shear-layer interfaces, e.g., as shown in the instantaneous flow field in Figure~\ref{fig:con_wz_re100}. In Figure~\ref{fig:con_local_Re_re100}, we further realized that a HSP situates in the neighborhood of low local Reynolds numbers $Re_{local}$. Recollecting the discussions in Section~\ref{sec:vonNeumman} and the results in Figure~\ref{fig:iso_flux}, we understand the region of a relatively low local Reynolds number makes a HSP less sensitive the background noises and is relatively appealing to its numerical stability. These low fluid momentum surrounding a HSP also represent the stagnant regions in wake where the fluid kinetic energy transfer of the shear layers is hindered.

The conclusions drawn above can be confirmed in the wake behind side-by-side cylinders too. In the side-by-side configurations, the adjacent cylinder could possibly exerts a strong gap-flow induced proximity interference to its counterpart and induces asymmetric shear-layer interaction in wake. In Figure~\ref{fig:flux_sbs_re47}, the cases at $Re=47$ is chosen for investigation, because this Reynolds number is close to the critical Reynolds number $Re \approx 48$ and the flow is sensitive to the disturbances. It is noticed in Figure~\ref{fig:flux_sbs_re47} that the evolution of the near-wake instability is directly associated with the value of fluid flux divergence ($-\nabla^2 p$) at a HSP in wake. Generally, as both cylinders are placed close to each other, the asymmetric shear-layer interactions develop in wake, and the near-wake instability gradually appears. Based on the discussions in Section~\ref{sec:vonNeumman}, we understand that these asymmetric shear-layer interactions can deteriorate the stability of a HSP in wake. Consequently, the instability of a limit periodic solution appears, the onset of vortex shedding. On the other hand, the near-wake instability becomes suppressed instead, while the cylinders' boundary layers start directly interfering each other, e.g., the case of $2\delta^{cyl} \approx 2.8$ in Figure~\ref{fig:flux_sbs_re47}. This phenomenon is also linked with abrupt drops of the fluid-flux divergence around a HSP and the corresponding fluctuation of the lift forces. This finding is subsequently confirmed at different Reynolds numbers, the cases of $2\delta^{cyl}_{100}$ and $2\delta^{cyl}_{150}$ in Figure~\ref{fig:flux_sbs_re47_re150}. Instead, the mean drag force of each cylinder keep increasing at small gap distances until $g/D \approx 2.0$ in this investigation, due to the formation of large recirculation regions. This is different from the dynamics of the near-wake instability, in which instability is suppressed when the boundary layer directly interact with each other. Moreover, Figure~\ref{fig:flux_sbs_re47_re150} also shows that the fluid-flux divergence at a HSP arise together with Reynolds number. If the strength of fluid-flux divergence at a HSP surges, a stronger pressure field is formed in the neighborhood and more fluid tends to diverge away from this point. It is possible that the fluid three dimensionality appears earlier in the gap flow at these gap ratios. This observation is supported with the findings in~\cite{liu2018dynamics}, in which the flow transition appears earlier in the gap flow of side-by-side cylinders. The detailed discussions of the fluid three dimensionality and the fluid-flux divergence at a HSP will be presented in Section~\ref{sec:three}.  

Similar to the findings in an isolated cylinder, the high fluid-flux divergence and pressure value are found around a HSP in Figure~\ref{fig:con_flux_re100td15} and~\ref{fig:con_p_re100td15}. In the wake behind side-by-side cylinders, the shear-layer interaction is further intensified and complicated. Nonetheless, similar to the case of an isolated cylinder in Figure~\ref{fig:con_wz_re100}, it is found that the streamline hyperbolic/saddle points are apparently formed along the shear-layer interfaces where $\partial v/\partial x - \partial u/\partial y = 0$. It can also be seen in Figure~\ref{fig:con_local_Re_re100td15} that each HSP is associated with a region of low local Reynolds number (low fluid inertia over the viscous force), from where the supply of fluid kinetic energy is cut, the shear layer gets strained and the vortices start shedding. These observations agree well with the findings in the case of an isolated cylinder.

In a near-wall configuration, if a cylinder is placed sufficient close to the wall, the wall will exerts a proximity interference to the wake and boundary layer of this cylinder. It results in an asymmetric shear layer interaction in wake and causes the onset of instability at a HSP in wake, as shown in Figure~\ref{fig:flux_nearWall_re47}. It shows that the wake around a HSP becomes unstable and its $(-\nabla^2 p)^{max}$ and $C^{rms}_l$ values get amplified for $e/D \in [4,10]$ at $Re=47$. Similar to the gap-flow induced proximity interference, the mean drag force is amplified as the gap distance reduces. However, different from the side-by-side arrangements, both the near-wake instability and the mean drag force are suppressed as soon as the boundary layer of cylinder directly interact with the wall boundary layer at $\delta^{wall}+\delta^{cyl} \approx 3.8$, $Re=47$ and $L_u = 16D$ in Figure~\ref{fig:flux_nearWall_re47}. This wall-induced regulation effect is also confirmed at different Reynolds numbers in Figure~\ref{fig:flux_nearWall_re47_re150}. As proven in Section~\ref{sec:vonNeumman}, if the shear-layer interaction around a HSP is imbalanced, e.g., the asymmetric wake caused by the wall-induced proximity interference in Figure~\ref{fig:flux_nearWall_re47} and Figure~\ref{fig:flux_nearWall_re47_re150}, the HSP in wake becomes unstable. In all cases of an isolated, side-by-side or a near-wall cylinder, the fluid-flux divergence at a HSP surges together with the Reynolds number. However, the HSP in wake of side-by-side cylinders manifest a larger $-\nabla^2 p$ value than that of a near-wall cylinder over different gap distances  at the same Reynolds number. This difference of maximum $-\nabla^2 p$ value becomes more prominent as the Reynolds number increases.
\begin{figure} \centering
	\begin{subfigure}[b]{0.5\textwidth}	
		\centering
		\hspace{-25pt}\includegraphics[trim=0.1cm 0.1cm 0.1cm 0.1cm,scale=0.25,clip]{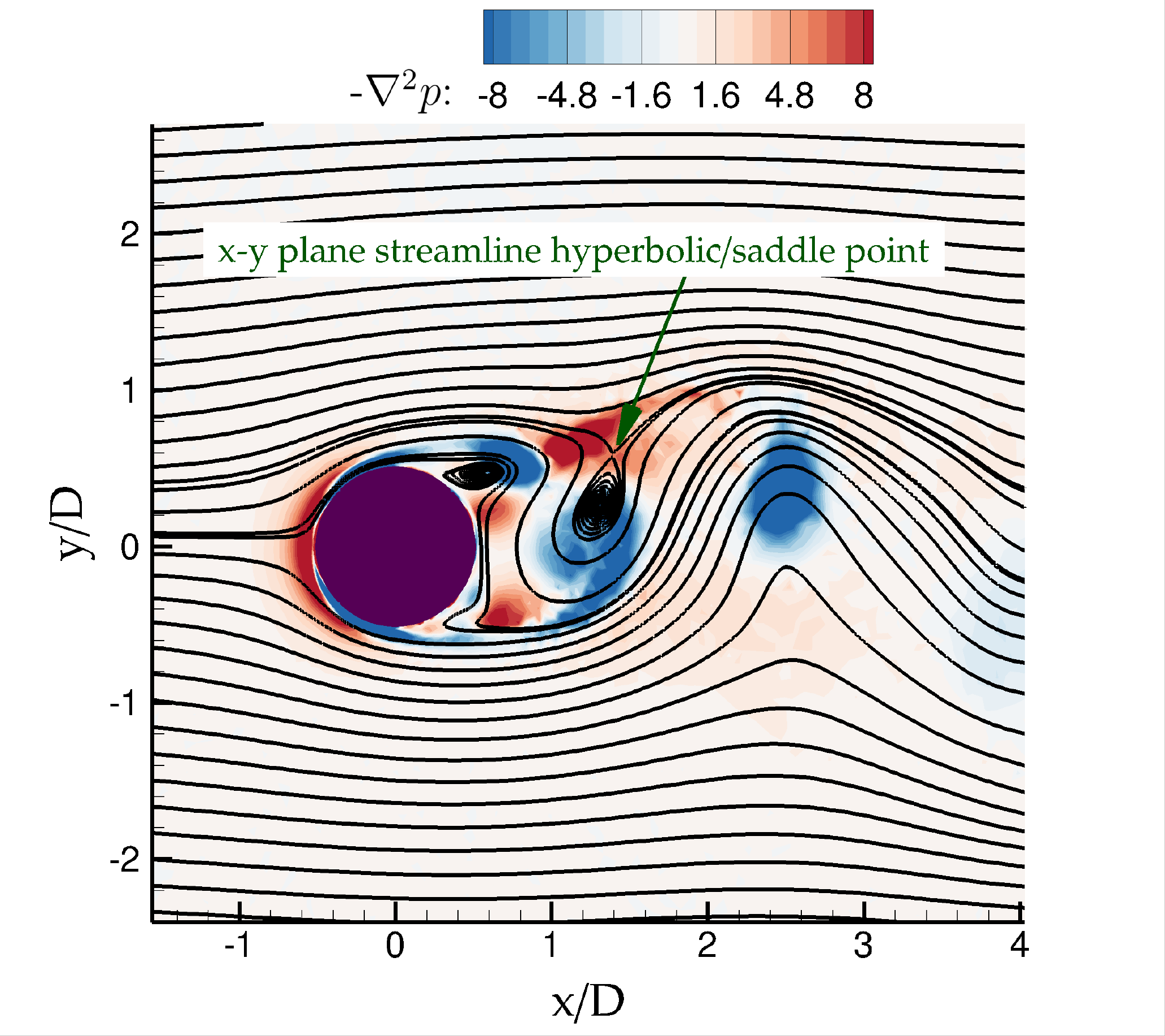}
		\caption{}
		\label{fig:re500_flux}
	\end{subfigure}%
	\begin{subfigure}[b]{0.5\textwidth}
		\centering
		\hspace{-25pt}\includegraphics[trim=0.1cm 0.1cm 0.1cm 0.1cm,scale=0.25,clip]{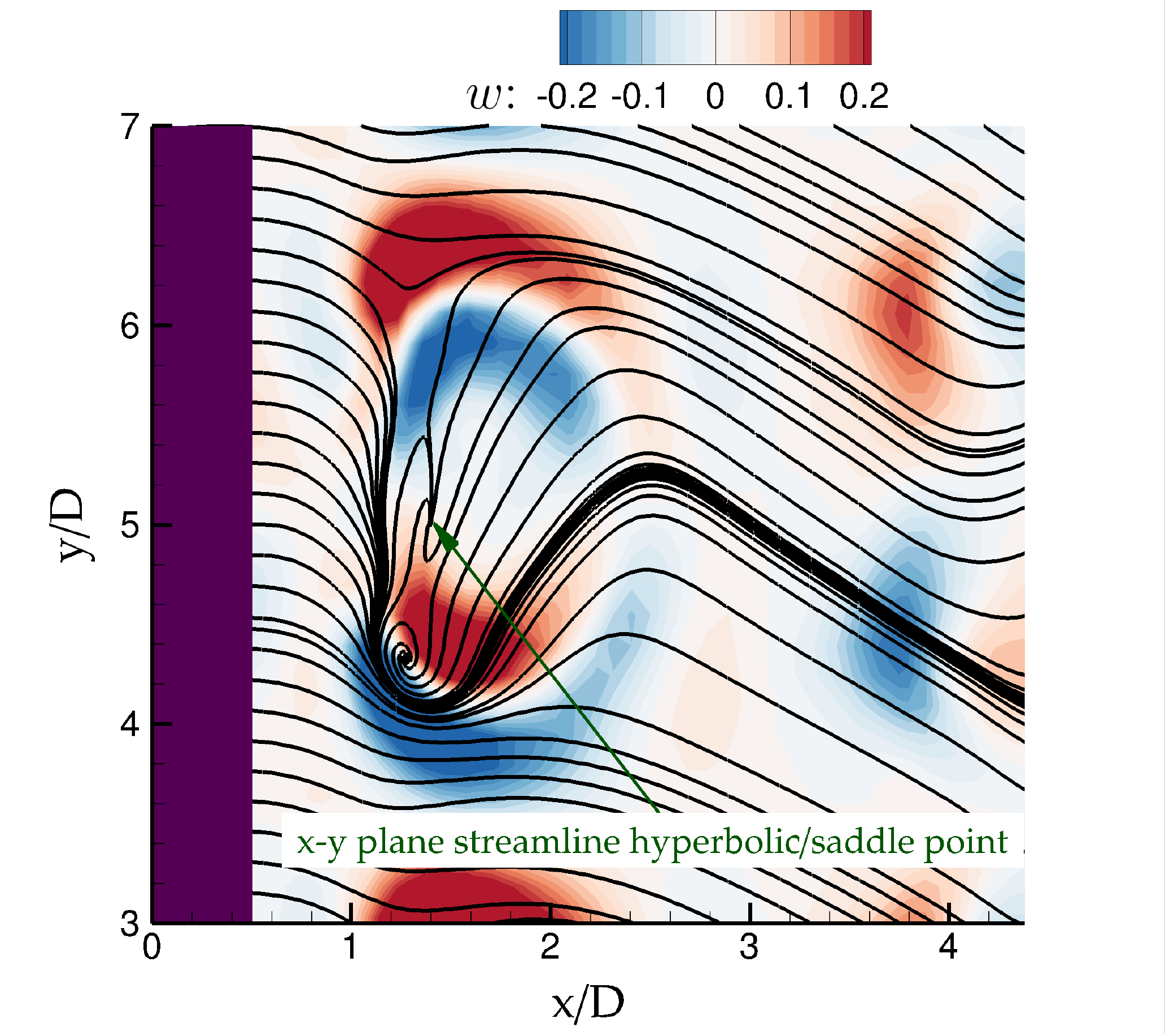}
		\caption{}
		\label{fig:re500_w}
	\end{subfigure}
	\caption{Wake behind a circular cylinder at $Re=500$, $L/D = 10$, $\Delta t=0.05$ and $tU/D = 3/4 T$: (a) a $x$-$y$ plane streamline singular hyperbolic point on the $z/D = 5.0$ plane and colored with $-\nabla^2 p$ contour and (b) a $x$-$y$ planar HSP on the $y/D = 0.5855$ plane and colored with $w$ contour. This HSP appears at $\bm{x}/D = (1.393,0.586,5.0)$. }
	\label{fig:three}
\end{figure}
\begin{figure} \centering
	\begin{subfigure}[b]{0.5\textwidth}	
		\centering
		\hspace{-25pt}\includegraphics[trim=0.1cm 0.1cm 0.1cm 0.1cm,scale=0.3,clip]{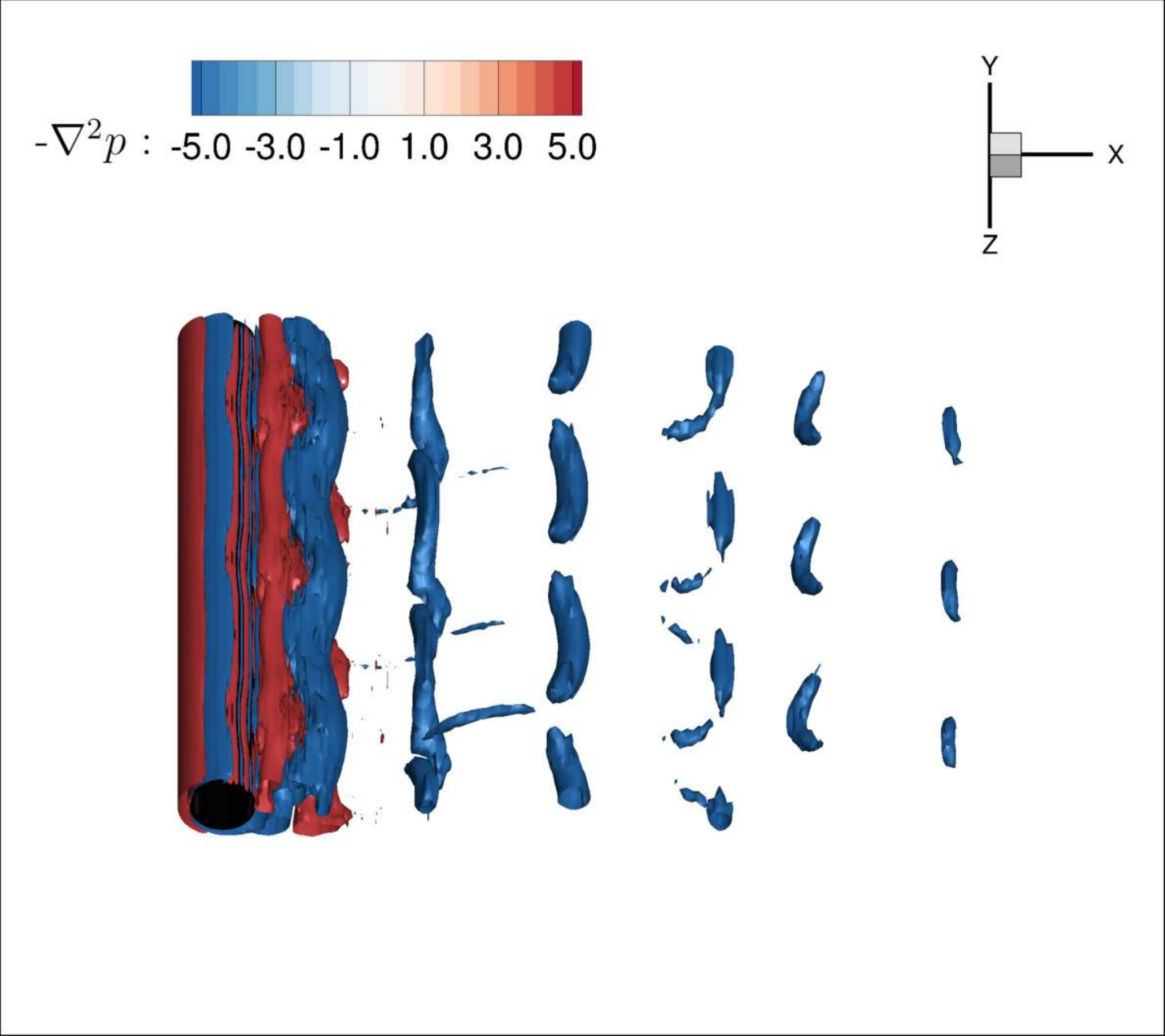}
		\caption{}
		\label{fig:flux_re500_275}
	\end{subfigure}%
	\begin{subfigure}[b]{0.5\textwidth}
		\centering
		\hspace{-25pt}\includegraphics[trim=0.1cm 0.1cm 0.1cm 0.1cm,scale=0.3,clip]{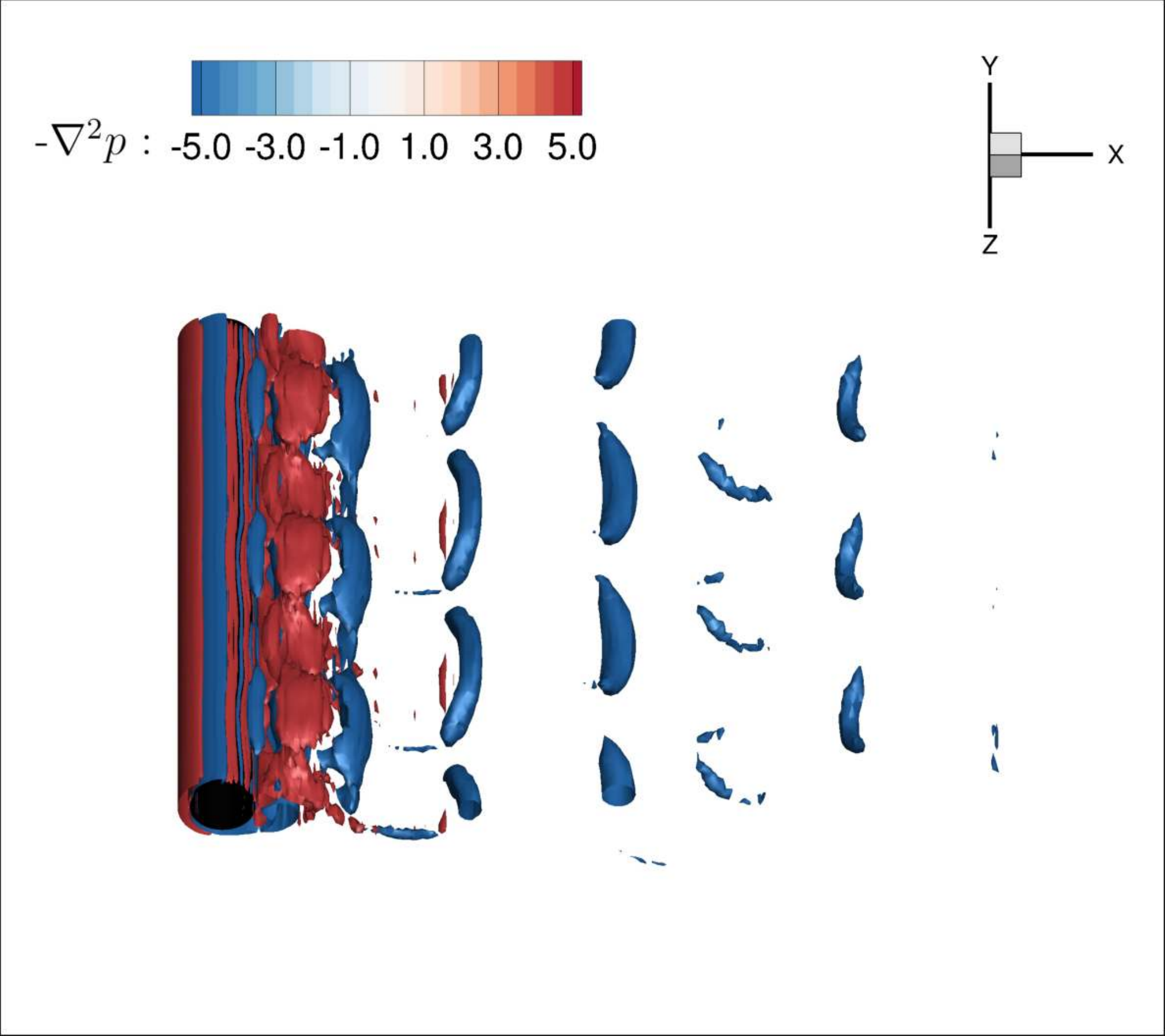}
		\caption{}
		\label{fig:flux_re500_300}
	\end{subfigure}
	\begin{subfigure}[b]{0.5\textwidth}	
		\centering
		\hspace{-25pt}\includegraphics[trim=0.1cm 0.1cm 0.1cm 0.1cm,scale=0.3,clip]{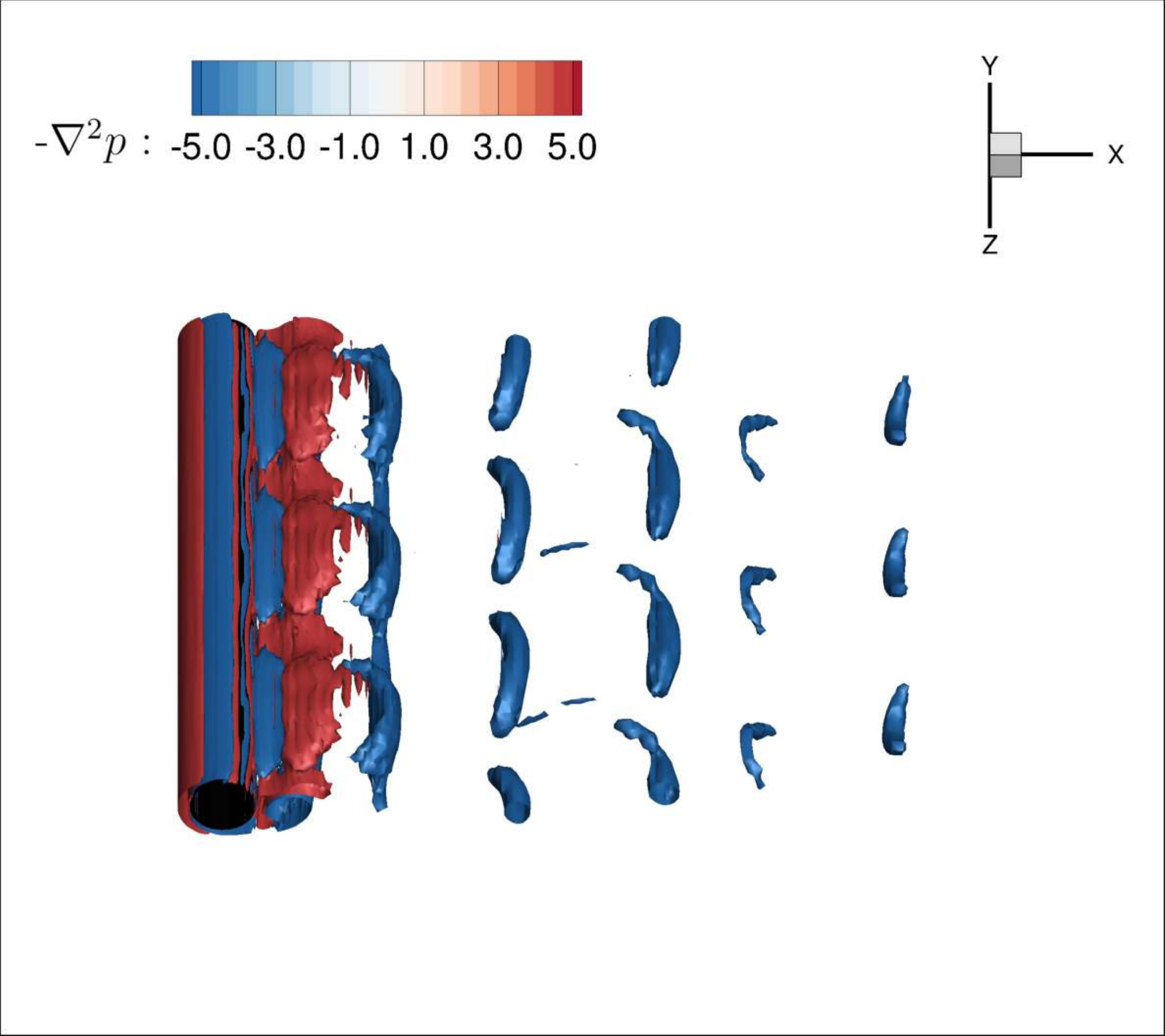}
		\caption{}
		\label{fig:flux_re500_325}
	\end{subfigure}%
	\begin{subfigure}[b]{0.5\textwidth}
		\centering
		\hspace{-25pt}\includegraphics[trim=0.1cm 0.1cm 0.1cm 0.1cm,scale=0.3,clip]{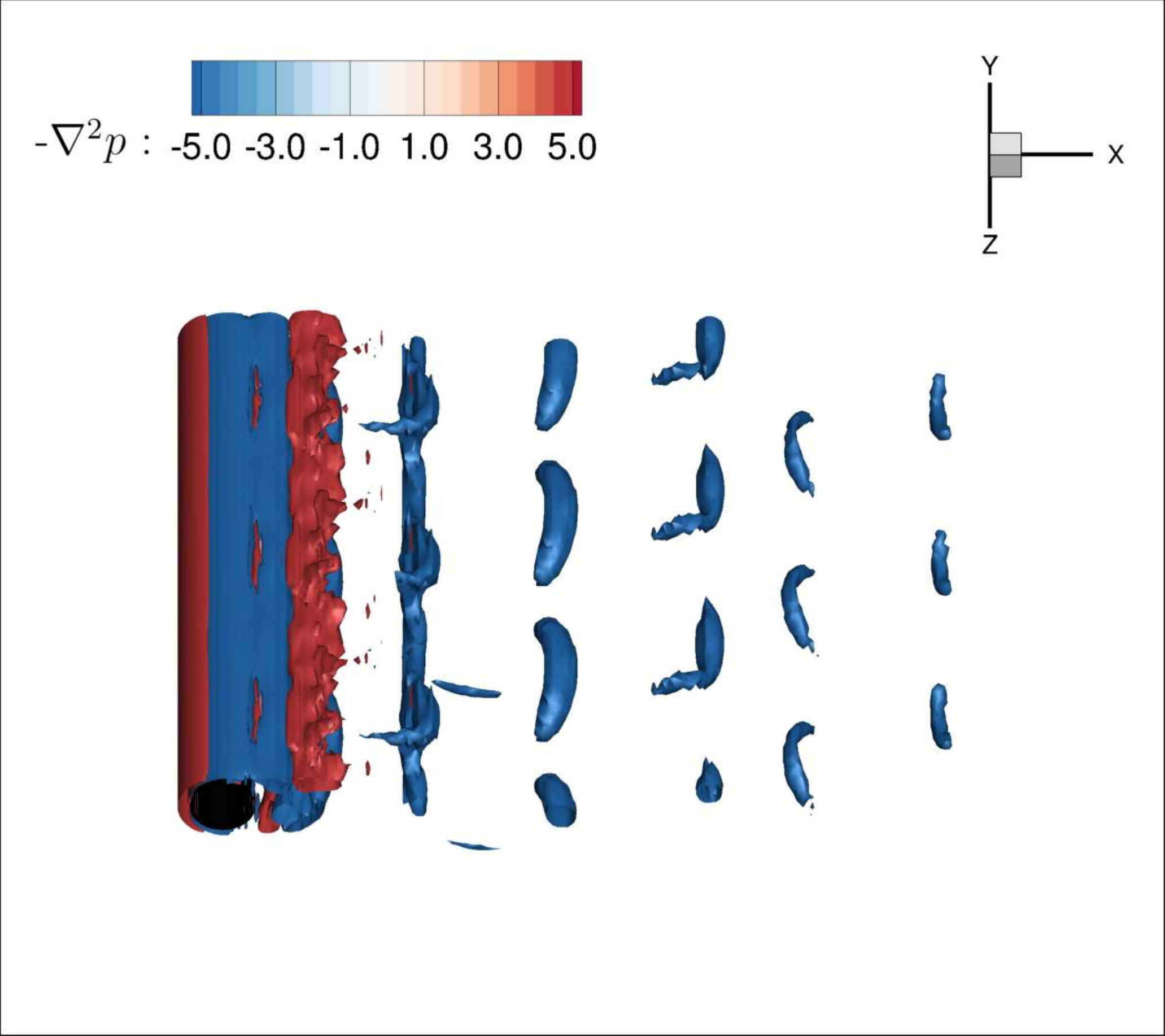}
		\caption{}
		\label{fig:flux_re500_350}
	\end{subfigure}
	\caption{Wake behind a circular cylinder at $Re=500$, $L/D = 10$ and $\Delta t=0.05$ visualized and colored using the divergence of pressure force: (a) $tU/D = 1/4 T$; (b) $tU/D = 2/4 T$; (c) $tU/D = 3/4 T$ and (d) $tU/D = 4/4 T$. The $x$-$y$ planar HSP appears at $tU/D = 3/4 T$ and $z/D = 1.5, 5$ and $8.5$, at where the large positive values of fluid-flux divergence are accumulated.}
	\label{fig:flux_re500}
\end{figure}
\begin{figure} \centering
	\begin{subfigure}[b]{0.5\textwidth}	
		\centering
		\hspace{-25pt}\includegraphics[trim=0.1cm 0.1cm 0.1cm 0.1cm,scale=0.3,clip]{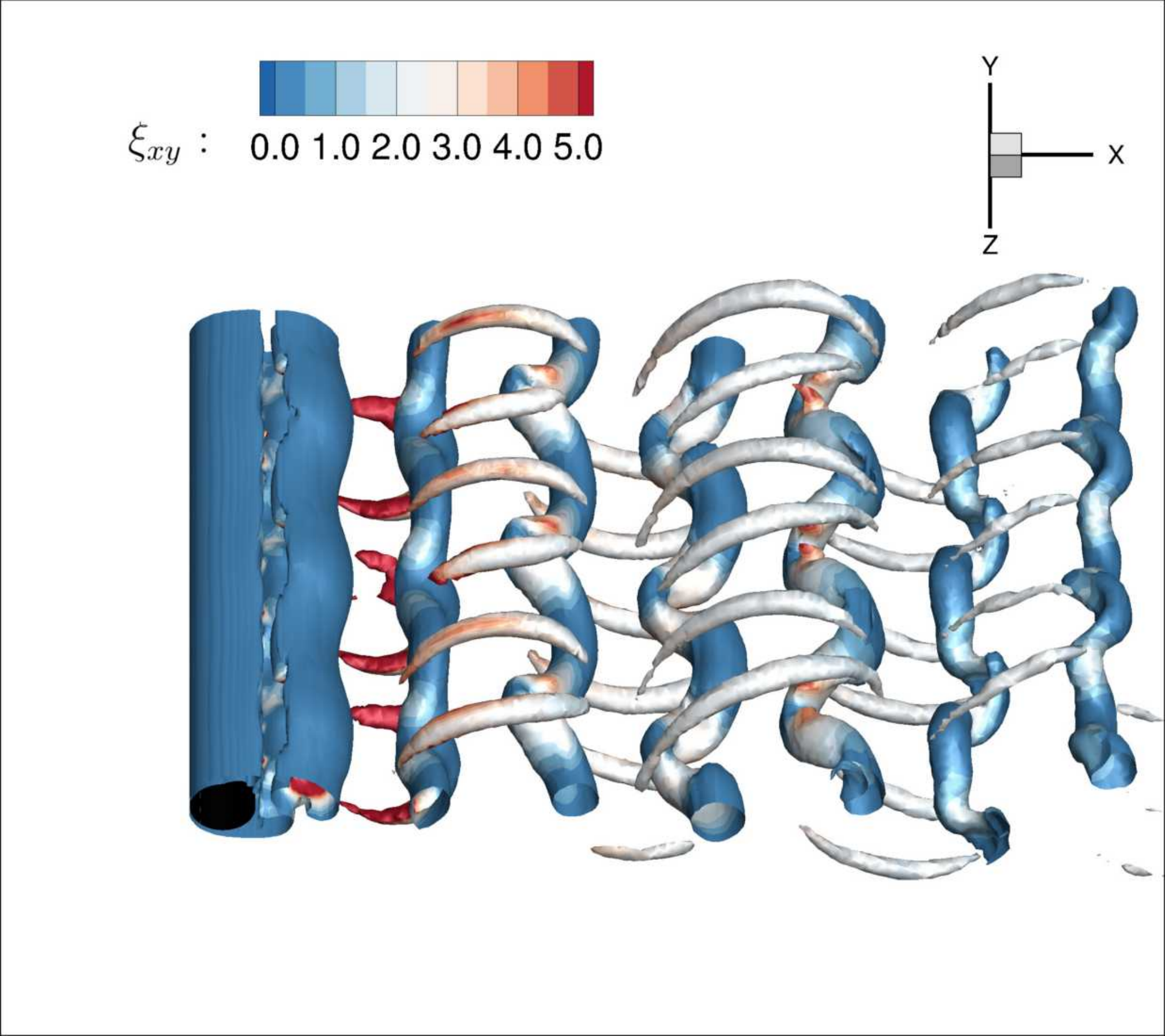}
		\caption{}
		\label{fig:en_re500_275}
	\end{subfigure}%
	\begin{subfigure}[b]{0.5\textwidth}
		\centering
		\hspace{-25pt}\includegraphics[trim=0.1cm 0.1cm 0.1cm 0.1cm,scale=0.3,clip]{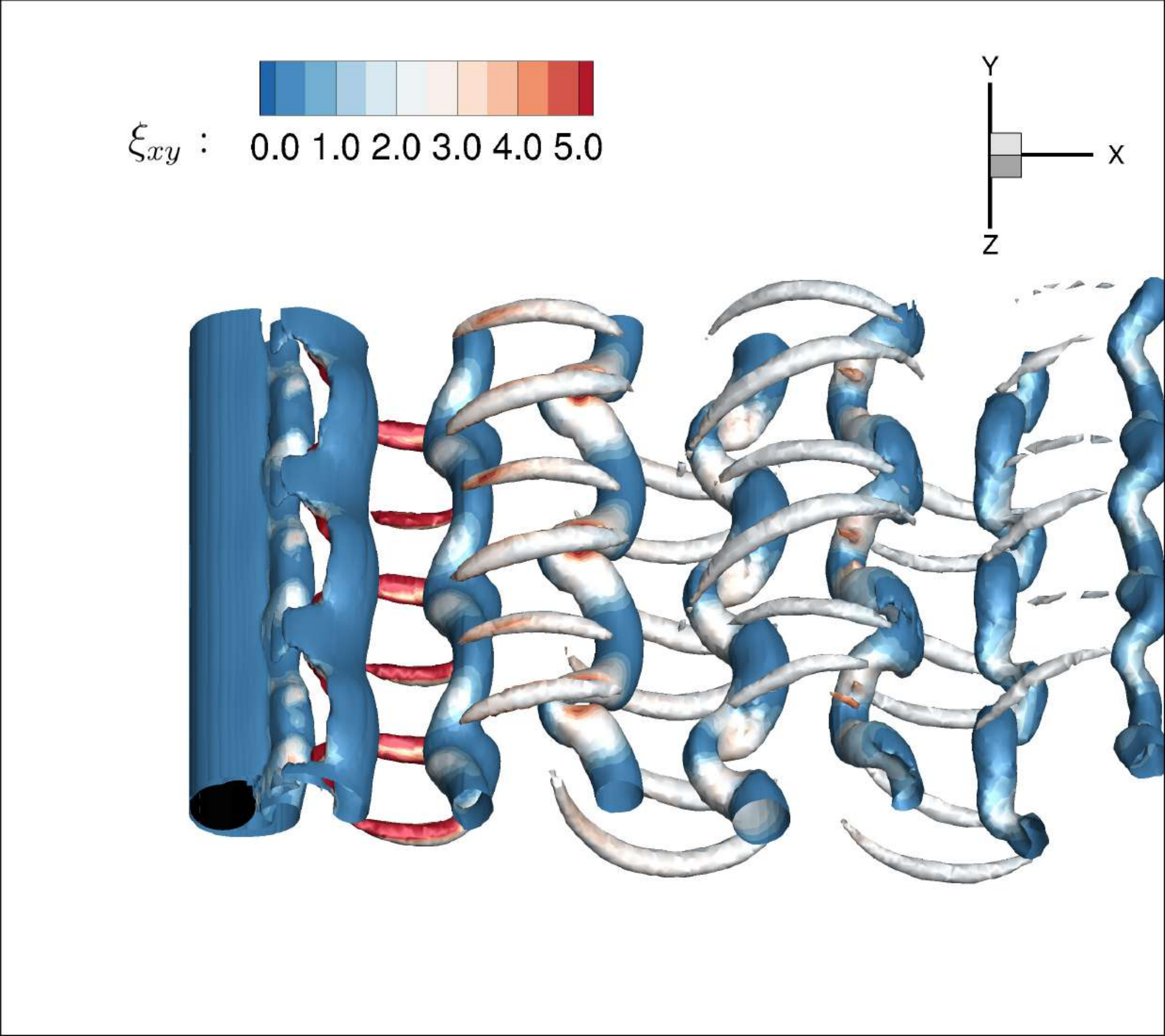}
		\caption{}
		\label{fig:en_re500_300}
	\end{subfigure}
	\begin{subfigure}[b]{0.5\textwidth}	
		\centering
		\hspace{-25pt}\includegraphics[trim=0.1cm 0.1cm 0.1cm 0.1cm,scale=0.3,clip]{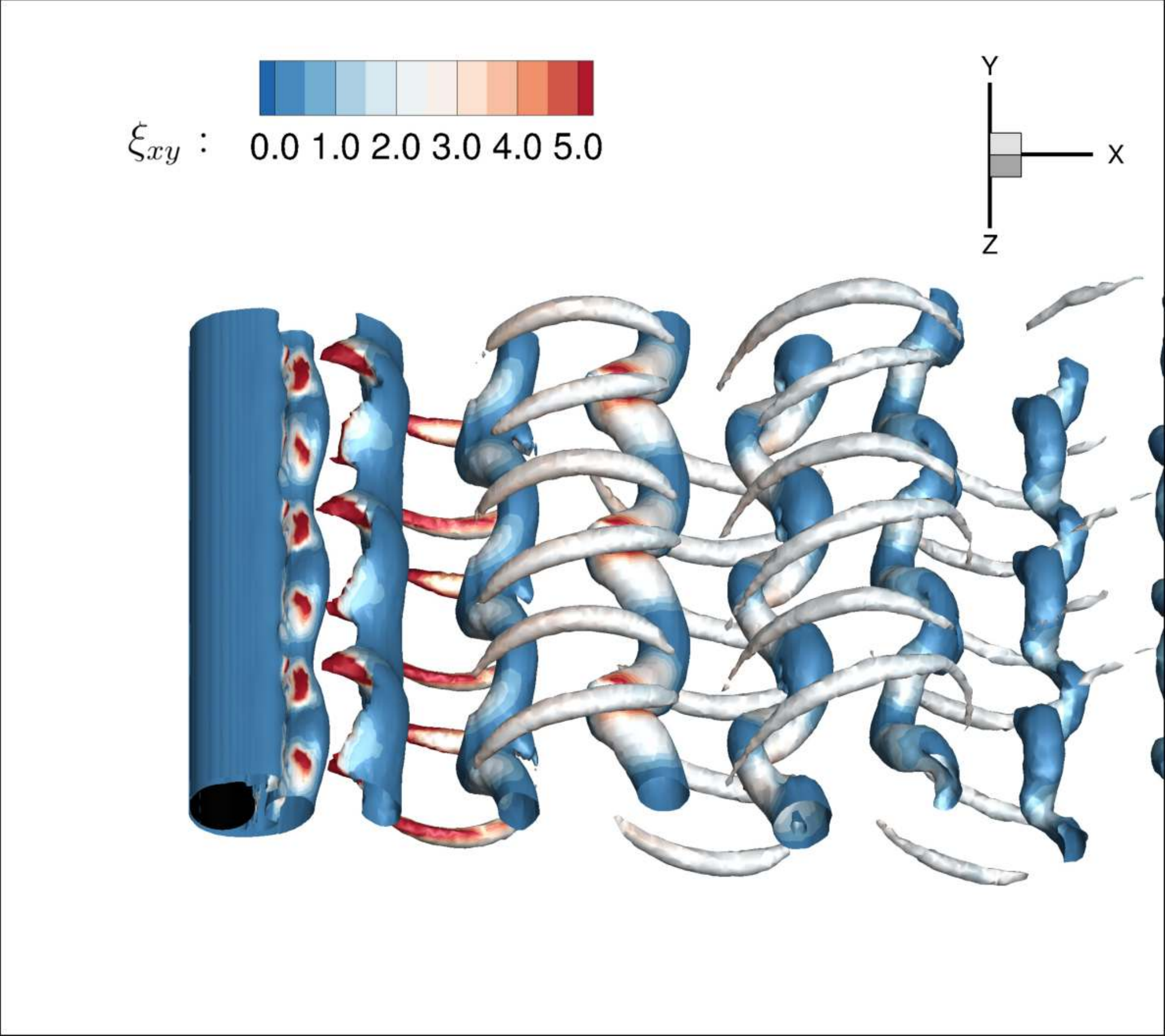}
		\caption{}
		\label{fig:en_re500_325}
	\end{subfigure}%
	\begin{subfigure}[b]{0.5\textwidth}
		\centering
		\hspace{-25pt}\includegraphics[trim=0.1cm 0.1cm 0.1cm 0.1cm,scale=0.3,clip]{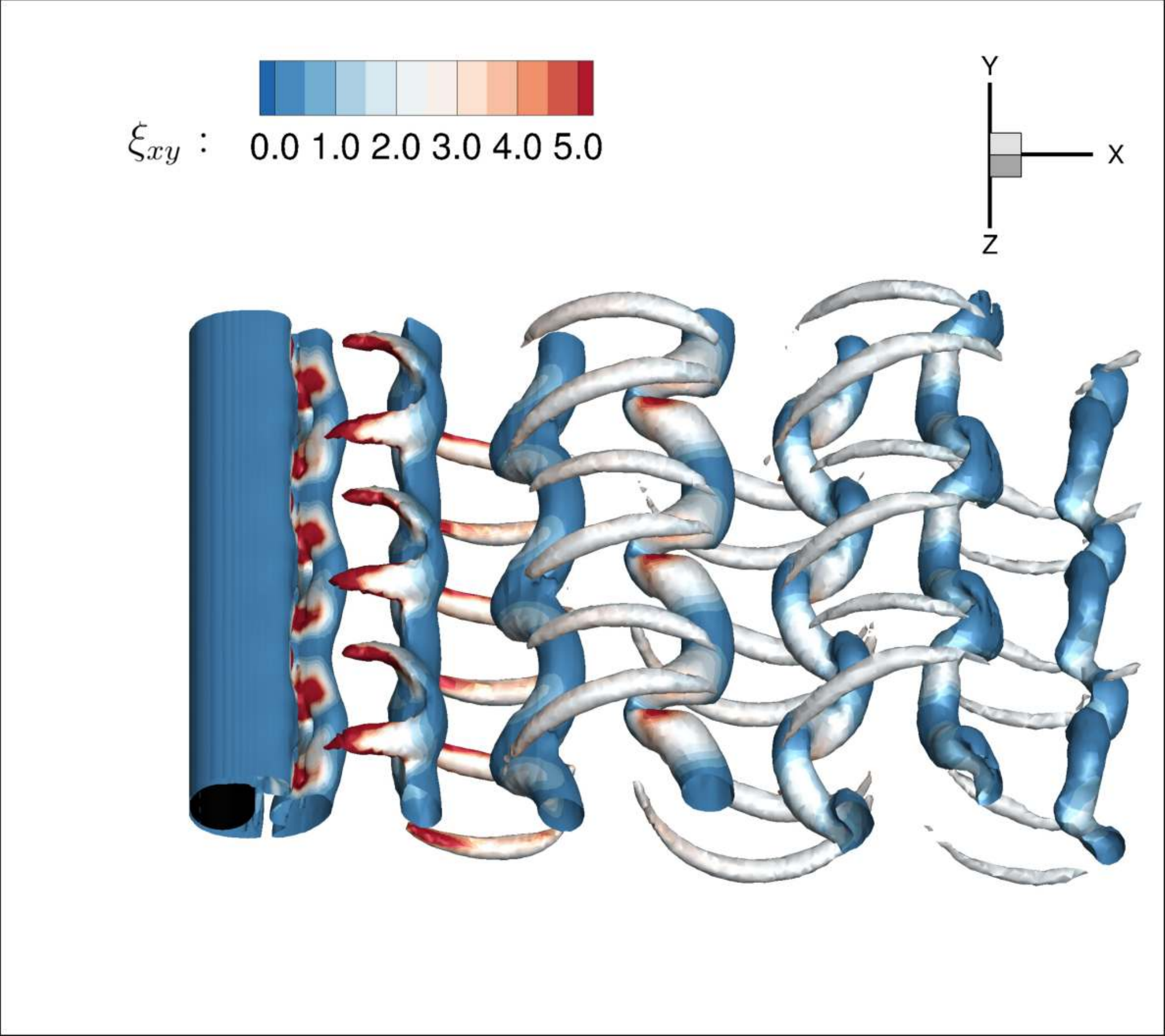}
		\caption{}
		\label{fig:en_re500_350}
	\end{subfigure}
	\caption{Wake behind a circular cylinder at $Re=500$, $L/D = 10$, $\Delta t=0.05$ visualized using Q criterion ($Q=0.5$)  and colored with secondary enstrophy ($\xi_{xy}$): (a) $tU/D = 1/4 T$; (b) $tU/D = 2/4 T$; (c) $tU/D = 3/4 T$ and (d) $tU/D = 4/4 T$. $\xi_{xy}$ is the secondary enstrophy to quantify three-dimensional streamwise vorticity clusters. The $x$-$y$ planar HSP appears at $tU/D = 3/4 T$ and $z/D=1.5, 5$ and $8.5$, at where $\xi_{xy}$ is intensified.}
	\label{fig:en_re500}
\end{figure}
\begin{figure} \centering
	\begin{subfigure}[b]{0.5\textwidth}	
		\centering
		\hspace{-25pt}\includegraphics[trim=0.1cm 0.1cm 0.1cm 0.1cm,scale=0.3,clip]{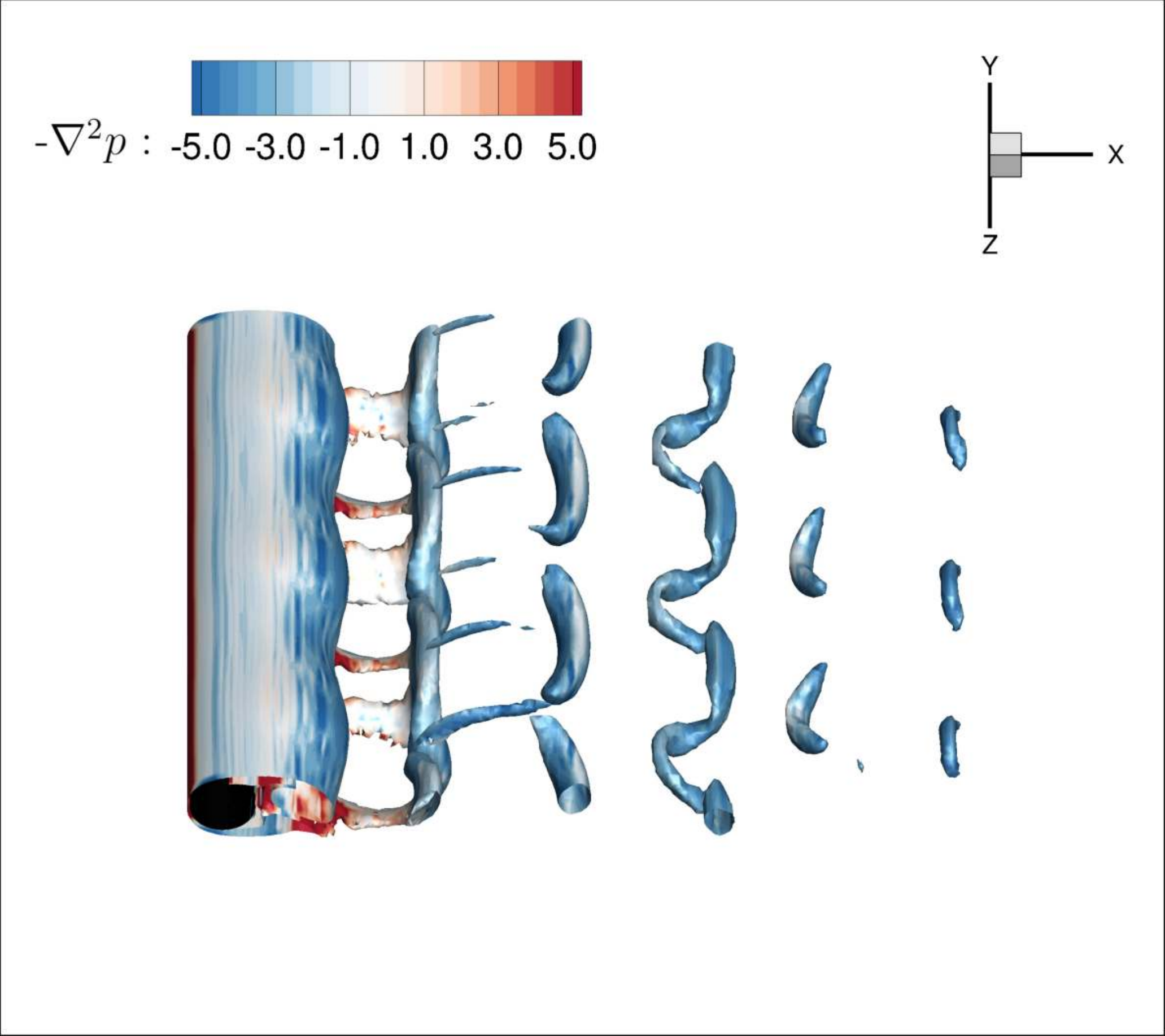}
		\caption{}
		\label{fig:vort_re500_275}
	\end{subfigure}%
	\begin{subfigure}[b]{0.5\textwidth}
		\centering
		\hspace{-25pt}\includegraphics[trim=0.1cm 0.1cm 0.1cm 0.1cm,scale=0.3,clip]{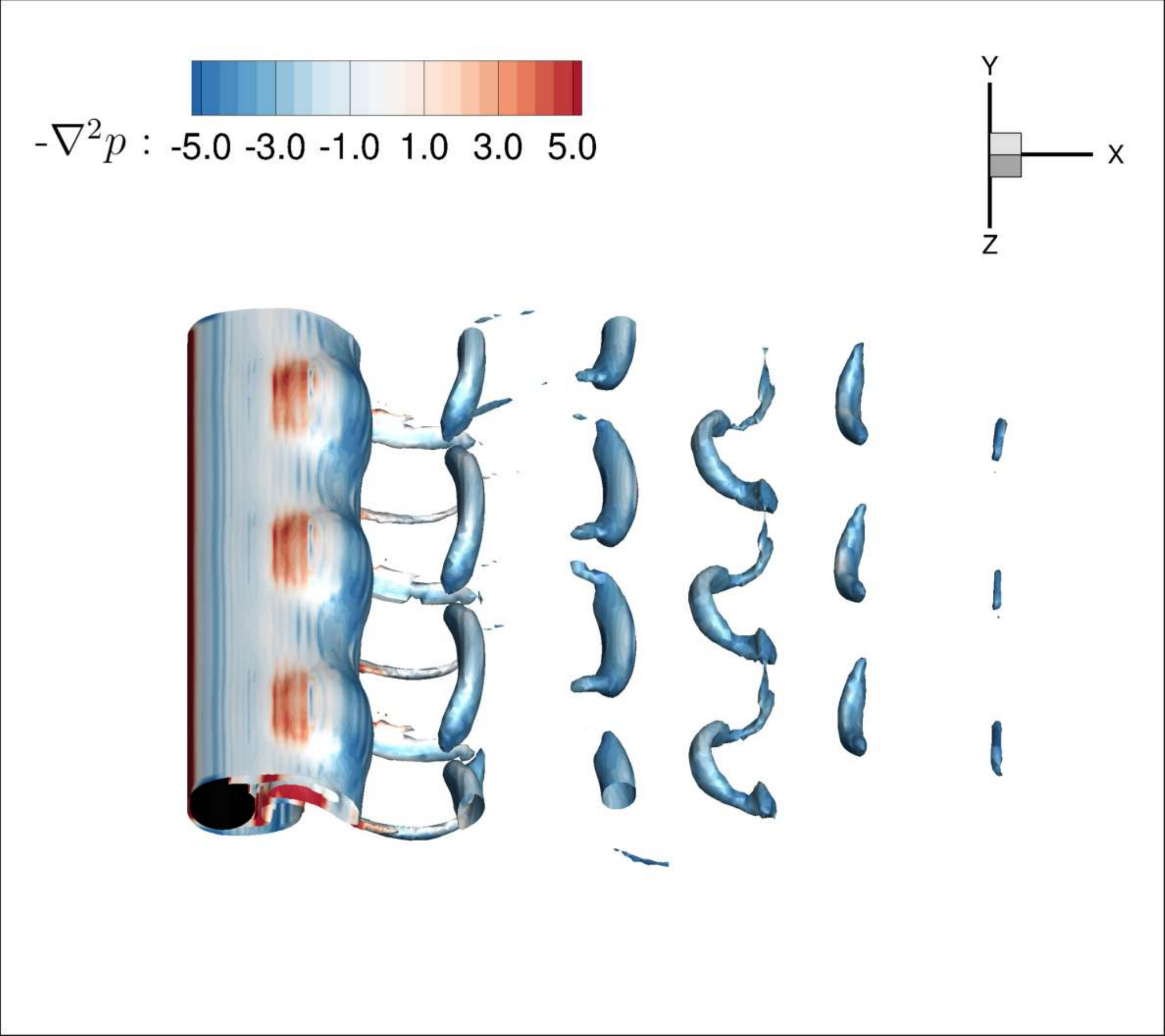}
		\caption{}
		\label{fig:vort_re500_300}
	\end{subfigure}
	\begin{subfigure}[b]{0.5\textwidth}	
		\centering
		\hspace{-25pt}\includegraphics[trim=0.1cm 0.1cm 0.1cm 0.1cm,scale=0.3,clip]{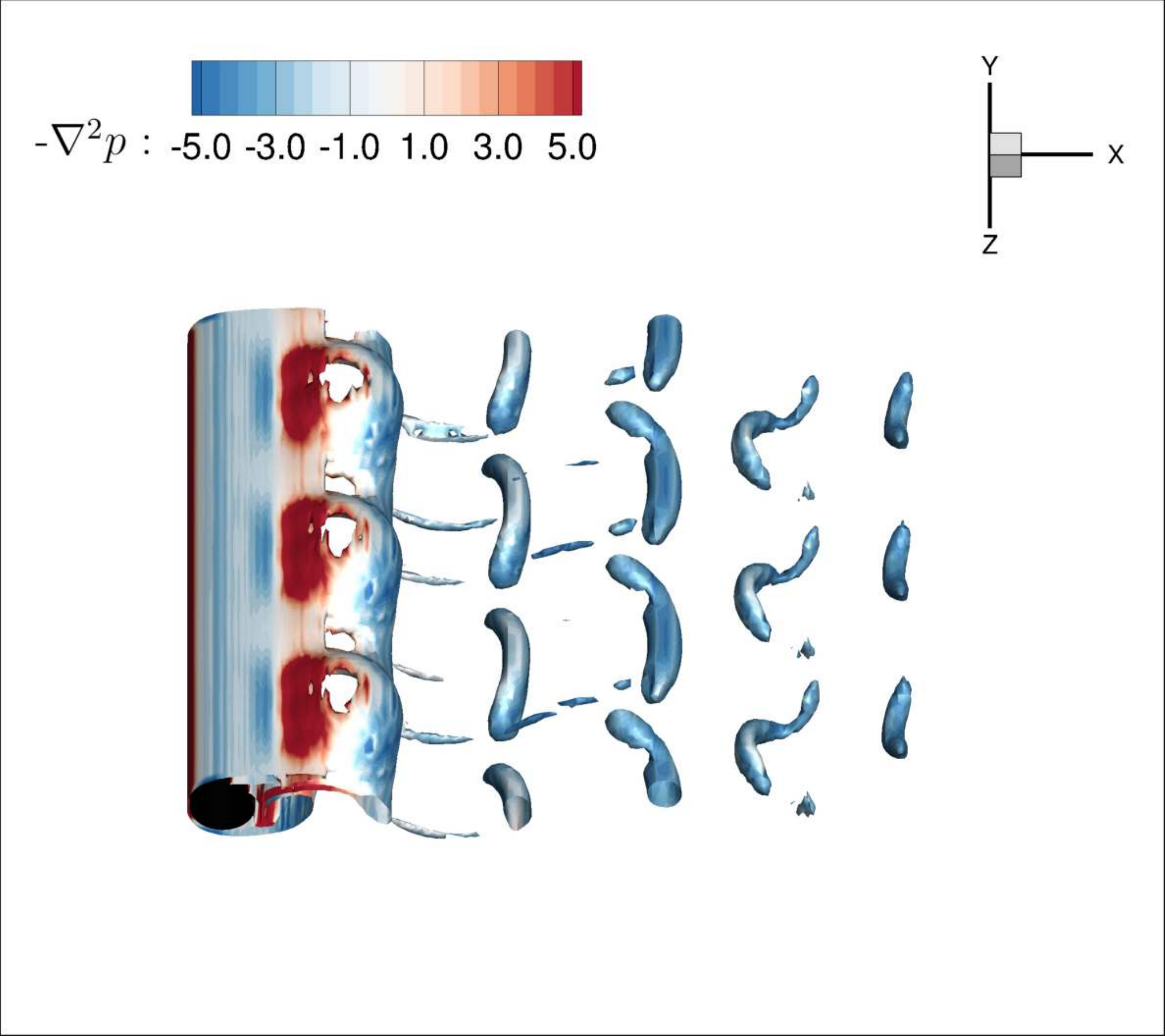}
		\caption{}
		\label{fig:vort_re500_325}
	\end{subfigure}%
	\begin{subfigure}[b]{0.5\textwidth}
		\centering
		\hspace{-25pt}\includegraphics[trim=0.1cm 0.1cm 0.1cm 0.1cm,scale=0.3,clip]{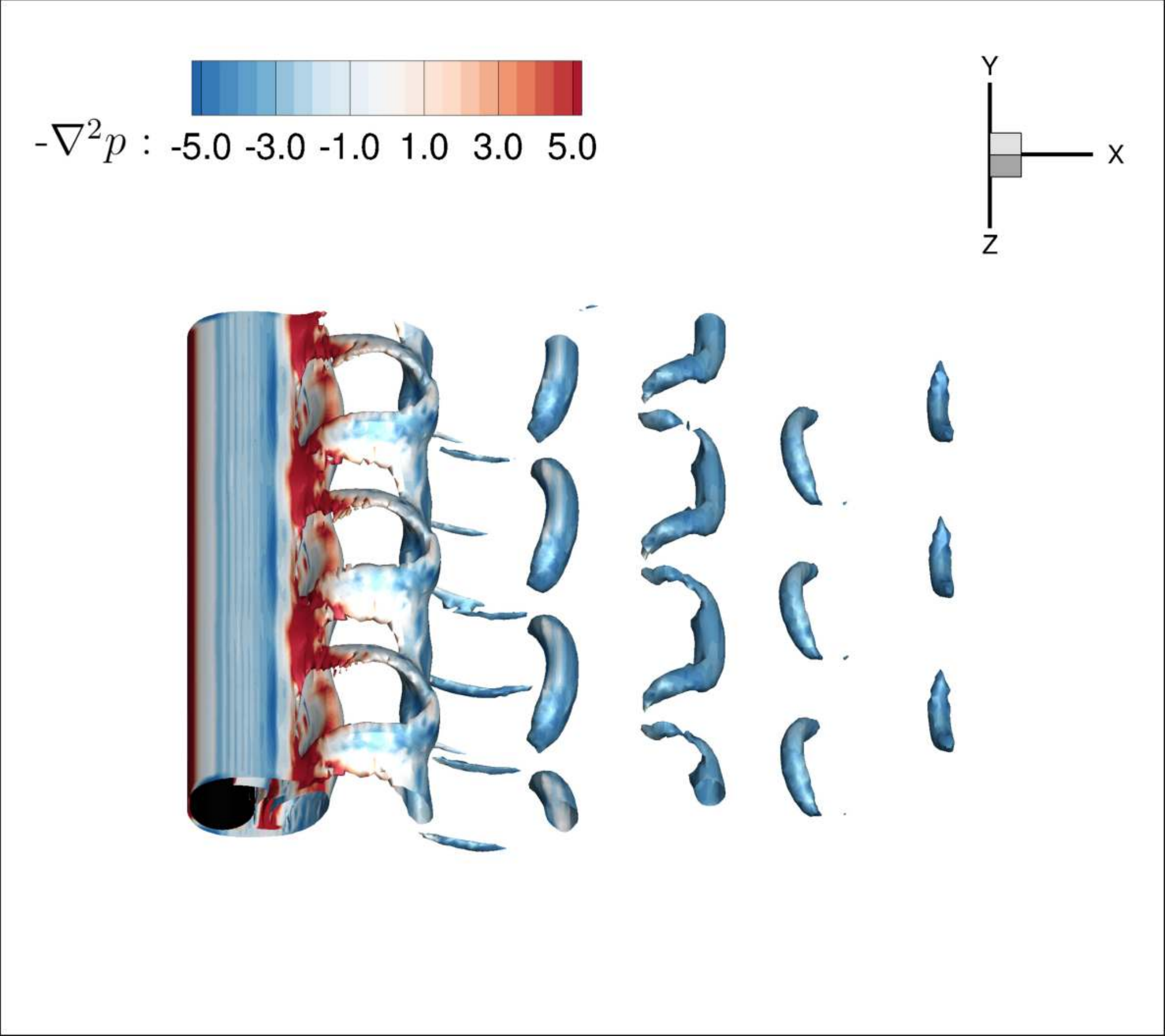}
		\caption{}
		\label{fig:vort_re500_350}
	\end{subfigure}
	\caption{Wake behind a circular cylinder at $Re=500$, $L/D = 10$, $\Delta t=0.05$ visualized using vorticity magnitude ($|\bm{\omega}|$) and colored with the divergence of pressure force: (a) $tU/D = 1/4 T$; (b) $tU/D = 2/4 T$; (c) $tU/D = 3/4 T$ and (d) $tU/D = 4/4 T$. The $x$-$y$ planar HSP appears at $3/4 T$ and $z/D = 1.5, 5$ and $8.5$, where is vorticity-free ($|\bm{\omega}| \approx 0$) and surrounded with high the fluid-flux divergence in all directions.}
	\label{fig:vort_re500}
\end{figure}

Similar to the findings in the isolated and side-by-side cylinders, the strong fluid-flux divergence and pressure can be observed around the HSP in Figure~\ref{fig:con_flux_re100e20} and~\ref{fig:con_p_re100e20}. An apparent adverse pressure gradient is imposed along the shedding shear layers in wake in Figure~\ref{fig:con_p_re100e20}, where a prominent streamline divergence is observed. The HSP is formed along the shear-layer interface in Figure~\ref{fig:con_wz_re100e20} and associated with relatively low fluid inertia in its neighborhood in Figure~\ref{fig:con_local_Re_re100e20}. a HSP is accompanied with a stagnant zone in the unsteady wake and cut the fluid kinetic energy supply of shedding shear layers in wake.

\subsection{Three dimensionality of a singular streamline hyperbolic/saddle critical point} \label{sec:three}
In Section~\ref{sec:prove}, it is analytically proven that a HSP is meta-stable for the balanced shear-layer interaction, becomes sensitive to noises at high Reynolds number and intrinsically linked with the net positive divergence of pressure force/fluid flux. The fluid is not only associated with a stagnant zone, but also inhibited from the rotation about its own axis at a HSP on its phase plane in wake. What's more, since the fluid pressure is a scalar, the net positive divergence of pressure force projects fluid flux in all directions, including the third-dimensional fluid flux. As result, the projection of a third-dimensional fluid flux is a very plausible path for the fluid flow. Although the analysis of a HSP can shed light upon the fluid three dimensionality, it is still insufficient to comprehensively explain the subtle mechanism of the chaotic turbulence and further investigations are in demand in future. In this section, we primarily focus on the investigation of the relationship between a singular streamline hyperbolic/saddle point in wake and fluid three dimensionality.    

To this end, the classical benchmark case of a uniform flow over a circular cylinder at $Re=500$ is chosen as an example. Figure~\ref{fig:Q_re500} shows the wake behind a cylinder within one vortex shedding period visualized using Q-criterion. It can be seen that the three-dimensional vortex braids start developing prominently from the primary vortex roll in Figure~\ref{fig:Q_re500_325} and shedding downstream in Figure~\ref{fig:Q_re500_350}. As we cut through a horseshoe vorticity cluster at $z/D=5.0$ in Figure~\ref{fig:Q_re500_325}, a $x$-$y$ planar HSP is situated right in the center of the horseshoe vorticity cluster at $\bm{x}/D = (1.393,0.586,5.0)$ in Figure~\ref{fig:re500_flux}. A high fluid-flux divergence is found accumulating around this HSP. The $x$-$z$ plane contours in Figure~\ref{fig:re500_w} show that this HSP becomes a source point of fluid flux in the third dimension (along $z$ axis), where the fluid is projecting into the third direction with high $z$-component velocity ($w$). This finding concretely support the conclusion drawn in Section~\ref{sec:ppe} and Section~\ref{sec:cylinder}. The fluid becomes stagnant, vorticity-free and induces the third-dimensional fluid flux at a HSP.
As documented by~\cite{williamson1996three}, the spanwise wavy modes of the primary vortex rolls (unstable mode A and mode B) are formed at different stages in flow transition. As long as there is any misalignment of HSP along the span of a cylinder, the prominent fluid flux can be observed projecting from a HSP into the third dimension, as demonstrated in Figure~\ref{fig:re500_w}. This finding provides the explicit evidence how a planar streamline hyperbolic point is intimately linked with the mechanism of fluid three dimensionality. The subtle linkage between a HSP and the fluid-flux divergence becomes one of the crucial factors to induce the fluid flux in the third dimension.

The iso-surfaces of the fluid-flux divergence is further visualized in Figure~\ref{fig:flux_re500} to further support our discussions. The spanwise discontinuity of the fluid-flux divergence is apparently observed between each HSP in Figure~\ref{fig:flux_re500_325}, at the moment when a HSP manifest its maximum $-\nabla^2 p$ value. These spanwise discontinuities or misalignment of fluid-flux divergence makes the three-dimensional fluid flux more significant. In Figure~\ref{fig:en_re500}, the secondary enstrophy ($\xi_{xy}$) is used to color the vorticity clusters. The value of $\xi_{xy}$ provides an indication of quantifying the intensity of three-dimensionality in flow. It is found that magnitude of $\xi_{xy}$ becomes tremendously intensified around each HSP. As time goes by, the HSP disappears in Figure~\ref{fig:flux_re500_350} and the distribution of fluid-flux divergence gradually becomes uniform along the primary vortex roll. This observation supports the proven relationship between a singular streamline hyperbolic point and fluid three dimensionality. As the vorticity clusters shed further downstream, the intensity of $\xi_{xy}$ is attenuated further.

In Section~\ref{sec:vonNeumman}, a HSP is proven to be meta-stable at balanced shear strains, where the magnitude of vorticity is significantly small $|\bm{\omega}| \approx 0$. A shear-layer interface or a stagnant point in flow field are typical places of this kind. This conclusion is evidently confirmed by the iso-surface of $|\bm{\omega}|$ in Figure~\ref{fig:vort_re500}. These figures depict the wake behind an isolated cylinder at $Re=500$ for one vortex shedding period. As the vorticity clusters evolve in near wake, Figure~\ref{fig:vort_re500_325} shows that a vorticity-free zone is formed right at each HSP. This vorticity-free zone inhibits the rotation of fluid along its axis and is accompanied with a large positive fluid-flux divergence, around where the fluid flux is projected away in all direction. As time goes by, these vorticity-free zones propagate, cut kinetic energy supply of shear layers, segregate the vorticity clusters and induce the three-dimensional horseshoe vorticity clusters in near wake, as shown in Figure~\ref{fig:vort_re500_350}.   

The findings in this section firmly support the conclusions drawn in Section~\ref{sec:prove} and Section~\ref{sec:cylinder}. These results show that a HSP is intimately associated with the net positive divergence of pressure force/fluid flux, which induces adverse pressure gradients in near wake. What's more, a HSP moves along the shear-layer interface, forms a stagnant vorticity-free zone and cut the kinetic energy supply of the shear layers. The fluid three dimensionality in near wake is tremendously amplified around a HSP with high values of $\xi_{xy}$. As long as an individual HSP appears in the flow field, a strong HSP is capable of being a source point projecting the fluid flux in all directions. These distinct characteristics of a HSP make itself a very unstable factor in the thre-dimensional flow. Especially, its behavior and influence on the flow along the normal direction to its phase plane become even more unpredictable in the fully turbulent flow at very high Reynolds numbers.  

\section{Concluding remarks} \label{sec:con}
The numerical stability of a HSP and its relationship with the net positive divergence of pressure force/fluid flux were rigorously investigated in this article. It was found that a HSP is merely meta-stable for the balanced shear-layer interaction, sensitive to Reynolds number and intrinsically related to the Poisson Pressure Equation/fluid-flux divergence. The value of fluid-flux divergence at a HSP surges together with the Reynolds number. Since a HSP is intrinsically associated with the net positive divergence of pressure force/fluid flux, it projects fluid flux in all directions, including the third dimension normal to its phase plane. As result, the strong third-dimensional fluid flux at a HSP was potentially formed at high Reynolds numbers, which was quantified by the value of secondary enstrophy ($\xi_{xy}$) in this investigation. Furthermore, it was also found that it formed a zone of relatively high pressure in wake, imposed adverse pressure gradients to the shedding shear layers in wake and deteriorated the near-wake stability. In multi-body systems, e.g., side-by-side and near-wall cylinders, the adjacent counterpart could exerted an intensive proximity interference, induced complicated and asymmetric shear-layer interactions and excited the instability of a HSP in wake. However, a direct interaction of the boundary layers at very small gap distances could suppressed/regulated the instability of a HSP in wake too. What's more, a HSP was found surrounded by a stagnant vorticity-free zone and moving along the shear-layer interfaces in wake. It inhibited the fluid curling about its own axis, cut the kinetic energy supply of shear layers in wake and developed streamwise braids in its vicinity.

\bibliographystyle{jfm}
\bibliography{jfm-instructions}

\end{document}